\documentstyle[11pt]{article}
\def\setfonts{%
\font\frbig=eufm10 scaled\magstephalf
\font\frscr=eufm8 scaled\magstephalf
\font\frscrscr=eufm8
\newfam\frfam
\textfont\frfam=\frbig
\scriptfont\frfam=\frscr
\scriptscriptfont\frfam=\frscrscr
\def\fr{\fam\frfam}

\font\openbig=msbm10 scaled\magstephalf
\font\openscr=msbm8 
\font\openscrscr=msbm8
\newfam\openfam
\textfont\openfam=\openbig
\scriptfont\openfam=\openscr
\scriptscriptfont\openfam=\openscrscr
\def\open{\fam\openfam}

\font\ssfbig=cmss10 scaled\magstephalf
\font\ssfscr=cmss8 
\font\ssfscrscr=cmss8
\newfam\ssffam
\textfont\ssffam=\ssfbig
\scriptfont\ssffam=\ssfscr
\scriptscriptfont\ssffam=\ssfscrscr
\def\ssf{\fam\ssffam}
}

\def\dimtwenty{2000}
\def\dimfifteen{1500}
\def\dimtwelve{1200}
\def\dimten{1000}

\makeatletter
\newdimen\normalarrayskip
\newdimen\minarrayskip
\normalarrayskip\baselineskip
\minarrayskip\jot
\newif\ifold \oldtrue \def\new{\oldfalse}
\def\arraymode{\ifold\relax\else\displaystyle\fi}

\def\@arrayskip{\ifold\baselineskip\z@\lineskip\z@
  \else
  \baselineskip\minarrayskip\lineskip2\minarrayskip\fi}
\def\@arrayclassz{\ifcase \@lastchclass \@acolampacol \or
\@ampacol \or \or \or \@addamp \or
 \@acolampacol \or \@firstampfalse \@acol \fi
\edef\@preamble{\@preamble
 \ifcase \@chnum
  \hfil$\relax\arraymode\@sharp$\hfil
  \or $\relax\arraymode\@sharp$\hfil
  \or \hfil$\relax\arraymode\@sharp$\fi}}
\def\@array[#1]#2{\setbox\@arstrutbox=\hbox{\vrule
  height\arraystretch \ht\strutbox
  depth\arraystretch \dp\strutbox
  width\z@}\@mkpream{#2}\edef\@preamble{\halign \noexpand\@halignto
\bgroup \tabskip\z@ \@arstrut \@preamble \tabskip\z@ \cr}%
\let\@startpbox\@@startpbox \let\@endpbox\@@endpbox
 \if #1t\vtop \else \if#1b\vbox \else \vcenter \fi\fi
 \bgroup \let\par\relax
 \let\@sharp##\let\protect\relax
 \@arrayskip\@preamble}
\@addtoreset{equation}{section}
\newcounter{@sc}
\newcounter{@scp}
\newcounter{@t}
\newlength{\@x}
\newlength{\@xa}
\newlength{\@xb}
\newlength{\@y}
\newlength{\@ya}
\newlength{\@yb}
\newsavebox{\@pt}

\def\bezier#1(#2,#3)(#4,#5)(#6,#7){\c@@sc#1\relax
  \c@@scp\c@@sc \advance\c@@scp\@ne
  \@xb #4\unitlength \advance\@xb -#2\unitlength \multiply\@xb \tw@
  \@xa #6\unitlength \advance\@xa -#2\unitlength
      \advance\@xa -\@xb \divide\@xa\c@@sc
  \@yb #5\unitlength \advance\@yb -#3\unitlength \multiply\@yb \tw@
  \@ya #7\unitlength \advance\@ya -#3\unitlength
      \advance\@ya -\@yb \divide\@ya\c@@sc
  \setbox\@pt\hbox{\vrule height\@halfwidth  depth\@halfwidth
   width\@wholewidth}\c@@t\z@
   \put(#2,#3){\@whilenum{\c@@t<\c@@scp}\do
      {\@x\c@@t\@xa \advance\@x\@xb \divide\@x\c@@sc \multiply\@x\c@@t
       \@y\c@@t\@ya \advance\@y\@yb \divide\@y\c@@sc \multiply\@y\c@@t
       \raise \@y \hbox to \z@{\hskip \@x\unhcopy\@pt\hss}%
       \advance\c@@t\@ne}}}
\makeatother

\def\theequation{\thesection.\arabic{equation}}

\setfonts

\def\lvm{\leavevmode\hbox to\parindent{\hfill}}
\def\req#1{(\ref{#1})}

\def\hw{highest-weight}
\def\conv{\ket{\rm conv}}

\def\ddsc{dense $\cG/\cQ$-descend\-ant}
\def\dQdsc{dense $\cQ$-descendant}
\def\dGdsc{dense $\cG$-descendant}

\def\BE{\begin{equation}}
\def\EE{\end{equation} }
\def\BA{\begin{array}} 
\def\EA{\end{array}}
\def\L{\left}
\def\R{\right}

\def\bar{\overline}
\def\frac#1#2{\mathchoice{{\textstyle{{#1}\over{#2}}}}{{#1\over#2}}%
  {{#1\over#2}}{{#1\over#2}}}
\def\ket#1{\mathchoice{%
{\left|{#1}\right\rangle}}{|{#1}\rangle}{|{#1}\rangle}{|{#1}\rangle}}
\def\kettop#1{\mathchoice{{\left|{#1}\right\rangle}_{\rm top}}%
  {|{#1}\rangle_{\rm top}}{|{#1}\rangle_{\rm top}}{|{#1}\rangle_{\rm top}}}
\def\ketM#1{\mathchoice{{\left|{#1}\right\rangle}}{|{#1}\rangle}%
  {|{#1}\rangle}{|{#1}\rangle}}
\def\ketSL#1{\left|{#1}\right\rangle_{\SL2}}

\def\longar{
\begin{picture}(60,3)
\put(1,2){\vector(1,0){58}}
\end{picture}
}

\def\N#1{N\!=\!#1}
\def\SL#1{s\ell(#1)}
\def\tSL#1{{\widehat{s\ell}}(#1)}
\def\tSSL#1#2{{\widehat{s\ell}}(#1|#2)}

\def\hell{\widehat{\ell}}

\def\mm{\cal}
\def\smm{\fr}

\def\mC{{\mm C}}
\def\mM{{\mm M}}

\def\mN{{\mm N}}
\def\mQ{{\mm Q}}

\def\mU{{\mm U}}
\def\mV{{\mm V}}

\def\smU{{\smm U}}
\def\smV{{\smm V}}

\def\half{{\textstyle{1\over2}}}

\def\tp{\widetilde{p}}

\def\cA{{\cal A}}

\def\cE{{\cal E}}

\def\cG{{\cal G}}
\def\cH{{\cal H}}

\def\cK{{\cal K}}
\def\cL{{\cal L}}

\def\cQ{{\cal Q}}

\def\cU{{\cal U}}

\def\cX{{\cal X}}

\def\oA{{\open A}}
\def\oB{{\open B}}
\def\oC{{\open C}}

\def\oN{{\open N}}
\def\oQ{{\open Q}}
\def\oX{{\open X}}
\def\oY{{\open Y}}
\def\oZ{{\open Z}}

\def\ctop{{\ssf c}}
\def\Ctop{{\ssf C}}
\def\htop{{\ssf h}}
\def\jplus{{\ssf j}^+}
\def\jminus{{\ssf j}^-}
\def\jtop{{\ssf j}}
\def\ttop{{\ssf t}}
\def\hplus{{\ssf h}^+}
\def\hminus{{\ssf h}^-}
\def\ellm{{\ssf l}}
\def\theell{{\ssf l}}
\def\hcc{{\ssf h}_{\rm cc}}
\def\lcc{{\ssf l}_{\rm cc}}
\def\ellch{{\ssf l}_{\rm ch}}

\def\hcm{{\ssf h}_{\rm cm}}
\def\ellcm{{\ssf l}_{\rm cm}}
\def\hpm{{\ssf h}^\pm}

\def\NPB{Nucl.\ Phys.\ B}
\def\PLB{Phys.\ Lett.\ B}
\def\MPLA{Mod.\ Phys.\ Lett.\ A}

\def\IJMPA{Int.\ J.\ Mod.\ Phys.\ A}

\newtheorem{lemma}{Lemma}[section]

\newtheorem{thm}[lemma]{Theorem}

\newtheorem{dfn}[lemma]{Definition}

\newtheorem{prop}[lemma]{Proposition}

{\par\medskip}

\newenvironment{prf}{\nopagebreak\par\smallskip
  \noindent{{\samepage\sc
      Proof.}}~\nolinebreak}{\nolinebreak$\Box$\par\medskip} 

\begin{document}
\hfuzz=1pt
\thispagestyle{empty}
\addtolength{\baselineskip}{1pt}

\setcounter{page}{1}

\begin{flushright}
  {\tt hep-th/9704111}
\end{flushright}

\thispagestyle{empty}

\begin{center}
  {\Large{\sc The Structure of Verma Modules over
      the\\[6pt] $N=2$ Superconformal Algebra}}\\[16pt]
  {\large A.~M.~Semikhatov and I.~Yu.~Tipunin}\\[4pt]
  {\small\sl Tamm Theory Division, Lebedev Physics
    Institute, Russian Academy of Sciences }
\end{center}

\addtolength{\baselineskip}{-2pt}

\noindent\hbox to.05\hsize{\hfill}
\parbox{.9\hsize}{\footnotesize We classify degeneration patterns of
  Verma modules over the $\N2$ superconformal algebra in two
  dimensions.  Explicit formulae are given for singular vectors that
  generate maximal submodules in each of the degenerate cases. The
  mappings between Verma modules defined by these singular vectors are
  {\it embeddings\/}; in particular, their compositions never vanish.
  As a by-product, we also obtain general formulae for $\N2$
  subsingular vectors.}

{\footnotesize
  \tableofcontents}

\addtolength{\baselineskip}{2pt}

\section{Introduction}\lvm
In this paper, we describe the structure of submodules and singular
vectors in Verma modules over the $\N2$ superconformal algebra in two
dimensions --- the $\N2$ supersymmetric extension of the Virasoro
algebra~\cite{[Ade]}. This algebra underlies the construction of $\N2$
strings~\cite{[Ade],[Marcus],[FT],[OV23]} (with its possible role in
the M-theory proposed in \cite{[KM]}), and, on the other hand, is
realized on the world-sheet of any non-critical string
theory~\cite{[GS2], [BLNW]}.  Other (non-exhaustive) references on the
$\N2$ superconformal algebra and $\N2$ models in conformal field
theory and string theory
are~\cite{[BFK],[EHy],[SS],[DvPYZ],[G],[IK],[KS],[EG],[Ga]}.

The $\N2$ algebra, however, hasn't been very privileged in several
respects, first of all because it is not an affine Lie algebra.  It
does not admit a root system enjoying all the properties of root
systems of affine Lie algebras, hence, in particular, there is no
canonical triangular decomposition. As a result, there is no canonical
way to impose `highest-weight'-type conditions on a vacuum vector
(hence, on singular vectors) in representations of the algebra. Trying
to follow the {\it formal\/} analogy with the case of affine Lie
algebras (e.g., $\tSL2$) and imitating the \hw{} conditions imposed
there leads to several complications, if not inconsistencies, with the
definition and properties of $\N2$ Verma modules. These complications
are related to the fact that there exist two different types of
Verma-like modules, and, while modules of one type can be submodules
of the other, the converse is not true. The definition of singular
vectors carried over from the affine Lie-algebra case does not
distinguish between the two types of submodules.

Among other facts pertaining to the $\N2$ algebra, let us note that
the different {\it sectors\/} of the algebra (the Neveu--Schwarz and
Ramond ones) are isomorphic, which is in contrast to the $\N1$ case.
This is due to the $\N2$ spectral flow~\cite{[SS]}. Thus, there is
`the only' $\N2$ algebra\footnote{With the exception of a somewhat
  exotic `twisted sector', where one of the fermions has half-integer,
  and the other, integer, modes, {\it which we do not touch upon in
    this paper}.} and its isomorphic images under the spectral flow.
However, the basis in the algebra can be chosen in different ways,
since the presence of the $U(1)$ current allows one to change the
energy-momentum tensor by the derivative of the current; the algebras
that appear in different contexts are in fact
isomorphic~\cite{[Ey],[W-top]} to one and the same $\N2$ algebra.
Finally, even the terminology used in the $\N2$ representation theory
does not appear to be unified, which may again be related to the fact
that the situation which is familiar from the affine Lie algebras does
not literally carry over to $\N2$.

\medskip

The object of our study is possible degenerations (reducibility
patterns) of $\N2$ Verma modules, i.e., the structure of submodules in
these modules.  This is more involved than in more familiar cases of
the Virasoro algebra and the standard Verma modules over the affine
algebra~$\tSL2$, due to two main reasons.  First, the $\N2$ algebra
has rank~3, which gives its modules more possibilities to degenerate.
Second, as we have already mentioned, there are two different types of
$\N2$ Verma-like modules that have to be distinguished clearly; we
follow refs.~\cite{[ST3],[FST]} in calling them the {\it
  topological\/} and {\it massive\/} Verma modules (in a different
terminology, the first ones are chiral, while the second ones are
tacitly understood to be `{\it the\/}' $\N2$ Verma modules).  The
topological Verma modules appearing as submodules are {\it twisted},
i.e., transformed by the spectral flow.  Ignoring the existence of two
types of modules and trying to describe degenerations of $\N2$ Verma
modules in terms of only massive Verma (sub)modules results in an
incorrect picture, e.g. apparent relations in Verma modules would then
seem to exist, in contradiction with the definition of {\it Verma\/}
modules.  In the embedding diagrams known in the literature,
similarly, some singular vectors appear to vanish when constructed in
a module built on another singular vector --- in which case one can
hardly talk about {\it embedding\/} diagrams.  The actual situation is
that topological Verma submodules may exist in massive Verma modules,
with one extra annihilation condition being imposed on the \hw{}
vector of such submodules. For example, submodules generated by the
so-called charged singular vectors~\cite{[BFK]} are {\it always\/}
(twisted) topological. That this fundamental fact about the charged
singular vectors has not been widely appreciated, is because the
nature of submodules is obscured when one employs singular vectors
defined using a formal analogy with the case of affine Lie algebras.
In fact, any mapping from a massive Verma module into a topological
Verma module necessarily has a kernel that contains another
topological Verma submodule, which makes a sequence of such mappings
look more like a BGG-resolution rather than an embedding diagram.
Another reason why literally copying the definition of singular
vectors from the affine Lie algebras complicates the analysis of $\N2$
modules is that using such singular vectors entails {\it
  sub\/}singular vectors.

Generally, when considering representations of algebras of rank
$\geq3$, one should take care of whether a given singular vector or a
set of singular vectors generate a {\it maximal\/} submodule.  That a
submodule \,$\mU_2$ generated from all singular vectors in a Verma
module \,$\mU$ is not maximal means that there exists a proper
submodule \,$\mU_1\neq\mU_2$ such that
\,$\mU_2\subset\mU_1\subset\mU$. Then, the quotient module
\,$\mU/\mU_2$ contains a submodule, which in the simplest case would
be generated from one or several singular vectors (otherwise, the
story repeats).  However, these vectors are {\it not\/} singular in
\,$\mU$, i.e., before taking the quotient with respect to \,$\mU_2$.
They are commonly known as subsingular vectors.

The question of whether or not singular vector(s) generate a maximal
submodule is unambiguous for the affine Lie algebras, where the root
system determines a fixed set of operators from the algebra that are
required to annihilate a state in order that it be a singular vector.
As we have already remarked, such annihilation conditions are not
defined uniquely in the $\N2$ case. Since the significance of singular
vectors consists in providing a description of submodules, one should
thus focus one's attention on the structure of submodules in $\N2$
Verma modules. As regards singular vectors, then, one has two options:
either to fix the convention that singular vectors satisfy {\it
  some\/} annihilation conditions (for instance, those copied
literally from the case of affine Lie algebras) and then to find a
system of sub-, subsub-, \ldots -singular vectors that `compensate'
for the failure of the chosen singular vectors to generate maximal
submodules; or to try to define singular vectors in such a way that
they generate maximal submodules, in which case the structure of
submodules would be described without introducing subsingular vectors.

In what follows, we present a regular way to single out and to
explicitly construct those vectors that generate maximal submodules in
$\N2$ Verma modules\footnote{\label{foot:new}To avoid
  misunderstanding, let us point out explicitly that we do not claim,
  of course, that any maximal submodule in any $\N2$ Verma module
  would be generated from one singular vector; this cannot be the case
  already for the sum of two submodules. What we are saying is that
  any maximal submodule is necessarily generated from an appropriate
  number of singular vectors that we work with in this paper. Note
  that this is not the case whenever subsingular vectors exist.}.
They turn out to satisfy `twisted' annihilation conditions, i.e. those
given by a spectral flow transform~\cite{[SS],[LVW]} of the
annihilation conditions imposed on the \hw{} vector in the module.
With this definition of singular vectors, subsingular vectors become
redundant.  However, in view of controversial statements that have
been made regarding `subsingular vectors in $\N2$ Verma
modules'~\cite{[Doerr2],[GRR],[EG]}, we will also show how our
description can be adapted to provide a systematic way to construct
the states in $\N2$ Verma modules that are subsingular vectors once
singular vectors are defined by the conventional, `untwisted',
annihilation conditions. The general picture that emerges in this way
is very simple and can be outlined as follows.

Recall that, in modules over a $\oZ\times\oZ$-graded algebra, any
vector that satisfies `\hw' conditions is a member of the family of
extremal states which make up an {\it extremal diagram\/}\footnote{In
  a context similar to that of the present paper, (diagrams of)
  extremal vectors were introduced in~\cite{[FS]}. Their usefulness in
  representation theory, which was pointed out in~\cite{[FS]}, has
  been demonstrated in~\cite{[ST2],[ST3],[S-sl21sing],[FST]}.}  (see
Eqs.~\req{topdiag} and ~\req{massdiagramdouble} for the topological
and the massive Verma modules, respectively).  For the $\N2$ algebra,
a given singular vector that we consider satisfies twisted \hw{}
conditions with a certain integral twist $\theta$; this vector belongs
to the extremal diagram that consists of states satisfying twisted
\hw{} conditions with all integral twists. That of the extremal states
which satisfies the conventional, `untwisted', \hw{} conditions is the
conventional singular vector.  Now, it is the properties of the
extremal diagram that are responsible for whether or not all of the
extremal states generate the same submodule.  Generically, it is
irrelevant which of the representatives of the extremal diagram is
singled out as `{\it the\/}' singular vector. In the degenerate cases,
however, there do exist preferred representatives that generate the
maximal possible submodule, while other extremal states generate a
smaller submodule.  Moreover, there exists a systematic way to divide
$\N2$ extremal diagrams into those vectors that do, and those that do
not, generate maximal submodules.

Using singular vectors that generate maximal submodules, it is not too
difficult to classify all possible degenerations of $\N2$ Verma
modules, since the structure of {\it submodules\/} is still relatively
simple. However, it may become quite complicated to describe the same
structure in terms of a restricted set of singular vectors that
satisfy zero-twist annihilation conditions (and then, necessarily, in
terms of subsingular vectors).  The upshot is that, in the degenerate
cases where several singular vectors exist in the module, their
zero-twist representatives may lie in the section of the extremal
diagram separated from the vectors generating the maximal submodule by
a state that satisfies stronger \hw{} conditions and, thus, is a \hw{}
vector in a twisted topological Verma submodule
(diagrams~\req{subsingtop}, \req{onesided}, and~\req{gap}).  Depending
on the relative positions of such topological singular vectors in the
extremal diagram, therefore, a single picture in terms of extremal
diagrams breaks into several cases in some of which the conventional
singular vectors do, while in others do not, generate maximal
submodules.

A careful analysis of the type of submodules in $\N2$ Verma modules is
also crucial for correctly describing one particular degeneration of
massive Verma modules where a massive Verma submodule is embedded into
the direct sum of two twisted topological Verma submodules.  In the
conventional terms, the situation is described either as the existence
of two linearly independent singular vectors with identical quantum
numbers~\cite{[Doerr2]} or as the existence of a singular vector and a
subsingular vector with identical quantum numbers (the latter case was
missed in the conventional approach).  In more invariant and in fact,
much simpler terms, both these cases are described uniformly, as the
existence of two singular vectors that satisfy twisted \hw{}
conditions (diagram \req{overlap}). It follows that linearly
independent singular vectors belong then to two twisted {\it
  topological\/} Verma submodules.

To summarize the situation with $\N2$ subsingular vectors, they are
superfluous when it comes to classifying degenerations of $\N2$ Verma
modules.  Instead, our strategy is as follows. Given a submodule in
the $\N2$ Verma module, we consider the entire extremal diagram that
includes the vectors from which that submodule is generated. On the
extremal diagram, then, we point out the states that generate the
maximal submodule. These states, which turn out to satisfy twisted
\hw{} conditions, are the singular vectors we work with in this paper.
Restricting oneself to those extremal states (the conventional
singular vectors) that fail to generate maximal submodules and
classifying the `compensating' subsingular vectors is then an exercise
in describing {\it the same\/} structure of submodules in much less
convenient~terms. However, in order to make contact with the problems
raised in the literature, we indicate in each of the degenerate
cases\footnote{With one exception, where the classification of
  subsingular vectors would be too long in view of a large number of
  different cases of relative positions of the extremal diagrams
  describing the relevant submodules; classifying the subsingular
  vectors then remains a straightforward, although lengthy and
  unnecessary, exercise.}  why and how the conventional singular
vectors fail to generate maximal submodules; we then explicitly
construct the corresponding subsingular vectors.

By choosing the singular vectors that satisfy twisted \hw{}
conditions, we sacrifice the formal similarity with the case of Ka\v
c--Moody algebras, yet in the end of the day one observes~\cite{[FST]}
that the structure of $\N2$ Verma modules is equivalent to the
structure of certain modules over the affine $\tSL2$ algebra: there is
a functor from the category of `relaxed' $\tSL2$ Verma modules
introduced in~\cite{[FST]} to $\N2$ Verma modules. Restricted to the
standard $\tSL2$ Verma modules, this functor gives the twisted
topological Verma modules over the $\N2$ algebra.  The $\tSL2$
singular vectors (which do generate maximal submodules) correspond
then precisely to the $\N2$ singular vectors that we consider in this
paper.  This gives an intrinsic relation (actually,
isomorphism~\cite{[FST]}) between affine $\tSL2$ singular vectors and
the $\N2$ singular vectors satisfying the twisted \hw{} conditions.
Note also that the issue of subsingular vectors is normally not
considered for the $\tSL2$ algebra; combined with the equivalence
proved in~\cite{[FST]}, this clearly signifies, once again, that $\N2$
subsingular vectors are but an artifact of adopting the `zero-twist'
definition for singular vectors (as we explain in some detail after
diagram~\req{subsingtop}).

\medskip

In what follows, we thus define and systematically refer to singular
vectors that satisfy twisted `\hw{}' conditions (see
Definitions~\ref{defsingtop} and~\ref{defsingmass}); the twist-zero
singular vectors will be referred to as the conventional, `untwisted',
singular vectors (as we explain below, these can also be characterized
as the top-level representatives of the extremal diagrams). As we have
already mentioned, the two essentially different types of $\N2$ Verma
modules are called the massive and the topological ones.
`Highest-weight' conditions will be used without quotation marks from
now on.  Whenever we talk about subsingular vectors, these will of
course be understood in the setting where one restricts oneself to the
conventional definition of singular vectors. `State' is synonymous to
`vector'. When applied to representations, the term `twisted' means
`transformed by the spectral flow'.

\medskip

Our main results are the classification of degenerations of $\N2$
Verma modules and the general construction of $\N2$ singular vectors.
We develop a systematic description of all possible degenerations of
$\N2$ Verma modules using the extremal diagrams. This allows us to
describe the structure of submodules without invoking subsingular
vectors.  The formalism that we develop for the $\N2$ algebra (see
also~\cite{[ST3]}) is, at the same time, a natural counterpart of the
construction of singular vectors of affine Lie algebras
(see~\cite{[MFF],[FST]}). To make contact with the issues discussed in
the literature, we also show how the properties of the extremal
diagrams determine whether or not the conventional singular vectors
generate maximal submodules; when they do not, we give the general
construction of the corresponding subsingular vectors that arise in
the conventional approach.

\medskip

In Section~2, we fix our notation and review the properties of the
$\N2$ algebra and singular vectors in its Verma modules.  In
Section~3, we describe all the degenerate cases where more than one
singular vectors exist.

\section{Preliminaries}
\subsection{The $\N2$ algebra, spectral flow transform, and Verma
  modules\label{sec:Prelim}}\lvm The $\N2$ superconformal algebra
$\cA$ is taken in this paper in the following basis (see~\cite{[ST3]}
for a discussion of the choice of various bases (and moddings) in the
algebra):
\begin{equation}\new
  \begin{array}{lclclcl}
    \L[\cL_m,\cL_n\R]&=&(m-n)\cL_{m+n}\,,&\qquad&[\cH_m,\cH_n]&=
    &\frac{\Ctop}{3}m\delta_{m+n,0}\,,\\

    \L[\cL_m,\cG_n\R]&=&(m-n)\cG_{m+n}\,,&\qquad&[\cH_m,\cG_n]&=&\cG_{m+n}\,,
    \\
    \L[\cL_m,\cQ_n\R]&=&-n\cQ_{m+n}\,,&\qquad&[\cH_m,\cQ_n]&=&-\cQ_{m+n}\,,\\

    \L[\cL_m,\cH_n\R]&=&\multicolumn{5}{l}{-n\cH_{m+n}+\frac{\Ctop}{6}(m^2+m)
      \delta_{m+n,0}\,,}\\
    \L\{\cG_m,\cQ_n\R\}&=&\multicolumn{5}{l}{2\cL_{m+n}-2n\cH_{m+n}+
      \frac{\Ctop}{3}(m^2+m)\delta_{m+n,0}\,,}
  \end{array}\qquad m,~n\in\oZ\,.
  \label{topalgebra}
\end{equation}
The generators $\cL_m$, $\cQ_m$, $\cH_m$, and $\cG_m$ are the Virasoro
generators, the BRST current, the $U(1)$ current, and the spin-2
fermionic current respectively. $\cH$ is not primary; instead, the
commutation relations for the Virasoro generators are centreless.  The
element $\Ctop$ is central. Since it is diagonalizable in any
representations (at least in all those that we are going to consider),
we do not distinguish between $\Ctop$ and a number $\ctop\in\oC$,
which it will be convenient to parametrize as
\begin{equation}
  \ctop=3\,{t-2\over t}
  \label{ctop}
\end{equation}
with $t\in\oC\setminus\{0\}$.

The spectral flow transform~\cite{[SS],[LVW]} produces isomorphic
images of the algebra $\cA$.  When applied to the generators of
\req{topalgebra} it acts as
\begin{equation}
  {\cal U}_\theta:\new
  \begin{array}{rclcrcl}
    \cL_n&\mapsto&\cL_n+\theta\cH_n+\frac{\ctop}{6}(\theta^2+\theta)
    \delta_{n,0}\,,&{}&
    \cH_n&\mapsto&\cH_n+\frac{\ctop}{3}\theta\delta_{n,0}\,,\\
    \cQ_n&\mapsto&\cQ_{n-\theta}\,,&{}&\cG_n&\mapsto&\cG_{n+\theta}\,.
  \end{array}
  \label{U}
\end{equation}
For any $\theta\in\oC$, this gives the mapping
$\cU_\theta:\cA\to\cA_\theta$ of the $\N2$ algebra $\cA\equiv\cA_0$ to
an isomorphic algebra $\cA_\theta$, whose generators $\cL^\theta_n$,
$\cQ^\theta_n$, $\cH^\theta_n$, and $\cG^\theta_n$ (where
$\cX^\theta_n=\cU_\theta(\cX_n)$) satisfy the same relations as those
with $\theta=0$.  The family $\cA_\theta$ includes the Neveu--Schwarz
and Ramond $\N2$ algebras, as well as the algebras in which the
fermion modes range over $\pm\theta+\oZ$.  Spectral flow is an {\it
  auto\/}morphism when~$\theta\in\oZ$.

Next, consider Verma modules over the $\N2$ algebra.  As already
mentioned in the Introduction, there are two different types of $\N2$
Verma modules, the topological\footnote{The name has to do with the
  fact that the \hw{} vectors in these modules correspond to primary
  states existing when the $\N2$ algebra is viewed as the topological
  conformal algebra~\cite{[Ey],[W-top]}.}  and the massive ones. Since
each of these can be `twisted' by the spectral flow, we give the
definitions of twisted modules, the `untwisted' ones being recovered
by setting the twist parameter $\theta=0$. An important point,
however, is that submodules of a given `untwisted' module can be the
twisted modules (which is the case with submodules in topological
Verma modules and also with submodules determined by the `charged'
singular vectors).
\begin{dfn}  A vector satisfying the \hw{}
  conditions\,\footnote{Here and henceforth, $\oN=1,2,\ldots$, while
    $\oN_0=0, 1,2,\ldots$.}
  \begin{equation}
    \new\begin{array}{rclcrcl} \cL_m\ket{h,t;\theta}_{\rm top}&=&0\,,
      \quad
      m\geq1\,,\quad& \cQ_\lambda\ket{h,t;\theta}_{\rm top}&=&0\,,
      &\lambda\in-\theta+\oN_0\\
      \cH_m\ket{h,t;\theta}_{\rm top}&=&0\,,\quad m\geq1\,,&
      \cG_\nu\ket{h,t;\theta}_{\rm top}&=&0\,,&\nu\in\theta+\oN_0
    \end{array}\quad\theta\in\oZ\,,
    \label{twistedtophw}
  \end{equation}
  with the Cartan generators having the following eigenvalues
  \begin{equation}\new
    \begin{array}{rcl}
      (\cH_0+\frac{\ctop}{3}\theta)\,\ket{h,t;\theta}_{\rm
        top}&=&
      h\,\ket{h,t;\theta}_{\rm top}\,,\\
      (\cL_0+\theta\cH_0+\frac{\ctop}{6}(\theta^2+\theta))
      \,\ket{h,t;\theta}_{\rm top}&=&0
    \end{array}
    \label{Cartantheta}
  \end{equation}
  is called the twisted topological \hw{} state.
  Conditions~\req{twistedtophw} are called the twisted topological
  \hw\ conditions.
\end{dfn}
\begin{dfn}
  The twisted topological Verma module $\smV_{h,t;\theta}$ is freely
  generated from a twisted topological \hw{} state
  $\ket{h,t;\theta}_{\rm top}$ by
  $$
  \cL_{-m}\,,\ m\in\oN\,,\qquad \cH_{-m}\,,\ m\in\oN\,,\qquad
  \cQ_{-m-\theta}\,,\ m\in\oN\,, \qquad \cG_{-m+\theta}\,,\ m\in\oN\,.
  $$
\end{dfn}
We write $\ket{h,t}_{\rm top}\equiv\ket{h,t;0}_{\rm top}$ and
$\mV_{h,t}\equiv\smV_{h,t;0}$.

$\N2$ modules are graded with respect to $\cH_0$ (the charge) and
$\cL_0$ (the level). {\it Extremal vectors\/} in $\N2$ modules are
those having the minimal level for a fixed $\cH_0$-charge.
Associating a rectangular lattice with the bigrading, we have that
increasing the $\cH_0$-grade by 1 corresponds to shifting to the
neighbouring site on the left, while increasing the level corresponds
to moving down. The extremal vectors separate the lattice into those
sites that are occupied by at least one element of the module and
those that are not.

The extremal diagram of a topological Verma module reads (in the
$\theta=0$ case for simplicity)
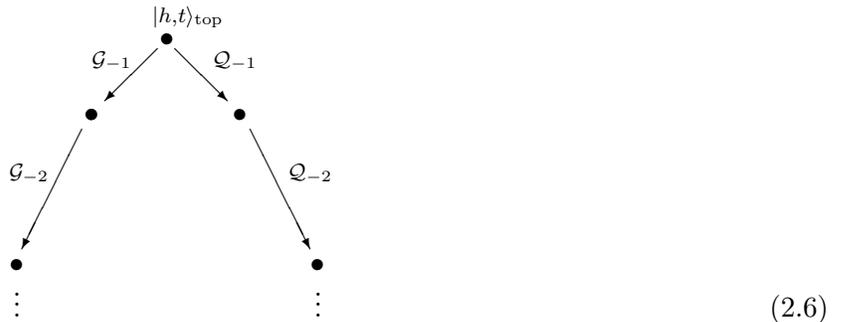
\begin{equation} \unitlength=1.00mm
  \begin{picture}(140,42)
    \put(50.00,6.00){
      \put(00.00,00.00){$\bullet$}
      \put(10.00,20.00){$\bullet$}
      \put(10.00,20.00){$\bullet$}
      \put(20.00,30.00){$\bullet$}
      \put(29.70,20.00){$\bullet$}
      \put(40.00,00.00){$\bullet$}
      \put(9.70,19.00){\vector(-1,-2){8}}
      \put(19.70,29.70){\vector(-1,-1){7}}
      \put(22.00,29.70){\vector(1,-1){7}}
      \put(32.00,19.00){\vector(1,-2){8}}
      \put(00.00,13.00){${}_{\cG_{-2}}$}
      \put(11.00,28.00){${}_{\cG_{-1}}$}
      \put(27.00,28.00){${}_{\cQ_{-1}}$}
      \put(37.00,13.00){${}_{\cQ_{-2}}$}
      \put(19.00,34.00){${}_{\ket{h,t}_{\rm top}}$}
      \put(00.50,-06.00){$\vdots$}
      \put(40.50,-06.00){$\vdots$}
      }
  \end{picture}
  \label{topdiag}
\end{equation}
Then, {\it all\/} the states in the module are inside the `parabola',
while none of the states from the module are associated with the
outside part of the plane. A characteristic feature of extremal
diagrams of topological Verma modules is the existence of a state that
satisfies stronger \hw{} conditions than the other states in the
diagram. Geometrically, this is a `cusp' point for the following
reasons. Assigning grade $-n$ to $\cQ_n$ and grade $n$ to $\cG_n$, we
see that every two adjacent arrows in the diagram represent the
operators whose grades differ by~1, except at the cusp, where they
differ by~2.  As we are going to see momentarily, the extremal
diagrams of submodules in a (twisted) topological Verma module have
`cusps' as well, these `cusp' points being the {\it topological
  singular vectors\/}:
\begin{dfn}\label{defsingtop}
  A topological singular vector in the (twisted) topological Verma
  module $\smV$ is any element of $\smV$ that is not proportional to
  the \hw\ vector and satisfies twisted topological \hw{} conditions
  (i.e., is annihilated by the operators $\cL_m$, $\cH_m$, $m\geq1$, \
  $\cQ_\lambda$, $\lambda\in-\theta+\oN_0$, and $\cG_\nu$,
  $\nu=\theta+\oN_0$ with some $\theta\in\oZ$).
\end{dfn}
The point is that the twist parameter $\theta$ that enters the \hw{}
conditions satisfied by the topological singular vector may be
different from the twist parameter of the module. One readily shows,
of course, that acting with the $\N2$ generators on a topological
singular vector defined in this way generates a {\it submodule\/}.

The next statement follows from the results of~\cite{[FST]}:
\begin{thm}[\cite{[FST]}]\label{structop:thm}
  Any submodule of a (twisted) topological Verma module is generated
  from either one or two topological singular vectors.
\end{thm}
This is directly parallel to the situation encountered in affine
$\SL2$ Verma modules --- which, in fact, is the statement
of~\cite{[FST]}, where a functor was constructed that maps
$\tSL2$-Verma modules to twisted topological Verma modules.  The
morphisms in a Verma modules category are singular vectors.  The
functor maps singular vectors in a $\tSL2$-Verma module to {\it
  topological\/} singular vectors and, thus, the assertion of the
Theorem follows from the well known facts in the theory of
$\tSL2$-Verma modules.  Thus, a maximal submodule of a topological
Verma module is either a twisted topological Verma module or a sum
({\it not\/} a direct one, of course) of two twisted topological Verma
modules. 
In what follows, we call a submodule {\it primitive\/} if it is not a
sum of two or more submodules.

\medskip

Next, consider the massive $\N2$ Verma modules.
\begin{dfn}
  A twisted massive Verma module $\smU_{h,\ell,t;\theta}$ is freely
  generated from a {\it twisted massive \hw{} vector\/}
  $\ket{h,\ell,t;\theta}$ by the generators
  \begin{equation}
    \cL_{-m}\,,~m\in\oN\,,\qquad \cH_{-m}\,,~m\in\oN\,,\qquad
    \cQ_{-\theta-m}\,,~m\in\oN_0\,,\qquad \cG_{\theta-m}\,,~m\in\oN\,,
    \label{verma}
  \end{equation}
  The twisted massive \hw{} vector $\ket{h,\ell,t;\theta}$ satisfies
  the following set of highest-weight conditions:
  \begin{equation}\new
    \begin{array}{rcl}
      \cQ_{-\theta+m+1}\,\ket{h,\ell,t;\theta}\kern-4pt&=&
      \kern-4pt\cG_{\theta+m}\,
      \ket{h,\ell,t;\theta}= \cL_{m+1}\,\ket{h,\ell,t;\theta}=
      \cH_{m+1}\,\ket{h,\ell,t;\theta}=0\,,~m\in\oN_0\,,\kern-40pt\\
      (\cH_0+\frac{\ctop}{3}\theta)\,\ket{h,\ell,t;\theta}\kern-4pt&=&
      \kern-4pt
      h\,\ket{h,\ell,t;\theta}\,,\\
      (\cL_0+\theta\cH_0+\frac{\ctop}{6}(\theta^2+\theta))\,
      \ket{h,\ell,t;\theta}
      \kern-4pt&=&\kern-4pt \ell\,\ket{h,\ell,t;\theta}\,.
    \end{array}
    \label{masshw}
  \end{equation}
\end{dfn}
Equations~\req{masshw} will be referred to as the {\it twisted massive
  \hw{} conditions}.  It is understood that the twisted massive \hw{}
vector does not satisfy the twisted {\it topological\/} \hw{}
conditions (i.e., $\cQ_{-\theta}\ket{h,\ell,t;\theta}\neq0$).  The
ordinary non-twisted case is obtained by setting $\theta=0$.  We
identify $\ket{h,\ell,t}\equiv\ket{h,\ell,t;0}$ and
$\mU_{h,\ell,t}\equiv\smU_{h,\ell,t;0}$.
When we say that a state in a Verma module satisfies (twisted)
massive \hw{} conditions, we will mean primarily the annihilation
conditions from~\req{masshw}.

An important property of the above definition is expressed by the
following Lemma, which underlies all the subsequent analysis.  The
Lemma (which follows by a straightforward calculation in the universal
enveloping algebra) is almost trivial, however we formulate it
explicitly because of its wide use in what follows.  Although we will
not refer to the Lemma explicitly, the reader should keep in mind that
it is implicit in almost all of our constructions.
\begin{lemma}
  If a state $\ket{\theta'}$ in a (twisted) massive Verma module
  satisfies the annihilation conditions~\req{masshw} with the
  parameter~$\theta$ equal to~$\theta'$ then the
  states~$\cG_{\theta'-N}\ldots\cG_{\theta'-1}\ket{\theta'}$,
  $N\geq1$, and~$\cQ_{-\theta'-N}\ldots\cQ_{-\theta'}\ket{\theta'}$,
  $N\geq0$, satisfy annihilation conditions~\req{masshw} with the
  parameter~$\theta$ equal to~$\theta'-N$ and~$\theta'+N+1$
  respectively.
\end{lemma}

The states referred to in the Lemma fill out the extremal diagram of
the massive Verma module. For $\theta=0$, it reads as
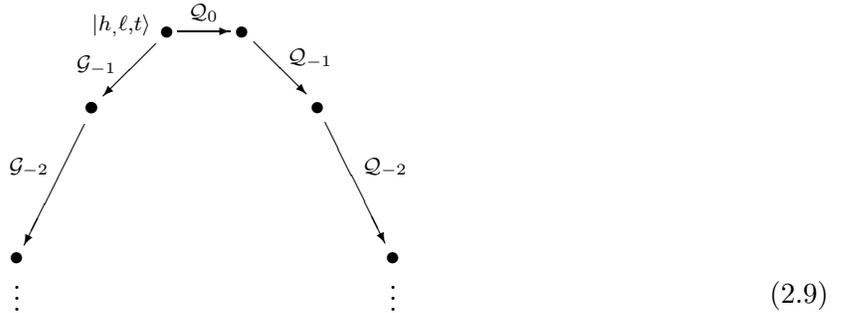
\begin{equation}
  \unitlength=1.00mm
  \begin{picture}(140,40)
    \put(50.00,05.00){
      \put(00.00,00.00){$\bullet$}
      \put(10.00,20.00){$\bullet$}
      \put(10.00,20.00){$\bullet$}
      \put(20.00,30.00){$\bullet$}
      \put(30.00,30.00){$\bullet$}
      \put(40.00,20.00){$\bullet$}
      \put(50.00,00.00){$\bullet$}
      \put(10.00,18.70){\vector(-1,-2){8}}
      \put(19.50,29.50){\vector(-1,-1){7}}
      \put(22.40,31.10){\vector(1,0){7}}
      \put(32.50,29.80){\vector(1,-1){7}}
      \put(42.00,19.00){\vector(1,-2){8}}
      \put(00.00,13.00){${}_{\cG_{-2}}$}
      \put(09.00,26.50){${}_{\cG_{-1}}$}
      \put(23.90,33.50){${}_{\cQ_{0}}$}
      \put(37.00,27.50){${}_{\cQ_{-1}}$}
      \put(47.00,13.00){${}_{\cQ_{-2}}$}
      \put(11.00,32.00){${}_{\ket{h_,\ell,t}}$}
      \put(00.50,-06.00){$\vdots$}
      \put(50.50,-06.00){$\vdots$}
      }
  \end{picture}
  \label{massdiagramdouble}
\end{equation}
While $\ket{h,\ell,t}$ satisfies the `untwisted' \hw{} conditions
($\theta=0$ in \req{masshw}), the Lemma tells us that the other states
in the extremal diagram satisfy twisted massive \hw{} conditions with
all~$\theta\in\oZ$.  The massive \hw{} vector must not be a `cusp
point' (i.e., it should not satisfy topological \hw{} conditions),
however other cusp points may appear in the diagram depending on the
\hw{} parameters~$(h,\ell,t)$. Whenever this happens, there is a
twisted topological submodule in the massive Verma module.  The
singular vectors appearing in the extremal diagrams of massive Verma
modules are called ``charged'' for historical reasons~\cite{[BFK]}.
\begin{dfn}
  The charged singular vector in a massive Verma module $\mU$ is any
  vector that satisfies twisted topological \hw{}
  conditions~\req{twistedtophw} (with whatever $\theta\in\oZ$) and
  belongs to the extremal diagram of the module.
\end{dfn}
An example of a charged singular vector is given in the following
diagram, where the twisted topological \hw{}
conditions~\req{twistedtophw} with $\theta=2$ are satisfied by the
extremal state at the point $C$:
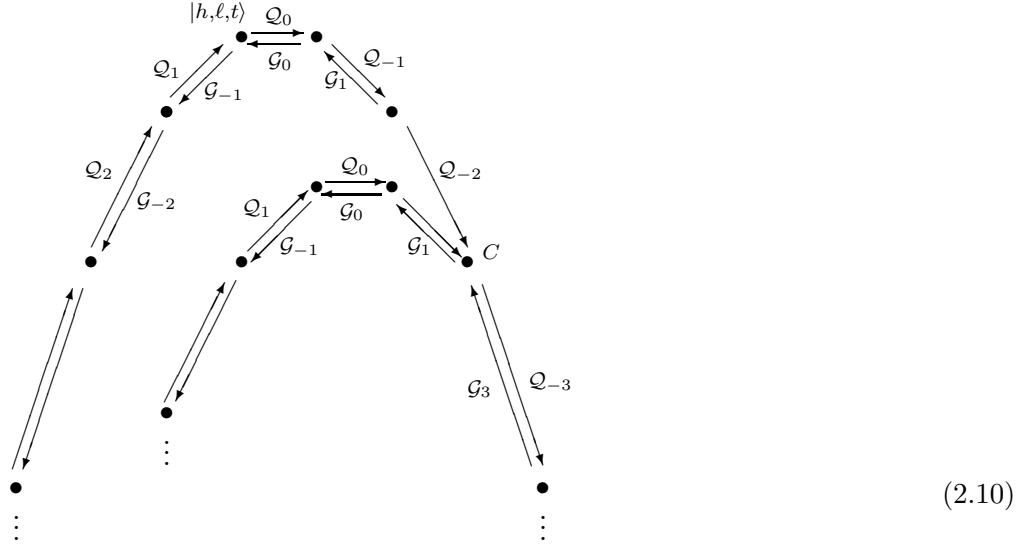
\begin{equation}
  \unitlength=1.00mm
  \begin{picture}(140,67)
    \put(30.00, 01.00){
      \put(00.00,00.00){$\bullet$}
      \put(10.00,30.00){$\bullet$}
      \put(20.00,50.00){$\bullet$}
      \put(20.00,50.00){$\bullet$}
      \put(30.00,60.00){$\bullet$}
      \put(40.00,60.00){$\bullet$}
      \put(50.00,50.00){$\bullet$}
      \put(60.00,30.00){$\bullet$}
      \put(70.00,00.00){$\bullet$}
      \put(50.00,40.00){$\bullet$}
      \put(40.00,40.00){$\bullet$}
      \put(30.00,30.00){$\bullet$}
      \put(20.00,10.00){$\bullet$}
      \put(11.00,33.00){\vector(1,2){8}}
      \put(20.50,48.50){\vector(-1,-2){8}}
      \put(00.50,03.50){\vector(1,3){8}}
      \put(09.90,27.50){\vector(-1,-3){8}}
      \put(21.50,53.00){\vector(1,1){7}}
      \put(29.70,59.00){\vector(-1,-1){7}}
      \put(32.20,61.50){\vector(1,0){7}}
      \put(38.70,59.95){\vector(-1,0){7}}
      \put(43.00,60.00){\vector(1,-1){7}}
      \put(49.00,52.00){\vector(-1,1){7}}
      \put(53.00,49.00){\vector(1,-2){8}}
      \put(63.00,28.00){\vector(1,-3){8}}
      \put(69.50,03.80){\vector(-1,3){8}}
      \put(59.30,31.00){\vector(-1,1){7.7}}
      \put(52.50,39.40){\vector(1,-1){7.7}}
      \put(42.10,41.60){\vector(1,0){8}}
      \put(49.55,39.98){\vector(-1,0){8}}
      \put(31.80,32.50){\vector(1,1){8}}
      \put(40.20,39.00){\vector(-1,-1){8}}
      \put(30.20,28.50){\vector(-1,-2){8}}
      \put(20.70,12.30){\vector(1,2){8}}
      \put(10.00,43.00){${}_{\cQ_2}$}
      \put(17.00,37.00){${}^{\cG_{-2}}$}
      \put(19.00,56.50){${}_{\cQ_{1}}$}
      \put(26.00,51.50){${}^{\cG_{-1}}$}
      \put(33.90,63.50){${}_{\cQ_{0}}$}
      \put(34.50,55.50){${}^{\cG_{0}}$}
      \put(47.00,58.00){${}_{\cQ_{-1}}$}
      \put(42.00,53.00){${}^{\cG_{1}}$}
      \put(57.00,43.00){${}_{\cQ_{-2}}$}
      \put(69.00,15.00){${}_{\cQ_{-3}}$}
      \put(61.00,12.00){${}^{\cG_{3}}$}
      \put(53.00,31.00){${}^{\cG_{1}}$}
      \put(44.00,35.50){${}^{\cG_{0}}$}
      \put(44.00,43.50){${}_{\cQ_{0}}$}
      \put(31.00,38.00){${}_{\cQ_{1}}$}
      \put(36.00,31.00){${}^{\cG_{-1}}$}
      \put(24.00,62.00){${}^{\ket{h,\ell,t}}$}
      \put(63.00,30.00){${}^{C}$}
      \put(00.50,-06.00){$\vdots$}
      \put(70.50,-06.00){$\vdots$}
      \put(20.50,04.00){$\vdots$}
      }
  \end{picture}
  \label{branchdiag}
\end{equation}

\bigskip

\noindent
As a result, no operator inverts the action of $\cQ_{-2}$, while each
of the other arrows is inverted {\it up to a scalar factor\/} by
acting with the opposite mode of the other fermion.  Thus, the
extremal diagram branches at the `topological points', and the crucial
fact is that, once we are on the inner parabola, we can never leave
it: none of the operators from the $\N2$ algebra map onto the
remaining part of the big parabola from the small one, or, in other
words, the inner diagram in corresponds to an $\N2$~{\it submodule}.
The general construction for the charged singular vectors is already
obvious from the above remarks, and it will be given in
Eqs.~\req{ECh}.

The `topological' nature of submodules generated from charged singular
vectors can be concealed if one allows submodules to be generated only
from the conventional singular vectors, i.e.\ those that satisfy
precisely the same \hw{} conditions as the \hw{} conditions satisfied
by the \hw{} vector of the module.  These conventional singular
vectors do not in general coincide with the `cusp' of the extremal
diagram of the topological submodule.  However, it is the existence of
this `cusp' that determines several crucial properties of the
submodule.

\medskip

Besides (twisted) topological Verma submodules, massive Verma modules
may have submodules of the same, `massive', type. These have to be
clearly distinguished from the topological ones.  The following
definition will allow us to single out massive Verma modules.
\begin{dfn}
  Let $\ket{Y}$ be a state in an $\N2$ Verma module that satisfies
  twisted massive \hw{} conditions with some $\theta\in\oZ$. Then
  $\ket{X}$ is said to be a \ddsc{} of~$\ket{Y}$ if either \
  $\ket{X}=\alpha\,\cG_{\theta-N}\,\ldots\,\cG_{\theta-1}\,\ket{Y}$,
  $N\in\oN$, \ or \
  $\ket{X}=\alpha\,\cQ_{-\theta-M}\,\ldots\,\cQ_{-\theta}\,\ket{Y}$,
  $M\in\oN_0$, where $\alpha\in\oC$, $\alpha\neq0$.
\end{dfn}
Those extremal states that do {\it not\/} generate the entire massive
Verma module necessarily have a vanishing \ddsc.  In~\req{branchdiag},
for example, a part of the states on the extremal diagram generate
only a submodule of the massive Verma module.  Thus, in order to
correctly define singular vectors that generate massive Verma
submodules in a massive Verma module, one has to avoid vanishing \ddsc
s of the singular vector.  This is formalized in the following
definition.
\begin{dfn}\label{defsingmass}
  A representative of a massive singular vector in the massive Verma
  module $\mU_{h,\ell,t}$ is any element of $\mU_{h,\ell,t}$ such that
  \begin{itemize}
    \addtolength{\parskip}{-6pt}
  \item[i)] it is annihilated by the operators $\cL_m$, $\cH_m$,
    $m\in\oN$, \ $\cQ_\lambda$, $\lambda\in-\theta+\oN$, and
    $\cG_\nu$, $\nu=\theta+\oN_0$ with some $\theta\in\oZ$,
    \pagebreak[3]

  \item[ii)] none of its \ddsc s vanish,

  \item[iii)] the \hw{} state $\ket{h,\ell,t}$ is {\it not\/} one of
    its descendants.
  \end{itemize}
\end{dfn}
The meaning of the definition is that any representative of a massive
singular vectors should generate an extremal diagram of the same type
as the extremal diagram \req{massdiagramdouble}. On the other hand,
vectors that do generate a given massive submodule can be chosen in
different ways, and we thus talk about {\it representatives\/} of a
massive singular vector.

\medskip

In the conventional approach, the \hw{} conditions imposed on any
singular vector read
\begin{equation}
  \cQ_{\geq1}\approx\cG_{\geq0}\approx
  \cL_{\geq1}\approx\cH_{\geq1}\approx0\,.
  \label{conventional}
\end{equation}
This selects the {\it top-level\/} (in accordance with the diagrams
being drawn `upside-down') representative of the extremal diagram of
the submodule. These conventional, `untwisted', singular vectors will
thus be called top-level representatives.  In~\cite{[BFK],[Doerr2]},
conditions~\req{conventional} apply equally to the representatives of
the charged and the massive singular vectors in our nomenclature. As
regards the charged singular vectors, choosing the top-level
representative conceals the fact that the submodule is of a different
nature than the module itself; ignoring this then shows up in a number
of `paradoxes' when analyzing degenerations~of~the~module.

In the general position, the massive singular vectors are equivalent
to the `uncharged' singular vectors in the conventional approach,
since these generate the same submodule. In the degenerate cases,
however, the conventional, top-level, singular vectors may not
generate the entire submodule generated from some other states on the
same extremal diagram.  This depends on the properties of the extremal
diagram of the submodule, which change when there appears a charged
singular vector, i.e., when one of the extremal states in the diagram
happens to satisfy twisted {\it topological\/} \hw{} conditions.  The
conventional representatives of singular vectors may then be separated
by such topological points from those sections of the extremal diagram
which generate the maximal submodule.

Our strategy is to define and explicitly construct singular vectors
that lie in the `safe' sections of the extremal diagrams (those from
which maximal submodules are generated).  As we have mentioned, this
eliminates the notion of subsingular vectors.  Describing the
structure of $\N2$ modules in this way appears to be more transparent
and in any case much more economical, considering a fast proliferation
of cases describing the subsingular vectors that have to be introduced
whenever the conventional, top-level, singular vectors lie in the
`wrong' section of the extremal diagram of the submodule. However,
given the analysis that follows, it is a straightforward exercise to
classify all such cases (and explicitly construct the subsingular
vectors) by looking at how the extremal diagram is divided into
different sections by the topological singular vectors.

\smallskip

In the next subsection, we develop the algebraic formalism that allows
us to construct singular vectors. The reader who is interested only in
the degeneration patterns may skip to Subsection~\ref{subsec:codim1}
and Section~\ref{sec:hicodim}.

\subsection{The algebra of continued operators\label{AlgI}}\lvm
In order to explicitly construct singular vectors, we follow
ref.~\cite{[ST3]} in making use of `continued' operators that
generalize the \ddsc{}s to the case of non-integral (in fact, complex)
$\theta$.

The new operators $g(a,b)$ and $q(a,b)$ can be thought of as a
continuation of the products of modes
$\cG_a\,\cG_{a+1}\ldots\cG_{a+N}$ and
$\cQ_a\,\cQ_{a+1}\ldots\cQ_{a+N}$, respectively, to a complex number
of factors. In particular, whenever the {\it length\/} $b-a+1$ of \ 
$g(a,b)$ or $q(a,b)$ is a non-negative integer, the corresponding
operator becomes, by definition, the product of the corresponding
modes:
\begin{equation}
  g(a,b)=\prod_{i=0}^{L-1}\cG_{a+i}\,,\quad q(a,b)=
  \prod_{i=0}^{L-1}\cQ_{a+i}\,,
  \quad{\rm iff}\quad L\equiv b-a+1=0,1,2,\ldots
  \label{integrallength}
\end{equation}
(in the case where $L=0$, the product evaluates as 1).

We now postulate a number of properties of the new operators in such a
way that these properties become identities whenever the operators
reduce to elements of the universal enveloping algebra. This is
analogous to the well-known story about complex exponents in the
construction of~\cite{[MFF]}.

To begin with, the idea of a `{\it dense\/}' filling with fermions is
formalized in the rules
\begin{equation}\new
  \begin{array}{rcl}
    g(a,b-1)\,g(b,\theta-1)\,\ket{\theta}_{g}&=&
    g(a,\theta-1)\,\ket{\theta}_{g}\,,\\
    q(a,b-1)\,q(b,-\theta-1)\,\ket{\theta}_{q}&=&
    q(a,-\theta-1)\,\ket{\theta}_{q}\,,
  \end{array}
  \quad a,b,\theta\in\oC\,,
  \label{Glue}
\end{equation}
where $\ket{\theta}_{g}$ is any state that satisfies
$\cG_{\theta+n}\ket{\theta}_{g}=0$ for $n\in\oN_0$, and
$\ket{\theta}_{q}$, similarly, satisfies
$\cQ_{-\theta+n}\ket{\theta}_{q}=0$ for $n\in\oN_0$.

Under the spectral flow transform~\req{U}, the operators $g(a,b)$ and
$q(a,b)$ behave in the manner that is also inherited from the
behaviour of the products~\req{integrallength}:
\begin{equation}
  {\cal U}_\theta:\new
  \begin{array}{rcl}g(a,b)&\mapsto&g(a+\theta,b+\theta)\,,\\
    q(a,b)&\mapsto&q(a-\theta,b-\theta)\,.
  \end{array}
  \label{Ugq}
\end{equation}

Further properties of the new operators originate in the fact that,
the $\N2$ generators $\cQ$ and $\cG$ being fermions, they satisfy the
vanishing formulae such as, e.g.,
$\cG_{n}\cdot\prod_{i=a}^{a+N}\cG_{i}=0$, $N\in\oN_0$, $a\leq n\leq
a+N$.  For complex values of the parameters, we impose
\begin{equation}
  \cG_a\,g(b,c)=0\,,\qquad
  \cQ_a\,q(b,c)=0\,,\quad a-b\in\oN_0\quad\mbox{\rm and}\quad
  (a-c\not\in\oN\quad\mbox{\rm or}\quad b-c-1\in\oN).
  \label{rightkill}
\end{equation}
Similarly, the `left-hand' annihilation properties are expressed by
the relations
\begin{equation}
  g(a,b)\,\cG_c = 0\,,\qquad q(a,b)\,\cQ_c = 0\,,
  \quad b-c\in\oN_0\quad\mbox{\rm and}\quad
  (a-c\not\in\oN\quad\mbox{\rm or}\quad a-b-1\in\oN)\,.
  \label{leftkill}
\end{equation}

The formulae to commute the continued operators with the bosons
$\cL_{\geq1}$ and $\cH_{\geq1}$ read
\begin{equation}\new
  \begin{array}{rcl}
    \left[\cK_p,\, g(a,b)\right]\kern-4pt&=&\kern-4pt
    \sum_{l=0}^{d(p,a,b)}g(a,b-l-1)
    \left[\cK_p\,,\,\cG_{b-l}\right]\,\cG_{b-l+1}\ldots \cG_b\,,
    \quad p\in\oN\,,\\
    \left[\cK_p,\, q(a,b)\right]\kern-4pt&=&\kern-4pt
    \sum_{l=0}^{d(p,a,b)}q(a,b-l-1)
    \left[\cK_p\,,\,\cQ_{b-l}\right]\,\cQ_{b-l+1}\ldots \cQ_b\,,
  \end{array}\label{LRight}
\end{equation}
where $\cK=\cL$ or $\cH$, and
\begin{equation}
  d(p,a,b)=\left\{\new
    \begin{array}{ll}
      b-a\,,&p-b+a\in\oN_0\quad\mbox{\rm and}\quad b-a+1\in\oN_0\,,\\
      p-1\,,  &{\rm otherwise}.
    \end{array}\right.
  \label{ddd}
\end{equation}
The main point here is that, even though the length $b-a+1$ may not be
an integer, there is always an integral number of terms on the RHS of
\req{LRight}.

Similarly, applying the $g$ and $q$ operators changes the eigenvalues
of $\cL_0$ and $\cH_0$, which can be expressed by the commutation
relations
\begin{equation}\new
  \begin{array}{rclcrcl}
    {[}\cL_0,\,g(a,b)]&=&-\half(a+b)(b-a+1)\,g(a,b)\,,&{}&
    [\cH_0,\,g(a,b)]&=&(b-a+1)\,g(a,b)\,,\\
    {[}\cL_0,\,q(a,b)]&=&-\half(a+b)(b-a+1)\,q(a,b)\,,&{}&
    [\cH_0,\,q(a,b)]&=&(-b+a-1)\,q(a,b)\,.
  \end{array}
  \label{L0H0}
\end{equation}

Further annihilation properties with respect to the fermionic
operators are as follows:
\begin{equation}
  \cQ_{-\theta+n}\,g(\theta,-1)\,
  \ket{h,\ell,t}=0\,,\quad\theta\in\oC\,,\quad n\in\oN\,,
  \label{(3.9)}
\end{equation}
while
\begin{equation}
  \cQ_{-\theta}\,g(\theta,-1)\,\ket{h,\ell,t}=
  2(\ell+\theta h-\frac{1}{t}({\theta}^2+\theta))
  \,g(\theta+1,-1)\,\ket{h,\ell,t}\,.
  \label{cross2new}
\end{equation}
It is understood here that the operators acting from the left of
$g(\theta,\theta'-1)$ or $q(-\theta,-\theta'-1)$ are the $\N2$
generators from \req{topalgebra} subjected to the spectral flow
transform $\cU_\theta$.  Similarly, for the $q$-operators, we have the
following properties:
\begin{equation}\new
  \begin{array}{rcl}
    \cG_{\theta+n}\,q(-\theta,0)\,\ket{h,\ell,t}&=&0\,,\quad n\in\oN\,,\\
    \cG_{\theta}\,q(-\theta,0)\,\ket{h,\ell,t}&=&
    2(\ell+\theta h-\frac{1}{t}(\theta^2+\theta))q(-\theta+1,0)\,
    \ket{h,\ell,t}\,.
  \end{array}
  \label{massiveanih}
\end{equation}
For the `continued' {\it topological\/} \hw{} states we have, in the
same manner,
\begin{eqnarray}
  \cQ_{-\theta'}\,g(\theta',\theta-1)\,\ket{h,t;\theta}_{\rm top}&=&
  2(\theta'-\theta)(h+\frac{1}{t}(\theta-\theta'-1))
  \,g(\theta'+1,\theta-1)\,\ket{h,t;\theta}_{\rm top}
  \label{cross2}\\
  \cG_{\theta'}\,q(-\theta',-\theta-1)\,\ket{h,t;\theta}_{\rm top}&=&
  2(\theta'-\theta)(h+1+\frac{1}{t}(\theta-\theta'-1))
  \,q(\theta'+1,\theta-1)\,\ket{h,t;\theta}_{\rm top}
  \label{cross1}
\end{eqnarray}

The formulae to commute the negative-moded $\cH$ and $\cL$ operators
through $q(a,b)$ and $g(a,b)$ read
\begin{equation}\new
  \begin{array}{rcl}
    \left[g(a,b),\,\cK_p\right]\kern-4pt&=&\kern-6pt
    \sum_{l=0}^{d(-p,a,b)}\cG_a\ldots \cG_{a+l-1}
    \left[\cG_{a+l}\,,\,\cK_p\right]\,g(a+l+1,b)\,,\\
    \left[q(a,b),\,\cK_p\right]\kern-4pt&=&\kern-6pt
    \sum_{l=0}^{d(-p,a,b)}\cQ_a\ldots \cQ_{a+l-1}
    \left[\cQ_{a+l}\,,\,\cK_p\right]\,q(a+l+1,b)\,,
  \end{array}
  \label{LLeft}
\end{equation}
where $d(p,a,b)$ is given by~\req{ddd}.  As before, $\cK=\cH$ or
$\cL$.

\medskip

The formulae postulated for $g$ and $q$ make up a consistent set of
algebraic rules (in particular, they are consistent with operator
associativity and with the positive integral length
reduction~\req{integrallength}).  All the properties listed above make
it easy to show the following
\begin{lemma}\label{gqonhw}\mbox{}\nopagebreak

  {\rm I}.~A massive \hw{} state maps under the action of operators
  $\;g\;$ and $\;q\;$ into the states \ $g(\theta,-1)\ket{h,\ell,t}$
  and $q(-\theta,0)\ket{h,\ell,t}$ that satisfy the following
  annihilation conditions:
  \begin{equation}\new
    \begin{array}{rcl}
      \cL_m\,g(\theta,-1)\,\ket{h,\ell,t}&=&0\,,\quad m\in\oN\,,\\
      \cH_m\,g(\theta,-1)\,\ket{h,\ell,t}&=&0\,,\quad m\in\oN\,,\\
      \cG_a\,g(\theta,-1)\,\ket{h,\ell,t}&=&0\,,\quad a\in\theta+\oN_0\,,\\
      \cQ_a\,g(\theta,-1)\,\ket{h,\ell,t}&=&0\,,\quad a\in-\theta+\oN\,,
    \end{array}
    \label{rightkillg}
  \end{equation}
  and
  \begin{equation}\new
    \begin{array}{rcl}
      \cL_m\,q(-\theta,0)\,\ket{h,\ell,t}&=&0\,,\quad
      m\in\oN\,,\\
      \cH_m\,q(-\theta,0)\,\ket{h,\ell,t}&=&0\,,\quad
      m\in\oN\,,\\
      \cG_a\,q(-\theta,0)\,\ket{h,\ell,t}&=&0\,,\quad
      a\in\theta+\oN\,,\\
      \cQ_a\,q(-\theta,0)\,\ket{h,\ell,t}&=&0\,,\quad
      a\in-\theta+\oN_0\,.
    \end{array}
    \label{rightkillq}
  \end{equation}

  {\rm II}.~The twisted topological \hw{} states are mapped under the
  action of $g$ and $q$ into the states that satisfy
  \begin{equation}\new
    \begin{array}{rclrcll}
      \cL_m\,g(\theta',\theta-1)\,\ket{h,t;\theta}_{\rm
      top}&=&0\,,\quad &
      \cL_m\,q(\theta',-\theta-1)\,\ket{h,t;\theta}_{\rm top}&=&0\,,&
      m\in\oN\,.\\
      \cH_m\,g(\theta',\theta-1)\,\ket{h,t;\theta}_{\rm top}&=&0\,,&
      \cH_m\,q(\theta',-\theta-1)\,\ket{h,t;\theta}_{\rm top}&=&0\,,&
      m\in\oN\,,\\
      \cG_a\,g(\theta',\theta-1)\,\ket{h,t;\theta}_{\rm top}&=&0\,,&
      \cG_a\,q(\theta',-\theta-1)\,\ket{h,t;\theta}_{\rm top}&=&0\,,&
      a\in-\theta'+\oN\,.\\
      \cQ_a\,g(\theta',\theta-1)\,\ket{h,t;\theta}_{\rm top}&=&0\,,&
      \cQ_a\,q(\theta',-\theta-1)\,\ket{h,t;\theta}_{\rm top}&=&0\,,&
      a\in\theta'+\oN_0\,,
    \end{array}
    \label{LHkill}
  \end{equation}
\end{lemma}
These equations allow us to relate the states satisfying the \hw{}
conditions with different twists, which is necessary for the
construction of singular vectors.

It is also useful to know the parameters (the corresponding $h$ and
$\ell$) of the vector obtained from $\ket{h, \ell, t;\theta}$ by the
action of a $q$- or a $g$-operator. These are described as follows: up
to a numerical coefficient, we have
\begin{equation}\new
  \begin{array}{rcl}
    g(\theta',\theta-1)\,\ket{h,\ell,t;\theta}\kern-4pt&\sim&\kern-4pt
    \ket{h',\ell',t;\theta'}\,,\\
    h'\kern-4pt&=&\kern-4pt h+\frac{2}{t}(\theta-\theta')\,,\\
    \ell'\kern-4pt &=&\kern-4pt
    \ell+(\theta'-\theta)(h-\frac{1}{t}(\theta'-\theta+1))\,,
  \end{array}
  \label{hnew}
\end{equation}
and
\begin{equation}\new
  \begin{array}{rcl}
    q(-\theta',-\theta)\ket{h,\ell,t;\theta}\kern-4pt &\sim&\kern-4pt
    \ket{h'',\ell'',t;\theta'+1}\,,\\
    h''\kern-4pt &=&\kern-4pt h +\frac{2}{t}(\theta-\theta'-1)\,,\\
    \ell''\kern-4pt &=&\kern-4pt
    \ell+(\theta'-\theta + 1)(h-\frac{1}{t}(\theta'-\theta+2))\,.
  \end{array}
  \label{lnewq}
\end{equation}
Note that whenever $\ell+(\theta'-\theta)h -
\frac{1}{t}((\theta'-\theta)^2 + \theta' - \theta) = 0$,
Eqs.~\req{massiveanih} allow us to show that, in addition
to~\req{lnewq},
\begin{equation}
    q(-\theta',-\theta)\ket{h,\ell,t;\theta}\sim
    \kettop{h + \frac{2}{t}(\theta-\theta')-1,t;\theta'}\,.
    \label{htopnew}
\end{equation}
In what follows, the above formulae will be used to construct the
general expressions for singular vectors in $\N2$ Verma modules.

\subsection{Singular vectors in codimension
  $1$\label{subsec:codim1}}\lvm
In the general position, there are no singular vectors in Verma
modules.  Singular vectors can appear in codimension~1, when there
is~1 relation between parameters of the \hw{} state. This is
considered in the present subsection, while the cases of a higher
codimension, where several singular vectors coexist in the module, are
considered in the next section. We begin with the topological Verma
modules.  As we are going to see, this case is also crucial for the
massive Verma modules, since the analysis of the latter reduces, to a
considerable degree, to the analysis of certain topological Verma
modules.
\begin{thm}\mbox{}\label{topsing:thm}\nopagebreak

  {\rm I.}~A singular vector exists in the topological Verma module
  $\mV_{h,t}$ if and only if $h=\hplus(r,s,t)$ or $h=\hminus(r,s,t)$,
  where
  \begin{equation}\new
    \begin{array}{l}
      \hplus(r,s,t)=-\frac{r-1}{t}+s-1\,,\\
      \hminus(r,s,t)=\frac{r+1}{t}-s\,,
    \end{array}
    \quad r,s\in\oN\,.
    \label{hplusminus}
  \end{equation}

  {\rm II.} All singular vectors in the topological Verma module
  $\mV_{\hpm(r,s,t),t}$ over the $\N2$ superconformal algebra are
  given by the explicit construction:
  \begin{eqnarray}
    \ket{E(r,s,t)}^+\kern-4pt&=&\kern-4pt
    g(-r,(s-1)t-1)\,q(-(s-1)t,r-1-t)\,\ldots{}
    g((s-2)t-r,t-1)\,q(-t,r-1-t(s-1))\nonumber\\
    {}&{}&\qquad\qquad
    {}\cdot g((s-1)t-r,-1)\,\kettop{\hplus(r,s,t),t}\,,\label{Tplus}\\
    \ket{E(r,s,t)}^-\kern-4pt&=&\kern-4pt
    q(-r, (s-1) t - 1)\,g(-(s-1)t, r - t - 1)\,\ldots
    q((s-2) t - r, t-1) \, g(-t, r - 1 - (s-1) t)\nonumber\\
    {}&{}&{}\qquad\qquad{}\cdot
    q((s-1) t - r, -1)\,\kettop{\hminus(r,s,t),t}\,,\label{Tminus}
  \end{eqnarray}
  where $r,s\in\oN$ and the factors in the first line of each formula
  are
  \begin{eqnarray}
    g(-r - t - m t + s t, -1 + m t)\,q(-m t, r - 1 + m t - s t)\,,
    \qquad s-1\geq m\geq1\label{plusfact}\\
\noalign{\noindent\mbox{and}}
    q(-r - t - m t + s t, -1 + m t)\,g(-m t, r - 1 + m t - s t)\,,
    \qquad
    s-1\geq m\geq1\,,\label{minusfact}
  \end{eqnarray}
  respectively.  The $\ket{E(r,s,t)}^\pm$ singular vectors satisfy
  twisted topological \hw\ conditions with the twist parameter
  $\theta=\mp r$, are on the level $rs+\half r(r-1)$ over the
  corresponding topological highest-weight state, and have the
  relative charge~$\pm r$.
\end{thm}
In what follows, we will need singular vector {\it operators\/}
$\cE^\pm(r,s,t)$ such that
$$\ket{E(r,s,t)}^\pm=\cE^\pm(r,s,t)\,\ket{\htop^\pm(r,s,t),t}_{\rm
  top}\,.
$$
In a direct analogy with the well-known affine Lie algebra
case~\cite{[MFF],[Mal]}, ``all singular vectors'' applies literally to
non-rational $t$, while for rational~$t$, a singular vector may be
given already by a subformula of Eqs.~\req{Tplus}, \req{Tminus} as
soon as that subformula (obtained by dropping several $g$- and
$q$-operators from the left) produces an element of the Verma module.

To avoid a possible misunderstanding, let us point out once again that
a given submodule may be generated from a state other than the
singular vectors we work with (in the present case, other than the
topological singular vectors). This is completely similar to the
situation in the standard case of (affine) Lie algebras, where it {\it
  is\/} possible to generate a given Verma submodule from some
vectors other than the \hw{} state of the submodule. However, any such
vector is a descendant of the \hw{} vector and, in this sense,
considering it as a `singular vector' is unnecessary (and, often,
inconvenient). An essential point about singular vectors~\req{Tplus},
\req{Tminus} is that the corresponding submodules can be {\it
  freely\/} generated from these vectors.
\begin{prf}
  Part {\rm I} was conjectured in~\cite{[S-sing]} and proved
  in~\cite{[FST]} as an immediate consequence of the equivalence
  result obtained there.  The construction of singular vectors in
  Part~{\rm II} is borrowed from~\cite{[ST2]}, while the fact that
  these are all singular vectors follows again from~\cite{[FST]}.  The
  scheme to evaluate the singular vectors as elements of the
  topological Verma module can be outlined as follows.  Consider, for
  definiteness,~\req{Tminus}. This can be rewritten as
  $$
  \ket{E(r,s,t)}^-=
  q(-r, (s-1) t - 1)\,
  \cE^{+,r-(s-1) t}(r,s-1,t)\,
  q((s-1) t - r, -1)\,\kettop{\hminus(r,s,t),t}\,,
  $$
  where $\cE^{+,\theta}(r,s-1,t)$ is the spectral flow transform of
  the singular vector operator. Now, assuming that this operator is
  already expressed in terms of modes of $\cL$, $\cH$, $\cG$, and
  $\cQ$, we shall prove that $\ket{E(r,s,t)}^-$ is a polynomial in
  $\cL_{\leq-1}$, $\cH_{\leq-1}$, $\cG_{\leq-1}$, and $\cQ_{\leq-1}$
  acting on $\kettop{\hminus(r,s,t),t}$. To this end, we
  use~\req{Glue} to rewrite $q(-r, (s-1) t - 1)\cE^{+,r-(s-1)
    t}(r,s-1,t)$ as $q(-r, (s-1) t - r - 1)\cQ_{(s-1) t - r}\ldots
  \cQ_{(s-1) t - 1}\, \cE^{+,r-(s-1) t}(r,s-1,t)$ and observe that all
  of the operators $\cQ_{(s-1) t - r}$, \ldots, $\cQ_{(s-1) t - 1}$
  annihilate the state $q((s-1) t - r,
  -1)\cdot\kettop{\hminus(r,s,t),t}$ in accordance
  with~\req{(3.9)}--\req{cross1}. After commuting these operators to
  the right, Eqs.~\req{leftkill} apply to $q(-r, (s-1) t - r - 1)$ and
  all of the remaining modes of~$\cQ$. Finally, we are left with a
  polynomial in the modes of only $\cL$ and $\cH$ between $q(-r, (s-1)
  t - r - 1)$ and $q((s-1) t - r, -1)$. Then, using Eqs.~\req{LLeft},
  we see that the two $q$-operators meet each other and are eliminated
  using Eqs.~\req{Glue} and~\req{integrallength}. Thus, we are left
  with a a polynomial in $\cL_{\leq-1}$, $\cH_{\leq-1}$,
  $\cG_{\leq-1}$, and $\cQ_{\leq-1}$ acting on
  $\kettop{\hminus(r,s,t),t}$.  This allows us to develop the
  induction argument, with the starting point being that in the {\it
    center\/} of each of the formulas~\req{Tplus} and~\req{Tminus},
  there is a $g$- or $q$-operator of the positive integral length~$r$,
  which therefore reduces to the product of modes according
  to~\req{integrallength}.
\end{prf}

\medskip

We now turn to singular vectors in massive $\N2$ Verma modules.  To a
given massive Verma module $\mU_{h,\ell,t}$ we associate four twisted
topological Verma modules whose \hw{} vectors are the ``continued''
states of the form of those entering \req{rightkillg} and
\req{rightkillq}.  Namely, let $\theta'$ and $\theta''=-\theta'+ht-1$
be two roots of the equation
\begin{equation}
  \ell=-\theta h+\frac{1}{t}(\theta^2+\theta)\,.
  \label{ell}
\end{equation}
Then, using Lemma~\ref{gqonhw} and Eqs.~\req{hnew} and~\req{lnewq}, it
is immediately verified that the states
\begin{equation}\new
  \begin{array}{l}
    g(\theta',-1)\ket{h,\ell,t}\,,\qquad
    q(-\theta',0)\ket{h,\ell,t}\,,\\
    g(\theta'',-1)\ket{h,\ell,t}\,,\qquad
    q(-\theta'',0)\ket{h,\ell,t}\,,
  \end{array}
  \label{toverma}
\end{equation}
formally satisfy the twisted topological \hw{}
conditions~\req{twistedtophw}, although possibly with a complex twist
parameter.

We will say, for brevity, that a \hw{} state admits a singular vector
if the corresponding singular vector exists in the module built on
that state and that a \hw{} state admits no singular vectors if no
singular vectors exist in the module.  As it turns out, all possible
degenerations of the massive Verma module $\mU_{h,\ell,t}$ occur
depending on whether and how many of states~\req{toverma} belong to
$\mU_{h,\ell,t}$ and/or admit a topological singular vector.  We now
introduce a stratification of the space of highest weights
$(h,\ell,t)$ controlled by the behaviour of vectors~\req{toverma}.  In
the subsequent sections, we consider each stratum in turn and study
the corresponding degenerations of massive Verma modules.  The
possible cases, whose labels ${\cal O}_{\rm xyz}$ indicate the
existence of typical (massive or charged) singular vectors, are as
follows:
\begin{enumerate}\addtolength{\parskip}{-6pt} \label{thelist}
\item[\rlap{\kern-20pt\underline{\rm codimension 0:}}]\nopagebreak
  
\item\label{codim0} none of states~\req{toverma} belong to
  $\mU_{h,\ell,t}$ and at least one of states~\req{toverma} admits no
  topological singular vectors;

\item[\rlap{\kern-20pt\underline{\rm codimension 1:}}]\nopagebreak
  
\item\label{codim1:m} ${\cal O}_{\rm m}${\rm :} one of the
  states~\req{toverma} admits precisely one topological singular
  vector, each of the other states~\req{toverma} admits at least one
  topological singular vector, while none of  states~\req{toverma}
  belong to $\mU_{h,\ell,t}$;
  
\item\label{codim1:c} ${\cal O}_{\rm c}${\rm :} one and only one of
  states~\req{toverma} belongs to $\mU_{h,\ell,t}$ and none of
  states~\req{toverma} admit a topological singular
  vector;\pagebreak[3]

\item[\rlap{\kern-20pt\underline{\rm codimension 2:}}]\nopagebreak
  
\item\label{codim2rat} ${\cal O}_{\rm mm}${\rm :} each of states
  \req{toverma} admits at least two distinct topological singular
  vectors, while none of states~\req{toverma} belong to
  $\mU_{h,\ell,t}$;\pagebreak[3]
  
\item\label{codim2cc} ${\cal O}_{\rm cc}${\rm :} precisely one of the
  states from each column in~\req{toverma} belongs to the
  module~$\mU_{h,\ell,t}$ and none of these two states admit a
  topological singular vector;
  
\item\label{codim2ct} ${\cal O}_{\rm cm}${\rm :} one of the states
  from~\req{toverma} belongs to the module $\mU_{h,\ell,t}$ and admits
  precisely one topological singular vector; none of
  states~\req{toverma} admit two different topological singular
  vectors; no two states from different columns in~\req{toverma}
  belong to $\mU_{h,\ell,t}$;

\item[\rlap{\kern-24pt\underline{\rm codimension 3:}}]\nopagebreak

\item\label{codim3rat} ${\cal O}_{\rm cmm}${\rm :} one of the states
  from~\req{toverma} belongs to the module~$\mU_{h,\ell,t}$ and admits
  at least two different topological singular vectors; no two states
  from different columns in~\req{toverma} belong to $\mU_{h,\ell,t}$;

\item\label{codim3last} ${\cal O}_{\rm ccm}${\rm :} precisely one of
  the states from each column in~\req{toverma} belongs to the
  module~$\mU_{h,\ell,t}$; each of these two states admits at least
  one topological singular vector.
\end{enumerate}
In what follows, we will often refer to cases
\ref{codim0}--\ref{codim3last} by saying that the \hw{} parameters
$(h,\ell, t)$ of \,$\mU_{h,\ell, t}$ belong to the
corresponding~${\cal O}_{\rm xyz}$.
\begin{lemma}\label{thm:2.37}
  The above cases \ref{codim0}--\ref{codim3last} divide the space of
  \hw{} parameters $(h,\ell, t)$ into a disjoint union.
\end{lemma}
\begin{prf}
  Observe, first of all, that each case in the above list is singled
  out by a combination of two conditions or their negations: that one
  of states~\req{toverma} belongs to the module~\,$\mU_{h,\ell, t}$
  and that a (necessarily topological) singular vector exists in the
  module built on one of states~\req{toverma}. The first condition
  means that the $\theta$ parameter is an integer of the appropriate
  sign such that formulae~\req{integrallength} apply and, thus, the
  corresponding state in an element of~$\mU_{h,\ell,t}$.  We find
  from~\req{ell} that the condition for this to be the case is $\ell =
  n(h+\frac{n-1}{t})$, $n\in\oZ$.  Next, whether or not a state
  from~\req{toverma} admits a topological singular vector is a matter
  of whether the corresponding $h'$ or $h''$ parameter determined
  according to \req{hnew} and~\req{lnewq} equals one of the
  $\htop^\pm$ from~\req{hplusminus}.  We see from~\req{ell} that this
  is the case if and only if $\ell =
  -\frac{t}{4}(h-\hminus(r,s,t))(h-\hplus(r,s+1,t))$, $r,s\in\oN$.
  Note that the two expressions for $\ell$ are precisely the zeros of
  the Ka\v c determinant~\cite{[BFK]}.
  
  The cases~\ref{codim0}--\ref{codim3last} do not overlap by
  construction; on the other hand, there are no other possible
  combinations of the two basic conditions, since such combinations
  (e.g., that {\it three distinct\/} states from~\req{toverma} belong
  to $\mU_{h,\ell, t}$, etc.) would either lead to an overdetermined
  system of equations on the parameters $h$, $t$, $\theta'$, and
  $\theta''$, which admits no solutions, or would contradict the
  embedding patterns of topological Verma modules, which are
  isomorphic~\cite{[FST]} to the embedding patterns of $\tSL2$ Verma
  modules.
\end{prf}

Unless one considers cases of codimension~2 or~3, there is no
discrepancy in the use of the term `charged' between the present paper
and the treatment of~\cite{[BFK]} (and similarly with the
correspondence `massive'--`uncharged'), in the sense that the
top-level representatives of singular vectors generate exactly the
same submodules as our singular vectors.  In the following two
Theorems, we take care not to slip down to a higher codimension and
recover the `charged' and the `massive' cases:
\begin{thm}\label{thm:charged}\mbox{}
  \nopagebreak

  {\rm I.}~The \hw{} of the massive Verma module \,$\mU_{h,\ell,t}$
  belongs to the set ${\cal O}_{\rm c}$ if and only if
  $\ell=\ellch(n,h,t)$, where
  \begin{eqnarray}
    \ellch(n,h,t)\kern-4pt&=&\kern-4pt n(h+\frac{n-1}{t})\,,
    \label{Lambdach}\\
    (n,h,t)\kern-4pt&\in&\kern-4pt(\oZ\times\oC\times\oC)\setminus
    \Bigl\{(n',s-\frac{2n'-1+r}{t'},t')\bigm|r,s\in\oZ,~r\neq0,~
    r\cdot s\geq0,~n'\in\oZ,~t'\in\oC\Bigr\}\,.\nonumber
  \end{eqnarray}

    {\rm II.}~Then, the massive Verma module $\mU_{h,\ellch(n,h,t),t}$
    contains precisely one submodule, which is generated from the
    charged singular vector
  \begin{equation}
    \ket{E(n,h,t)}_{\rm ch}=\left\{\kern-4pt\new
      \begin{array}{ll}
        \cQ_{n}\,\ldots\,\cQ_0\,\ket{h,\ellch(n,h,t),t}&n\leq0\,,\\
        \cG_{-n}\,\ldots\,\cG_{-1}\,\ket{h,\ellch(n,h,t),t}\,,&
        n\geq1\,.
      \end{array}\right.
    \label{ECh}
  \end{equation}
  Every such vector satisfies the twisted topological \hw{} conditions
  \req{twistedtophw} with $\theta=-n$ and, therefore, the submodule is
  isomorphic to a twisted topological Verma module.
\end{thm}
\begin{prf}
  The formula for $\ellch(n,h,t)$ is obvious from the proof of
  Lemma~\ref{thm:2.37}; the condition $\ell=\ellch(n,h,t)$ is
  equivalent to the fact that a solution of Eq.~\req{ell} is an
  integer ($\theta'\in\oZ$ or $\theta''\in\oZ$).  This reproduces the
  `charged' series of zeros of the Ka\v c determinant~\cite{[BFK]}.
  The excluded set $\oX(n,t)$ is that where other zeros of the Ka\v c
  determinant occur.  Finally, a straightforward calculation in the
  universal enveloping algebra shows that the state~\req{ECh} does
  satisfy the twisted topological \hw{} conditions, which completes
  the proof.
\end{prf}
The top-level representative of \req{ECh}, which reads as
\begin{equation}
  \label{top-level-charged}
  \ket{s(n,h,t)}_{\rm ch}=\left\{\new
    \begin{array}{ll}
      \cG_0\,\ldots\,\cG_{-n-1}\,\ket{E(n,h,t)}_{\rm ch}\,,
      &n\leq0\,,\\
      \cQ_1\,\ldots\,\cQ_{n-1}\,\ket{E(n,h,t)}_{\rm ch}\,,
      &n\geq1\,,
    \end{array}
  \right.
\end{equation}
is the conventional charged singular vector satisfying the conditions
given in~\cite{[BFK]}. Thus, the conventional charged singular vector
necessarily belongs to a twisted topological Verma submodule, and it
is the \hw{} vector of this submodule that we call the charged
singular vector $\ket{E(n,h,t)}_{\rm ch}$ in this paper.

Further, as regards the massive singular vectors, we have
\begin{thm}\mbox{}\nopagebreak
  
  {\rm I.~}~The \hw{} of the massive Verma module \,$\mU_{h,\ell,t}$
  belongs to the set ${\cal O}_{\rm m}$ if and only if
  $\ell=\ellm(r,s,h,t)$, where
  \begin{eqnarray}
    \ellm(r,s,h,t)\kern-4pt&=&\kern-4pt
    -\frac{t}{4}(h-\hminus(r,s,t))(h-\hplus(r,s+1,t))\,,\\
    (r,s,h,t)\kern-4pt&\in&\kern-4pt
    \Bigl(\oN\times\oN\times\oC\times(\oC\setminus\oQ)
    \bigcup\,\oY\Bigr)\setminus
      \Bigl\{(r',s',\pm s'-\frac{2n-1\pm r'}{t'},t')\!\Bigm|\!n\in\oZ,~
      r',s'\in\oN,~t'\in\oC\Bigr\}\,,\nonumber
  \end{eqnarray}
  where
  \begin{equation}
    \label{oY}
    \new\begin{array}{rcl}
      \oY&=&\Bigl\{(r',s',h',-\frac{p}{q})\!\Bigm|\!1\leq r'\leq p,~
      1\leq s'\leq q,~p,q\in\oN,~h'\in\oC\Bigr\}\,\bigcup\,{}
      \\{}&{}&\qquad
      \Bigl\{(r',1,h',-\frac{p}{q})\!\Bigm|\!p+1\leq r'\leq 2p,~
      p,q\in\oN,~h'\in\oC\Bigr\}
    \end{array}
  \end{equation}
  In this case, $\mU_{h,\ell,t}$ contains precisely one submodule,
  which is a massive Verma module.
  
  {\rm II.}~Then, the representatives of the massive singular vector
  in the massive Verma module $\mU_{h, \ellm(r, s, h, t),t}$ are given
  by
  \begin{eqnarray}
    \mbox{}\kern-25pt
    \ket{S(r,s,h,t)}^-\kern-8pt&=&\kern-7pt
    g(-rs,r+\theta^-(r,s,h,t)-1)\,
    \cE^{-,\theta^-(r,s,h,t)}(r,s,t)\,g(\theta^-(r,s,h,t),-1)\,
    \ket{h,\ellm(r,s,h,t),t}\,,
    \label{Sminus}\\
    \mbox{}\kern-25pt\ket{S(r,s,h,t)}^+\kern-8pt&=&\kern-7pt
    q(1-rs,r-\theta^+(r,s,h,t)-1)\,
    \cE^{+,\theta^+(r,s,h,t)}(r,s,t)\,q(-\theta^+(r,s,h,t),0)\,
    \ket{h,\ellm(r,s,h,t),t},
    \label{Splus}
  \end{eqnarray}
  where $\cE^{\pm,\theta}(r, s, t)$ are the topological singular
  vector {\sl operators\/} subjected to the spectral flow transform
  with parameter~$\theta$, and
  \begin{equation}\new
    \begin{array}{rclcl}
      \theta^-(r, s, h,t)
      &=&\frac{t}{2}(h-\hminus(r,s,t))\,,\\
      \theta^+(r, s, h,t)
      &=&\frac{t}{2}(h-1-\hplus(r,s,t))\,.
    \end{array}
    \label{theta1theta2}
  \end{equation}
  The RHSs of \req{Sminus} and \req{Splus} evaluate as elements of
  $\;\mU_{h, \ellm(r, s, h, t),t}$ and satisfy the twisted massive
  \hw{} conditions
  \begin{equation}\new
    \begin{array}{l}
      \cQ_{\geq1\mp rs}\,\ket{S(r,s,h,t)}^\pm=
      \cH_{\geq1}\,\ket{S(r,s,h,t)}^\pm=\cL_{\geq1}\,\ket{S(r,s,h,t)}^\pm=
      \cG_{\geq\pm rs}\,\ket{S(r,s,h,t)}^\pm=0\,,\\
      \cL_0\,\ket{S(r,s,h,t)}^\pm=
      \theell^\pm(r,s,h,t)\,\ket{S(r,s,h,t)}^\pm\,,\\
      \cH_{0}\,\ket{S(r,s,h,t)}^\pm=(h \mp r s)\,\ket{S(r,s,h,t)}^\pm
    \end{array}
    \label{hwsing}
  \end{equation}
  with
  \begin{equation}
    \theell^\pm(r,s,h,t)=
    \theell(r,s,h,t) + \half r s (r s + 2 \mp 1)\,.
    \label{theell}
  \end{equation}
  Either of the $\ket{S(r,s,h,t)}^\pm$ states generates the entire
  massive Verma submodule; in particular, all of the \ddsc s
  of~\req{Sminus} and~\req{Splus} are on the same extremal subdiagram
  (the extremal diagram of the submodule) and coincide up to numerical
  factors whenever they are in the same grade:
  \begin{equation}\new
    \begin{array}{ll}
      c_-(i, h, t)\,\cQ_{i+1-rs}\ldots\cQ_{rs}\,\ket{S(r,s,h,t)}^-=
      c_+(i, h, t)\,\cG_{rs-i}\ldots\cG_{rs-1}\,\ket{S(r,s,h,t)}^+\,,&
      i=0,\ldots,2rs\,,\kern-30pt\\
      c_-(i, h, t)\,\cG_{-rs+i}\ldots\cG_{-rs-1}\,\ket{S(r,s,h,t)}^-=
      c_+(i, h, t)\,\cG_{-rs+i}\ldots\cG_{rs-1}\,\ket{S(r,s,h,t)}^+\,,&
      i\leq-1\,,\\
      c_-(i, h, t)\,\cQ_{rs-i}\ldots\cQ_{rs}\,\ket{S(r,s,h,t)}^-=
      c_+(i, h, t)\,\cQ_{rs-i}\ldots\cQ_{-rs}\,\ket{S(r,s,h,t)}^+\,,&
      i\geq2rs+1\,,\kern-30pt
    \end{array}
    \label{compare}
  \end{equation}
  where $c_\pm(i,h,t)$ are ($r$- and $s$-dependent) polynomials in~$h$
  and~$t$.
\end{thm}
\begin{prf}
  A state $\kettop{h',t;\theta'}$ or $\kettop{h'',t;\theta''}$
  from~\req{toverma} admits a singular vector if and only
  if~\req{hplusminus} holds for the corresponding $h'$ or $h''$
  parameter determined according to \req{hnew} and~\req{htopnew}.
  Using~\req{ell}, we see that this is the case if and only if $\ell
  =\ellm(r,s,h,t)$, $r,s\in\oN$, which gives zeros of the Ka\v c
  determinant~\cite{[BFK]}.  Excluding the set $\oX(r,s,t)$ guarantees
  that this is the only zero. Further, a unique submodule can also
  occur for negative {\it rational\/} $t=-\frac{\tp}{q}$ provided $r$
  is sufficiently small (the `smallness' of $r$ depends on whether
  $s=1$ or $s\geq1$, since these cases correspond to different
  degenerations of the auxiliary topological Verma modules; the
  corresponding embedding diagrams are isomorphic~\cite{[FST]} to
  embedding diagrams of the $\tSL2$ Verma modules with negative
  rational $k+2=-\frac{\tp}{q}$ and with the same $r$ and $s$), whence
  Part~I follows.

  The fact that~\req{Sminus} and~\req{Splus} are elements of the Verma
  module~$\;\mU_{h, \ellm(r, s, h, t),t}$ follows similarly to how
  this was described in the proof of Theorem~\ref{topsing:thm} (in the
  present case, one consider the topological singular vector operators
  $\cE^{\pm}(r,s,t)$ as already expressed as polynomials in the modes,
  then one subjects these operators to the spectral flow transform
  with $\theta=\theta^\pm(r,s,h,t)$, and, finally, applies the
  formulae of Sec~\ref{AlgI}).  Formulae~\req{hwsing} follow
  from~\req{rightkillg}--\req{lnewq} and~\req{theta1theta2}.
  Equations~\req{theell} are obtained by applying~\req{L0H0} to
  explicit expressions~\req{Sminus} and~\req{Splus}.  Two singular
  vectors~\req{Sminus} and~\req{Splus} generate the same submodule
  because they are descendants of each other, as expressed by
  Eqs.~\req{compare}, which, in turn, follows by comparing with the
  theory of $\tSL2$ relaxed Verma modules by means of the direct and
  the inverse \hbox{functors constructed in~\cite{[FST]}}.
\end{prf}
The structure of~\req{Sminus} and~\req{Splus} reflects the property
stipulated in item~\ref{codim1:m} of the list on
page~\pageref{thelist}, that the corresponding topological \hw{} state
from~\req{toverma} admit a singular vector.  Namely, Eqs.~\req{Sminus}
and~\req{Splus} mean that we first map from the massive Verma module
$\mU_{h,\theell(r,s,h,t),t}$ either by $g(\theta^-(r,s,h,t),-1)$ or by
$q(-\theta^+(r,s,h,t),0)$ in such a way that the resulting state
satisfies twisted {\it topological\/} \hw{} conditions with the twist
parameters $\theta^\mp(r,s,h,t)$ respectively, which are the roots
of~\req{ell} with $\ell=\theell(r,s,h,t)$.  Even though $\theta^\mp(r,
s, h, t)$ are, in general, complex, we build spectral-flow-transformed
topological singular vectors on these states and, finally, map back to
the original module~$\mU_{h,\theell(r,s,h,t),t}$.

{}From the correspondence with the zeros of the Ka\v c determinant, we
also see that the massive Verma module \,$\mU_{h,\ell,t}$ is
irreducible if and only if conditions of item~\ref{codim0} of the list
on p.~\pageref{thelist} are satisfied.

\section{Submodules and singular vectors in
  codimension $\geq2$\label{sec:hicodim}}\lvm To proceed with the
degeneration patterns of $\N2$ Verma modules, we begin with
topological Verma modules, where 2 is the highest codimension, and
then consider codimensions~2 and~3 in the massive case.

\subsection{Topological Verma modules}\lvm
A further degeneration in the setting of Theorem~\ref{topsing:thm}
means that the parameter $t$ is rational, $t=p/q$.  This case is the
least interesting one {\it as regards the structure of submodules\/},
since the structure of topological Verma module
$\mV_{\htop^\pm(r,s,\frac{p}{q}),\frac{p}{q}}$ is
determined~\cite{[FST]} by the well-known structure of the Verma
module $\mM_{\jtop^\pm(r,s,\frac{p}{q}-2),\frac{p}{q}-2}$ over the
affine $\tSL2$ algebra, where
$\jplus(r,s,k)=\frac{r-1}{2}-(k+2)\frac{s-1}{2}$ and
$\jminus(r,s,k)=-\frac{r+1}{2}+(k+2)\frac{s}{2}$.  This applies to the
BGG resolution~\cite{[BGG]}, embedding diagrams~\cite{[RCW],[Mal]},
etc.

Recall that the Verma module $\mM_{j,k}$ over the~$\tSL2$ algebra
\begin{equation}\new
  \begin{array}{rcl}
    {[}J^0_m,\,J^\pm_n]&=&{}\pm J^\pm_{m+n}\,,\qquad
    [J^0_m,\,J^0_n]~=~{}\frac{K}{2}\,m\,\delta_{m+n,0}\,,\\
    {[}J^+_m,\,J^-_n]&=&{}K\,m\,\delta_{m+n,0} + 2J^0_{m+n}\,,
  \end{array}\quad m,n\in\oZ
  \label{sl2modes}
\end{equation}
(where the generator $K$ is central) is freely generated by the modes
$J^+_{\leq-1}$, $J^-_{\leq0}$, and $J^0_{\leq-1}$ from the \hw\ vector
$\ketSL{j,k}$ that satisfies the following \hw{} conditions:
\begin{equation}\new
  \begin{array}{rcl}
    J^+_{\geq0}\,\ketSL{j,k}&=&J^0_{\geq1}\,\ketSL{j,k}=
    J^-_{\geq1}\,\ketSL{j,k}=0\,,\\
    J^0_{0}\,\ketSL{j,k}&=&j\,\ketSL{j,k}\,,\qquad K\,\ketSL{j,k}=
    k\,\ketSL{j,k}\,,
  \end{array}
  \quad j,\,k\in\oC\,.
  \label{sl2hig}
\end{equation}
Singular vectors in $\mM_{j,k}$ are labelled by $r,s\in\oN$ and can be
of the `$+$' or `$-$' type. General formulae for these singular
vectors, which we denote as $\ket{{\rm MFF}(r,s,k)}^\pm$, can be found
in~\cite{[MFF]} or, in our present conventions,
in~\cite{[FST]}.
\begin{thm}[\cite{[FST]}]\label{sl2n2} For arbitrary $h\in\oC$ and
  $t\in\oC\setminus\{0\}$,
  \begin{enumerate}
    \addtolength{\parskip}{-8pt}
  \item the topological $\N2$ Verma module $\mV_{h,t}$ is irreducible
    if and only if the $\tSL2$ Verma module $\mM_{-\frac{t}{2}h,t-2}$
    is irreducible;

  \item the module $\mV_{h,t}$ has a submodule generated by a singular
    vector $\ket{E(r,s,t)}^\pm$, Eqs.~\req{Tplus} or \req{Tminus}, if
    and only if the module $\mM_{-\frac{t}{2}h,t-2}$ has a submodule
    generated by the singular vector $\ket{{\rm MFF}(r,s,t-2)}^\pm$
    respectively.
  \end{enumerate}
\end{thm}
Whenever the singular vector in $\mM_{-\frac{t}{2}h,t-2}$ has relative
$J^0_0$-charge $\pm r$, it is clear from formulae~\req{Tplus}
and~\req{Tminus} that the corresponding topological singular
vector in $\mV_{h,t}$ has relative charge $\mp r$ and satisfies the
twisted topological \hw{} conditions~\req{twistedtophw} with the twist
parameter $\theta=\pm r$.

Thus, the appearance of one or more singular vectors in a topological
$\N2$ Verma module can be read off from the corresponding $\tSL2$
Verma module.  In view of the correspondence at the level of {\it
  submodules}, it might seem puzzling that one can talk about
subsingular vectors in topological $\N2$ Verma modules, since these
are absent in $\tSL2$ Verma modules.  In fact, this apparent paradox
illustrates our general statement that, for the $\N2$ superconformal
algebra, subsingular vectors are an artifact of an ``inconvenient''
definition of singular vectors.  They have to be considered when one
restricts oneself to submodules generated only from the conventional,
top-level, $\N2$ singular vectors, which do not always generate
maximal submodules.  On the other hand, singular vectors~\req{Tplus}
and \req{Tminus}, which satisfy {\it twisted\/} topological \hw{}
conditions, allow one to work with maximal submodules, and it is these
singular vectors that are in $1:1$
correspondence~\cite{[S-sing],[FST]} with the $\tSL2$ singular
vectors.

In the conventional approach, on the other hand, subsingular vectors
occur in the topological Verma modules
$\mV_{\hpm(r,s,\ttop(r,s,n)),\ttop(r,s,n)}$, where
$\ttop(r,s,n)=\frac{n-r}{s}$ with $r$ greater than $n$.  Namely, we
have the following
\begin{prop}\label{prop:top} The quotient of the topological Verma
  module $\mV_{h,t}$ over the submodules generated by conventional
  singular vectors is reducible --- i.e., a subsingular vector exists
  in $\mV_{h,t}$ --- if and only if \ $t=\ttop(r,s,n)$,
  $h=\htop^\pm(r,s,\ttop(r,s,n))$, $1\leq n<r$. In the `$-$' case, for
  definiteness, the subsingular vector in
  $\mV_{\htop^-(r,s,\ttop(r,s,n)),\ttop(r,s,n)}$ is given by
  \begin{equation}
    \ket{\rm Sub}=
    \cG_0\ldots\cG_{r-n-1}\,\cG_{r-n+1}\ldots\cG_{r-1}\,
    \ket{E(r,s,\frac{n-r}{s})}^-
    \label{subsing:top}
  \end{equation}
  (where $\ket{E(r,s,t)}^-$ is the topological singular
  vector~\req{Tminus}).  This becomes singular in the quotient module
  $\mV_{\htop^-(r,s,\ttop(r,s,n)),\ttop(r,s,n)}/\mC$, where $\mC$ is
  the submodule generated from the top-level representative of the
  singular vector in~$\mV_{\htop^-(r,s,\ttop(r,s,n)),\ttop(r,s,n)}$,
  which is given by
  $\cG_0\ldots\cG_{r-1}\,\ket{E(r,s,\frac{n-r}{s})}^-$.
\end{prop}
\begin{prf}
  The parameters are such that the
  module~$\mV_{\hminus(r,s,\ttop(r,s,n)),\ttop(r,s,n)}$ contains at
  least two submodules
  \begin{equation}
    \mC'\stackrel{{}^{{}_{\ket{E(n,1,\frac{n-r}{s})}^{+,r}}}}{\longar}
    \mC
    \stackrel{{}^{{}_{\ket{E(r,s,\frac{n-r}{s})}^-}}}{\longar}
    \mV_{\hminus(r,s,\ttop(r,s,n)),\ttop(r,s,n)}\,,
  \end{equation}
  where $\mC\approx\mV_{\hminus(r,s,\ttop(r,s,n))-\frac{2r}{t},
    \ttop(r,s,n);r}$ and~$\mC'\approx\mV_{\hminus(r,s,\ttop(r,s,n))+
    \frac{2(n-r)}{t},\ttop(r,s,n);r-n}$, the arrows mean embeddings by
  means of the corresponding singular vectors, and $\ket{E(r, s,
    t)}^{\pm,\theta}$ are the topological singular vectors subjected
  to the spectral flow transform with parameter~$\theta$.  All we have
  to do in the conventional approach is to describe these submodules
  in terms of (submodules generated from) the top-level
  representatives of extremal diagrams.  Thus, consider the
  conventional singular vector
  \begin{equation}
    \conv=
    \cG_0\ldots\cG_{r-1}\,\ket{E(r,s,\frac{n-r}{s})}^-
  \end{equation}
  which satisfies \hw{} conditions~\req{masshw} with $\theta=0$.  It
  is clear that $\conv$ belongs to the submodule~$\mC'$ iff~$n<r$ and
  belongs to~$\mC$ iff~$n>r$.  Indeed, the \hw{} vector of~$\mC'$ is
  $\ket{\rm
    h.w.'}=\cG_{r-n}\ldots\cG_{r-1}\,\ket{E(r,s,\frac{n-r}{s})}^-$ and
  we have $\conv=\cG_0\ldots\cG_{r-n-1}\,\ket{\rm h.w.'}$
  whenever~$n<r$; on the other hand, $\ket{\rm
    h.w.'}=\cG_{r-n}\ldots\cG_{-1}\,\conv$ whenever~$n>r$. Hence, in
  the case where $n<r$, the module~$\mC'$ is generated from $\conv$ by
  the action of the $\N2$ generators, whereas in the~$n>r$ case,
  $\conv$ generates~$\mC$.  There exists the minimal submodule~$\mN$
  such that~$\mC\subset\mN$ (it is possible that
  $\mN\approx\mV_{\hminus(r,s,\ttop(r,s,n)),\ttop(r,s,n)}$).  Now, let
  us take the quotient of~$\mN$ over the submodule generated from all
  conventional singular vectors.  In the case where $n>r$, this is the
  quotient over the maximal submodule of~$\mN$, therefore the quotient
  is irreducible and, thus, there are no subsingular vectors
  in~$\mV_{\hminus(r,s,\ttop(r,s,n)),\ttop(r,s,n)}$.  On the other
  hand, in the case where~$n<r$, this quotient contains the \hw{}
  vector of~$\mC$ and, therefore, is reducible.

  The explicit formula~\req{subsing:top} is obvious from the analysis
  of extremal diagrams (see~\req{subsingtop} and below), however it
  can also be checked by direct calculations that the
  vector~\req{subsing:top} satisfies conventional \hw{} conditions
  (Eqs.~\req{masshw} with $\theta=0$) modulo descendants of~$\conv$:
  \begin{equation}
    \cH_1\,\ket{\rm Sub}=
    \cG_0\ldots\cG_{r-n-2}\,\cG_{r-n}\ldots\cG_{r-1}\,
    \ket{E(r,s,\frac{n-r}{s})}^-=
    (-1)^{(r-n-1)}\frac{r-n}{2 s(n+1)}\cQ_{-r+n+1}\,\conv\,,
    \label{subsing:H1}
  \end{equation}
  and similarly for the other annihilation conditions.  Finally,
    for modules~$\mV_{h,t}$ with $h$ and $t$ {\it not\/} as in the
    Proposition, the topological singular vectors and the conventional
    one are \ddsc s of each other.  Therefore, they generate the same
    submodule and there are no subsingular vectors
    in~those~cases.
\end{prf}
The above proof and the construction of subsingular vectors can be
illustrated in the following extremal diagram:
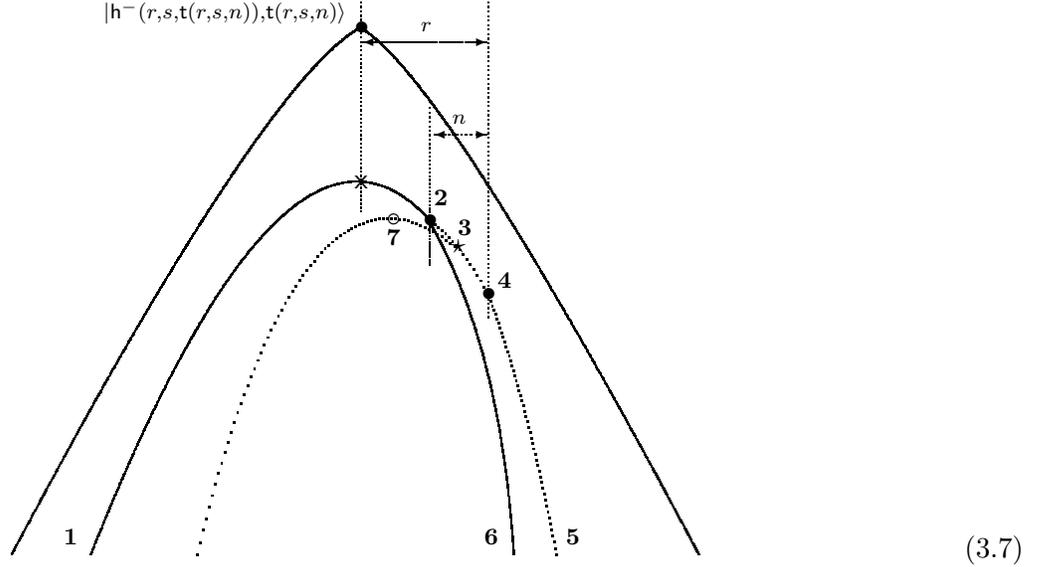
\begin{equation}
  \message{please wait...}
  \unitlength=1pt
  \begin{picture}(300,210)
    \bezier{\dimten}(20,1)(115,180)(152,200.5)
    \bezier{\dimten}(152,200.5)(185,180)(280,1)
    \put(150,198){$\bullet$}
    \put(2.3,0){
      \bezier{50}(150,210)(150,168)(150,130.8)
      }
    \put(55,200){${}^{\ket{\hminus(r,s,\ttop(r,s,n)),\ttop(r,s,n)}}$}
    \bezier{50}(155,195)(175,195)(198,195)
    \put(154,195){\vector(-1,0){1}}
    \put(199,195){\vector(1,0){1}}
    \put(175,195){${}^r$}
    \put(0,-35){
      \bezier{10}(182,195)(187,195)(198,195)
      \put(181,195){\vector(-1,0){1}}
      \put(199,195){\vector(1,0){1}}
      \put(187,195){${}^n$}
      }
    \put(28.3,0){
      \bezier{45}(150,170)(150,130)(150,110.8)
      }
    \put(50.3,0){
      \bezier{70}(150,210)(150,150)(150,90.8)
      }
    \put(150,139.8){${\ssf x}$}    
    \put(10,0){
      \put(188,97){$\bullet$}
      \put(166,125){$\bullet$}
      \put(176.5,115){$\star$}
      \bezier{\dimten}(40,1)(115,190)(169.5,126)
      {\linethickness{0.7pt}
        \bezier{15}(170,126)(180,118)(190,100)
        \bezier{50}(190,100)(205,60)(216,1)
        \bezier{70}(80,1)(120,165)(177.5,117.5)
        }
      \put(152,125.5){$\circ$}
      \bezier{\dimten}(168,127)(195,80)(200,1)
      }
    \put(40,5){\footnotesize{\bf1}}
    \put(180,133){\footnotesize{\bf2}}
    \put(189,122){\footnotesize{\bf3}}
    \put(204,102){\footnotesize{\bf4}}
    \put(230,5){\footnotesize{\bf5}}
    \put(199,5){\footnotesize{\bf6}}
    \put(162,118){\footnotesize{\bf7}}
  \end{picture}
  \label{subsingtop}
  \message{done}
\end{equation}
Here, the line {\bf 1}--{\bf 2}--{\bf 3}--{\bf 4}--{\bf 5} is the
extremal diagram\footnote{Here and in what follows, extremal diagrams
  of the type of \req{topdiag}, \req{massdiagramdouble}, etc., are
  shown schematically as parabolas, rather than `discrete
  approximations' thereof.} of~$\mC$.  The `cusp' of this diagram is
at {\bf 4}, i.e., this point represents topological singular
vector~\req{Tminus} that satisfies the twisted topological \hw{}
conditions with the twist parameter $\theta=r$.  Further, the
topological singular vector $\ket{E(n,1,\ttop(r,s,n))}^{+,r}$ in the
submodule is represented by the point {\bf 2}.  Then, consider any
state in the section of the extremal diagram of the submodule between
{\bf 1} and {\bf 2} (for instance, the conventional singular
vector~$\conv$ marked with a $\times$). The \dQdsc s of this state
terminate at~{\bf 2}: \ 
$\cQ_{n-r}\cdot\ket{E(n,1,\ttop(r,s,n))}^{+,r}=0$.  Applying instead a
one-lower mode of $\cQ$ to $\ket{E(n,1,\ttop(r,s,n))}^{+,r}$ and then
constructing dense $\cQ$ descendants, one spans the line {\bf 2}--{\bf
  6}, as
$\ldots\cQ_{n-r-2}\,\cQ_{n-r-1}\cdot\ket{E(n,1,\ttop(r,s,n))}^{+,r}$.
Thus neither the states around {\bf 1} on the solid inner parabola,
nor $\conv$ generate the maximal submodule, whereas the state
$\ket{E(r,s,\ttop(r,s,n))}^-$ at {\bf 4} does (in particular,
$\ket{E(n,1,\ttop(r,s,n))}^{+,r}$ is a dense $\cG$-descendant of
$\ket{E(r,s,\ttop(r,s,n))}^-$: ${\bf
  2}=\cG_{r-n}\,\ldots\,\cG_{r-1}\cdot{\bf~4}$).

Thus, the states {\bf 3}--{\bf 5}--\ldots\ (along with infinitely many
other, non-extremal, states), even though inside a proper submodule of
$\mV_{\hminus(r,s,\ttop(r,s,n)),\ttop(r,s,n)}$, are {\it not\/} in the
submodule generated by the top-level singular vector~$\conv$.  In the
quotient module over the topological Verma submodule generated
by~$\conv$ (or, {\it equivalently\/}, by the state
$\ket{E(n,1,\ttop(r,s,n))}^{+,r}$ at {\bf 2}), the state {\bf 3} (such
that ${\bf 2}=\cG_{r-n}\cdot\bf 3$) satisfies twisted topological
\hw{} conditions (with the twist parameter $\theta=r-n$).  Acting on
this state with $\cG_{r-n-1}$, $\cG_{r-n-2}$, \ldots, gives the state
$\ket{\rm Sub}$ at {\bf 7}, which satisfies the {\it untwisted\/}
massive \hw{} conditions as long as the state $\conv$ is factored
away.

\medskip

To avoid a possible misunderstanding, let us point out once again that
the structure of submodules described above is in fact the same as
that of the $\tSL2$ Verma
module~$\mM_{\jminus(r,s,\ttop(r,s,n)-2),\ttop(r,s,n)-2}$, with the
$\tSL2$ singular vectors corresponding to the topological singular
vectors that necessarily satisfy {\it twisted\/} topological \hw{}
conditions. On the contrary, the {\it conventional\/} $\N2$ singular
vectors are not the counterparts of the $\tSL2$ \hw{} states. This is
what leads one to observing subsingular vectors in the conventional
approach (whereas in the corresponding $\tSL2$ Verma module, there are
no reasons altogether to define singular vectors as a counterpart of
the $\N2$ top-level representatives of extremal diagrams, hence no
subsingular vectors there).

\subsection{Massive singular vectors in codimension $2$}\lvm
We now turn to massive Verma modules.  In codimension~$2$, three cases
from the list on page~\pageref{thelist} are arranged into the three
following Theorems (\ref{twomassive}, \ref{twocharge}, and
\ref{thm:codim2}), while Propositions \ref{prop:nosing},
\ref{prop:nosing2}, and \ref{prop:cm} are given in order to make
contact with the conventional description in terms of top-level,
untwisted, representatives of singular vectors and, accordingly, in
terms of subsingular vectors; we show why the subsingular vectors
appear and how they can be constructed explicitly.

The following observations are central for the subsequent
constructions:
\begin{lemma}\label{chargemass:lemma}
  Let \,$\mU\equiv\mU_{h,\ell,t}$ be a massive Verma module.
  \begin{itemize}
    \addtolength{\parskip}{-8pt}
  \item[i)] Let \,$\mU\supset\mU'$ and \,$\mU\supset\mC$, where
    \,$\mU'$ is a massive Verma submodule and \,$\mC$ is a twisted
    topological Verma module generated from a charged singular vector
    in \,$\mU$ such that for any twisted topological Verma module
    $\mC''$, $\mU\supset\mC''\supset\mC$, it follows that $\mC''=\mC$.
    Then there exists a twisted topological Verma module
    \,$\mC'
    =\mU'\cap\mC 
    \neq\{0\}$. Moreover, the embeddings~\,$\mC'\subset\mC$
    and~\,$\mC'\subset\mU'$ are given by a topological singular vector
    in~\,$\mC$ and by a charged singular vector in~\,$\mU'$
    respectively.

  \item[ii)] Conversely, if \,$\mU\supset\mU'$, where \,$\mU'$ is a
    massive Verma module, and \,$\mU'\supset\mC'$, where \,$\mC'$ is a
    submodule generated from a charged singular vector in \,$\mU'$,
    then \,$\mU\supset\mC$, where \,$\mC$ is a submodule generated
    from a charged singular vector in \,$\mU$. Moreover, \,$\mC$ is
    maximal ($\,\mU\supset\mC''\supset\mC\Longrightarrow\mC''=\mC$),
    \,$\mC'\subset\mU'\cap\mC$, and \,$\mC'$ is generated from a
    topological singular vector in the topological Verma
    module~\,$\mC$.

  \item[iii)] If \,$\mU\supset\mC'$, where $\mC'$ is a twisted
    topological Verma module, there exists a twisted topological Verma
    submodule $\mC\subset\mU$ such that the embedding is given by the
    charged singular vector, $\mC$ is maximal
    ($\mU\supset\mC''\supset\mC\Longrightarrow\mC''=\mC$), and
    $\mC'\subset\mC$ (with the embedding given by a topological
    singular vector).
  \end{itemize}
\end{lemma}
\begin{prf}  The Lemma can be illustrated by
  $$
  \new
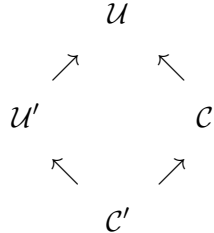
  \begin{array}{ccccc}
    {} & { }& \mU & {} & {}\\
    {}& \nearrow & {} & \nwarrow & {}\\
    \mU'\kern-6pt & {} & {} & {} & \kern-6pt\mC \\
    {}& \nwarrow & {} & \nearrow & {}\\
    {} & { }& \mC' & {} & {}
  \end{array}
  $$
  As regards item~i), let us assume the contrary, namely that
  $\mU'\cap\mC=\{0\}$. We then take the quotient $\mQ=\mU/\mC$,
  which is a twisted topological Verma module. It should contain all
  of the extremal states of the massive Verma module~$\mU'$, however
  some of these states are clearly outside the extremal diagram
  of~$\mQ$ according to their bigrading. Thus,
  $\mU'\cap\mC
  =\mC'\neq\{0\}$.

  Further, it follows from Theorem~\ref{structop:thm} that the
  embedding $\mC'\subset\mC$ is given by a topological singular vector
  in~$\mC$. In the extremal diagram of \,$\mU'$, choose a
  state~$\ket{\star'}$ from which all of the~$\mU'$ module is
  generated and consider the \ddsc{}~$\ket{\rm top'}$
  of~$\ket{\star'}$ with the minimal number of the $\cG$ or $\cQ$
  modes among those \ddsc s that belong to~$\mC'$. Such a state
  necessarily exists, since otherwise the quotient $\mU/\mC$ would
  contain extremal states of $\mU'$ that lie outside the module
  $\mU/\mC$.  The state~$\ket{\rm top'}$ satisfies the twisted
  topological \hw{} conditions and the module~$\mC'$ is generated
  from~$\ket{\rm top'}$.  Therefore,~$\ket{\rm top'}$ coincides with a
  topological singular vector in~$\mC$ and is at the same time a
  charged singular vector in~$\mU'$.  This completes the proof of~i).
  
  To prove ii), let us fix extremal states $\ket{\star}$ in $\mU$ and
  $\ket{\star'}$ in $\mU'$ such that $\mU$ and $\mU'$ are generated
  from $\ket{\star}$ and $\ket{\star'}$ respectively. The \hw{}
  vector~$\ket{\rm top'}$ of $\mC'$ is a \ddsc{} of~$\ket{\star'}$.
  Now, a charged singular vector exists in the extremal diagram of the
  submodule whenever $\theell^\pm(r,s,h,t)=\ellch(N,h \mp r s, t)$
  (see Eqs.~\req{Lambdach} and~\req{theell}), which implies a similar
  relation for the dimension $\ellm(r,s,h,t)$ of the massive Verma
  module $\mU_{h,\theell(r,s,h,t),t}$.  Assuming, for definiteness,
  that $\ket{\rm top'}$ is a \dGdsc{} of $\ket{\star'}$, we thus see
  that there exists a state $\ket{\rm top}$ such that it is a \dGdsc{}
  of $\ket{\star}$, satisfies twisted topological \hw{} conditions,
  and is not a \dGdsc{} of any other state satisfying twisted
  topological \hw{} conditions.  Let us consider the module $\mC$
  generated from~$\ket{\rm top}$ and take the quotient
  of~$\mQ=\mU/\mC$, which is a twisted topological Verma module.  By
  analyzing the bigradings, it is easy to see that some of the \dGdsc
  s of $\ket{\star'}$ lie outside the extremal diagram of~$\mQ$,
  therefore these \dGdsc s belong to $\mC$.  Therefore,
  $\mC\cap\mC'\neq\{0\}$. If $\mC\cap\mC'\neq\mC'$, the module $\mQ$
  contains the submodule $\mC'/(\mC\cap\mC')$ which is not a twisted
  topological {\it Verma\/} module. This contradicts
  Theorem~\ref{structop:thm}.  Therefore, $\mC\cap\mC'=\mC'$ and
  $\mC'\subset\mU'\cap\mC$, whence also follows the fact that the
  embedding $\mC'\subset\mC$ is given by the topological singular
  vector.

  The proof of~iii), which is rather tedious, is relegated to the
  Appendix.
\end{prf}

The Lemma is used in the following Theorem, which describes the
occurrence of (at least!) two different massive singular vectors in
the massive Verma module~\,$\mU_{h,\ell,t}$.  This is
case~\ref{codim2rat} of the list on page~\pageref{thelist} and it
corresponds to rational~$t$:
\begin{thm}\label{twomassive}
  The \hw{} of the massive Verma module \,$\mU_{h,\ell,t}$ belongs to
  the set ${\cal O}_{\rm mm}$ if and only if $\ell=\theell(r,s,h,t)$,
  where
  $$\new
    (r,s,h,t)\in\oN\times\oN\times\oC\times\oQ\setminus{}
    \Bigl(\Bigl\{(r,s,\pm s - \frac{2n - 1 \pm r}{p/q},\frac{p}{q})\!
    \Bigm|\!
    r,s\in\oZ,~n\in\oZ,~p\in\oZ,~q\in\oN\Bigr\}
    \bigcup\,\oY\Bigr)
  $$
  with $\oY$ as in~\req{oY}. \ Then,
  \begin{enumerate}\addtolength{\parskip}{-6pt}
  \item any primitive submodule of \,$\mU_{h,\ell,t}$ is a massive
    Verma module generated from the representative
    $\ket{S(a,b,h,\frac{p}{q})}^-$ of a massive singular vector, where
    $a,b\in\oN$ is a solution to $\ell=\theell(a,b,h,\frac{p}{q})$;
    equivalently, that submodule is also generated from the
    $\ket{S(a,b,h,\frac{p}{q})}^+$ representative of the massive
    singular vector with the same $a$ and~$b$.

  \item The structure of
    \,$\mU_{h,\theell(r,s,h,\frac{p}{q}),\frac{p}{q}}$ is determined
    by the structure of the topological Verma module\\
    $\mV_{\hminus(r,s,\frac{p}{q}),\frac{p}{q}}$ in the following way:
  \begin{enumerate}
    
  \item for any massive Verma submodule
    $\mU'\subset\mU_{h,\theell(r,s,h,\frac{p}{q}),\frac{p}{q}}$
    generated from a massive singular vector, there exists a submodule
    in $\mV_{\hminus(r,s,\frac{p}{q}),\frac{p}{q}}$ generated from a
    topological singular vector $e=\cE^\pm(a,b,\frac{p}{q})\,
    \ket{\hminus(r,s,\frac{p}{q}),\frac{p}{q})}_{\rm top}$,
    $a,b\in\oN$, such that $\mU'$ is generated from the massive
    singular vector
    \begin{equation}
    g(-ab,\mp a+\theta^-(a,b,h,\frac{p}{q})-1)\,
    \cE^{\pm,\theta^-(a,b,h,\frac{p}{q})}(a,b,\frac{p}{q})\,
    g(\theta^-(a,b,h,\frac{p}{q}),-1)\,
    \ket{h,\theell(r,s,h,\frac{p}{q}),\frac{p}{q}}\,;
      \label{usee}
    \end{equation}

  \item conversely, for any singular vector in \
    $\mV_{\hminus(r,s,\frac{p}{q}),\frac{p}{q}}$ of the form
    $e=\ket{E(a,b,\frac{p}{q})}^+$ with $a\geq1$, $b\geq2$, \ or
    $e=\ket{E(a,b,\frac{p}{q})}^-$ with $a,b\geq1$, there exists a
    massive singular vector constructed as in~\req{usee} that
    generates a massive Verma submodule
    $\mU'\subset\mU_{h,\theell(r,s,h,\frac{p}{q}),\frac{p}{q}}$;

  \item two different singular vectors $e_1$ and $e_2$ in
    \,$\mV_{\hminus(r,s,\frac{p}{q}),\frac{p}{q}}$ correspond in this
    way to the same massive Verma submodule
    \,$\mU'\subset\mU_{h,\theell(r,s,h,\frac{p}{q}),\frac{p}{q}}$ if
    and only if one of the $e_i$ is the
    $\ket{E(c\geq1,1,\frac{p}{q})}^+$ singular vector in the module
    generated from the other.
    \end{enumerate}
  \end{enumerate}
\end{thm}
(As before, $\cE^{\pm,\theta}(r,s,t)$ is the spectral flow transform
of the topological singular vector operator read off from~\req{Tplus}
and \req{Tminus}; {\it primitive\/} refers to a submodule that is not
a sum of other submodules.)\pagebreak[3]
\begin{prf}
  By definition, the \hw{} of the Verma module~$\mU_{h,\ell,t}$
  belongs to~${\cal O}_{\rm mm}$ whenever each of the
  states~\req{toverma} admits two topological singular vectors and
  none of the solutions to Eq.~\req{ell} is an integer:
  $\theta',\theta''\not\in\oZ$. The last condition reformulates as the
  constraint~$\frac{1}{2}\left(1-ht \pm \sqrt{4\ell
      t+(ht-1)^2}\right)\notin\oZ$. On the other hand, analysing all
  possible cases where the states~\req{toverma} have the specified
  number of singular vectors for negative rational $t=-\frac{p}{q}$
  requires analysing the embedding diagrams of the auxiliary
  topological Verma modules, these embedding diagrams being isomorphic
  to the standard embedding diagrams of $\tSL2$ Verma modules. This
  gives the values of $(r,s,h,t)$ as in the Theorem. Further, by the
  definition of~${\cal O}_{\rm mm}$, there are no charged singular
  vectors in~$\mU_{h,\ell,t}$, therefore taking into account
  Lemma~\ref{chargemass:lemma} we obtain that each submodule is
  generated from~\req{Sminus} as well as from~\req{Splus}.
  
  Part 2 becomes obvious from the explicit formulae for massive
  singular vectors~\req{Sminus} and~\req{Splus}.  Let us consider, for
  definiteness, Eq.~\req{Sminus}.  The part
  $g(\theta^-(r,s,h,\frac{p}{q}),-1)
  \cdot\ket{h,\ellm(r,s,h,\frac{p}{q}),\frac{p}{q}}$ of the formula
  represents the \hw{} vector of the twisted topological Verma module
  $\smV_{\hminus(r,s,\frac{p}{q}),\frac{p}{q};
    \theta^-(r,s,h,\frac{p}{q})}$.  Now, let us take any element
  $\ket{\nu}$ from the extremal diagram of the massive submodule which
  satisfies the twisted massive \hw{} conditions with the twist
  $\theta=\nu$.  The operator $g(\mp
  a+\theta^-(r,s,h,\frac{p}{q}),\nu-1)$ maps the state $\ket{\nu}$
  into the module $\smV_{\hminus(r,s,\frac{p}{q}),\frac{p}{q};
    \theta^-(r,s,h,\frac{p}{q})}$.  The image of $\ket{\nu}$ under
  this mapping is a twisted topological singular vector referred to in
  Part~2(a).  The fact that $e_1$ is
  the~$\ket{E(c\geq1,1,\frac{p}{q})}^+$ singular vector built on
  $e_2=\ket{E(a,b,\frac{p}{q})}^{\pm}$ means that $e_1=g(\mp a-c,\mp
  a-1)\,\ket{E(a,b,\frac{p}{q})}^{\pm}$. Then Part~2(c) follows from
  the identity
  \begin{eqnarray}
    &&g(-ab,a-c+\theta^-(a,b,h,\frac{p}{q})-1)\cdot{}
    \nonumber\\
    &&\qquad{}\cdot
    g(\mp a-c+\theta^-(a,b,h,\frac{p}{q}),
    \mp a+\theta^-(a,b,h,\frac{p}{q})-1)
    \cE^{\pm,\theta^-(a,b,h,\frac{p}{q})}(a,b,\frac{p}{q})
    \cdot{}\nonumber\\
    &&\qquad\qquad g(\theta^-(a,b,h,\frac{p}{q}),-1)\,
    \ket{h,\theell(r,s,h,\frac{p}{q}),\frac{p}{q}}
    =\\
    &&g(-ab,a+\theta^-(a,b,h,\frac{p}{q})-1)\,
    \cE^{\pm,\theta^-(a,b,h,\frac{p}{q})}(a,b,\frac{p}{q})\,
    g(\theta^-(a,b,h,\frac{p}{q}),-1)\,
    \ket{h,\theell(r,s,h,\frac{p}{q}),\frac{p}{q}}\,,\nonumber
  \end{eqnarray}
  where we used Eqs.~\req{Glue}. Let us point out once again that the
  constructions of the type of $g(-ab,\mp a+\theta-1)
  \cE^{\pm,\theta}(a,b,t) g(\theta,-1)\ket{h,\ell,t}$ in
  Eq.~\req{usee} evaluate as Verma module elements using the formulae
  of Sec.~\ref{AlgI}.
\end{prf}

If we recall that the structure of topological Verma modules is
equivalent~\cite{[FST]} to the structure of the standard $\tSL2$ Verma
modules, we see that the modules $\mU_{h,\ell,t}$ with
$(h,\ell,t)\in{\cal O}_{\rm mm}$, too, have essentially the same
(familiar) structure as the $\tSL2$ Verma modules with a rational
level $k=t-2$.

In the present case, restricting oneself to only top-level singular
vectors is innocuous\footnote{We remind the reader that, when we are
  talking about subsingular vectors, these are understood in the
  setting where the conventional definition of singular vectors is
  adopted, i.e., only top-level representatives of extremal diagrams
  are `allowed' to generate submodules.}:
\begin{prop}\label{prop:nosing} Under the conditions of
  Theorem~\ref{twomassive}, the quotient of
  \,$\mU_{h,\theell(r,s,h,\frac{p}{q}),\frac{p}{q}}$ with respect to
  the conventional singular vectors is irreducible, i.e., no
  subsingular vectors exist in the massive Verma module
  \,$\mU_{h,\theell(r,s,h,\frac{p}{q}),\frac{p}{q}}$.
\end{prop}
\begin{prf}
  Indeed, in this case there are no charged singular vectors
  in~$\mU_{h,\theell(r,s,h,\frac{p}{q}),\frac{p}{q}}$, therefore each
  of the extremal states of the submodule is a \ddsc{} of any other
  extremal state of the same submodule. Thus, each element of the
  extremal diagram of the submodule generates the same module.
\end{prf}

\medskip

Next is the case where \,$\mU_{h,\ell,t}$ contains two charged
singular vectors none of which are descendants of the other, i.e.\ the
extremal diagram contains two states that satisfy twisted topological
\hw{} conditions and lie on the different sides of the \hw{} vector.
That is, the extremal diagram has two branching points similar to that
in~\req{branchdiag}, but on the different sides of $\ket{h,\ell,t}$.
This is case~\ref{codim2cc} in the list on page~\pageref{thelist}:
\begin{thm}\label{twocharge}The \hw{} of the massive Verma module
  \,$\mU_{h,\ell,t}$ belongs to the set ${\cal O}_{\rm cc}$ if and
  only if $h=\hcc(n,m,t)$ and $\ell=\lcc(n,m,t)$, where
  \begin{eqnarray}
    \hcc(n,m,t)\kern-4pt&=&
    \kern-4pt\frac{1}{t}(1 - m - n)\,,\qquad\lcc(n,m,t)=
    -\frac{mn}{t}\,,\\
    (n,m,t)\kern-4pt&\in&\kern-4pt
    \Bigl(\oN\times(-\oN_0)\times(\oC\setminus\oQ)\Bigr)\bigcup\,
    \Bigl\{(n',m',-\frac{p}{q})\!\Bigm|\!n'\in\oN,~m'\in-\oN_0,~
    p,q\in\oN,~1\leq n'-m'\leq q\Bigr\}\,.\nonumber
  \end{eqnarray}
  Then the massive Verma module~$\mU_{\hcc(n,m,t),\lcc(n,m,t),t}$
  contains two twisted topological Verma submodules
  $\mC_1\approx\smV_{\frac{m-n-1}{t},t;-m}$ and
  $\mC_2\approx\smV_{\frac{n+1-m}{t},t;-n}$ generated from the charged
  singular vectors $\ket{E(m, \hcc(n,m,t), t)}_{\rm ch}$ and
  $\ket{E(n, \hcc(n,m,t), t)}_{\rm ch}$ respectively.  The maximal
  submodule in \,$\mU_{\hcc(n,m,t), \lcc(n,m,t), t}$ is
  $\,\mC_1\cup\,\mC_2$, and \,$\mC_1\cap\,\mC_2=0$.
\end{thm}
\begin{prf}
  The definition of the set ${\cal O}_{\rm cc}$ implies that both
  solutions $\theta'$ and $\theta''=-\theta'+ht-1$ of
  equation~\req{ell} are integers, whence the conditions
  $h=\hcc(n,m,t)$ and $\ell=\lcc(n,m,t)$ follow.  Then the singular
  vectors referred to in the theorem are all singular vectors in the
  module~$\mU_{h,\ell,t}$, since, in accordance with
  Lemma~\ref{chargemass:lemma}, any other submodule
  in~$\mU_{h,\ell,t}$ would have non-empty intersections with~$\mC_1$
  and~$\mC_2$, which would then be generated from
  singular vectors in~$\mC_1$ and~$\mC_2$.  But by the definition of
  the set ${\cal O}_{\rm cc}$, the modules~$\mC_1$ and~$\mC_2$ both
  are irreducible.
\end{prf}

This case is still harmless if one wishes to work with only the
top-level representatives of extremal diagrams of submodules.  Since
there are only two singular vectors in
$\,\mU_{\hcc(n,m,t),\lcc(n,m,t),t}$, we immediately obtain
\begin{prop}\label{prop:nosing2}
  Under the conditions of Theorem~\ref{twocharge}, the quotient of
  $\,\mU_{\hcc(n,m,t),\lcc(n,m,t),t}$ with respect to the conventional
  singular vectors is irreducible, i.e., there are no subsingular
  vectors in the massive Verma module
  \,$\mU_{\hcc(n,m,t),\lcc(n,m,t),t}$.
\end{prop}

\smallskip

The third possibility in codimension~2, as described in
case~\ref{codim2ct} of the list on page~\pageref{thelist}, is when one
of the states~\req{toverma} belongs to the original module
\,$\mU_{h,\ell,t}$, hence there is a charged singular vector in
\,$\mU_{h,\ell,t}$. The submodule $\mC$ generated from the charged
singular vector contains a singular vector.  One of the possibilities
is that this is a second charged singular vector in
\,$\mU_{h,\ell,t}$, situated on the same side from the \hw{} vector as
the first charged singular vector. Otherwise, the submodule generated
from the singular vector in $\mC$ corresponds to a massive Verma
submodule in \,$\mU_{h,\ell,t}$ in accordance with
Lemma~\ref{chargemass:lemma}.
\begin{thm}\label{thm:codim2}The \hw{} of the massive Verma module
  \,$\mU_{h,\ell,t}$ belongs to the set ${\cal O}_{\rm cm}$ if and
  only if $h=\hcm^\sigma(r,s, n, t)$, $\ell=\ellcm^\sigma(r,s, n, t)$,
  where $\sigma\in\{-,+\}$,
  \begin{eqnarray}
    \hcm^\pm(r, s, n, t)\kern-4pt&=&\kern-4pt
    \frac{1-2n}{t}\pm(s - \frac{r}{t})\,,
    \qquad
    \ellcm^\pm(r, s, n, t)=\frac{n}{t}(-n \pm(s t - r))\,,
    \label{bars}\\
    (\sigma,r,s,n,t)\kern-4pt&\in&\kern-4pt
    \Bigl(\{\pm\}\times\oN\times\oN\times\oZ\times(\oC\setminus\oQ)
    \Bigr)
    \bigcup\,\oA\bigcup\,\oB\,,\nonumber
  \end{eqnarray}
  where
  \begin{eqnarray}    
      \oA&=&\Bigl\{(\{+\},r,0,n,t)\!\Bigm|\!r\in\oN,~n\in\oN,~t\in\oC
      \setminus\oQ
      \Bigr\}
      \bigcup\,\Bigl\{(\{-\},r,0,n,t)\!
      \Bigm|\!r\in\oN,~n\in-\oN_0,~t\in\oC
      \setminus\oQ\Bigr\}\,,\nonumber
      \\
      \oB&=&\Bigl\{(\{+\},r,s,n,-\frac{p}{q})\!
      \Bigm|\!n\in\oN,~p,q\in\oN,~
      1\leq r\leq p,~0\leq s\leq q-1\Bigr\}
      \bigcup\,\label{oAoB}\\
      {}&{}&\Bigl\{(\{-\},r,s,n,-\frac{p}{q})\!\Bigm|\!n\in-\oN_0,~
      p,q\in\oN,~
      1\leq r\leq p,~0\leq s\leq q-1\Bigr\}\,.\nonumber
  \end{eqnarray}
  Then, \,$\mU_{\hcm^\pm(r,s, n, t),\ellcm^\pm(r,s, n, t),t}$ contains
  a twisted topological Verma submodule
  \,$\mC\approx\smV_{\hpm(r,s,t),t;-n}$ generated from the charged
  singular vector $\ket{E(n,\hcm^\pm(r, s, n, t),t)}_{\rm ch}$.
  
  For $s\neq0$, further, \,$\mU_{\hcm^\pm(r,s, n, t),\ellcm^\pm(r,s,
    n, t),t}$ contains a massive Verma submodule \ $\mU'$ generated
  from the massive singular vector $\ket{S(r,s,\hcm^\pm(r, s, n,
    t),t)}^+$ (if $n\geq1$) or $\ket{S(r,s,\hcm^\pm(r, s, n, t),t)}^-$
  (if $n\leq0$), where $r,s\geq1$.  The maximal submodule in
  \,$\mU_{\hcm^\pm(r,s, n, t),\ellcm^\pm(r,s, n, t),t}$ is
  \,$\mU'\cup\mC$.  The intersection \,$\mU'\cap\,\mC$ is a twisted
  topological Verma module generated from the topological singular
  vector in $\mC$ given by
  \begin{equation}\new
    \begin{array}{rcl}
      \ket{T}^\pm&=&\left\{\kern-5pt\new
        \begin{array}{ll}
          {\cal E}^{\pm,-n}(r, s + \half \pm \half, t)\,
          \ket{E(n,\hcm^\pm(r, s, n, t),t)}_{\rm ch}\,,&n\in\oN\,,\\
          {\cal E}^{\pm,-n}(r, s + \half \mp \half, t)\,
          \ket{E(n,\hcm^\pm(r, s, n, t),t)}_{\rm ch}\,,&n\in-\oN_0\,.
        \end{array}\right.
    \end{array}
    \label{innerhw}
  \end{equation}
  When $s=0$, the corresponding state \req{innerhw} is a topological
  singular vector in \,$\mC$ and, at the same time, a charged singular
  vector in~\,$\mU_{h,\ell,t}$.
\end{thm}
\begin{prf}
  The definition of ${\cal O}_{\rm cm}$ requires that precisely one of
  the solutions of Eq.~\req{ell} be an integer, whence the existence
  of submodule~$\mC$ follows. Further, by the definition of ${\cal
    O}_{\rm cm}$, the module~$\mC$ contains precisely one singular
  vector. This is vector~\req{innerhw}.  If this vector were
  $\ket{E^+(r,1,t)}$ for $n>0$ or $\ket{E^-(r,1,t)}$ for $n\leq0$, it
  would be a second charged singular vector in~\,$\mU_{h,\ell,t}$.
  Then, by the conditions of the Theorem and
  Lemma~\ref{chargemass:lemma}, there are no other submodules
  in~\,$\mU_{h,\ell,t}$. If singular vector~\req{innerhw} is not one
  of the above, we see from Lemma~\ref{chargemass:lemma} that any
  other submodule in~\,$\mU_{h,\ell,t}$ is a massive Verma module that
  has a nontrivial intersection with~$\mC$.  However, any submodule
  in~\,$\mU_{h,\ell,t}$ can intersect~$\mC$ over the submodule
  generated from the only singular vector~\req{innerhw} in~$\mC$.
  Thus, \,$\mU_{h,\ell,t}$ can contain only one massive submodule.
  Finally, the state $\ket{S(r,s,\hcm^\pm(r, s, n,t),t)}^+$ in the
  case of $n\geq1$ or $\ket{S(r,s,\hcm^\pm(r, s, n, t),t)}^-$ in the
  case of $n\leq0$ generates this submodule because the respective
  state satisfies the (twisted) massive \hw{} conditions and does not
  belong to~$\mC$ since~$\mC$ contains no states with the gradings as
  that of the {\it respective\/} $\ket{S(\ldots)}^\pm$
  state\footnote{On the other hand, for $n>0$ for example, the vector
    $\ket{S(r,s, \hcm^-(r, s, n, t), t)}^-$ belongs to the topological
    Verma submodule $\mC'$ generated from the \hw{}
    state~\req{innerhw} whenever~$n\leq r(s+1)$, in which case it is
    then a \dGdsc\ of~$\ket{T}^-$: \ $\ket{S(r,s,\hcm^-(r, s, n,
      t),t)}^-= \cG_{-rs}\,\ldots\,\cG_{r-n-1}\,\ket{T}^- $, and
    similarly for $n\leq0$ with ${+}\leftrightarrow{-}$.}.
\end{prf}

The situation described in the Theorem is illustrated in the following
extremal diagram (choosing, for definiteness, $n>0$ and the `$-$' case
in~\req{bars})
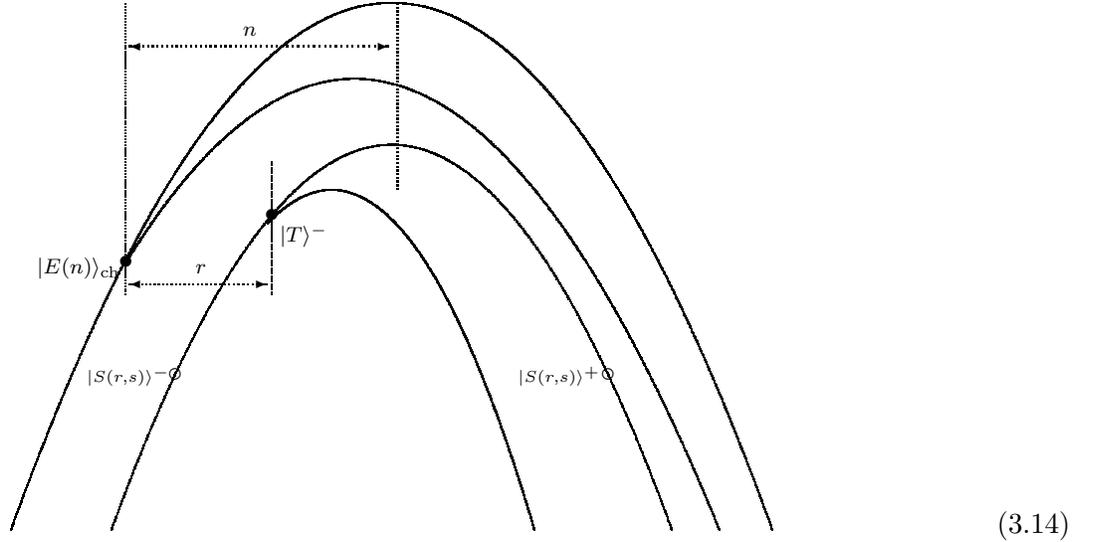
\begin{equation}
  \message{please wait...}
  \unitlength=1pt
  \begin{picture}(300,210)
    \put(104,106){${}^{\ket{T}^-}$}
    \bezier{\dimtwenty}(2,1)(146,400)(290,1)    
    \bezier{50}(148.3,200)(148.3,165)(148.3,130)
    \put(-103,0){\bezier{100}(148.3,200)(148.3,144)(148.3,90)}
    \bezier{30}(47,94)(70,94)(97,94)
    \put(48,94){\vector(-1,0){1}}
    \put(98,94){\vector(1,0){1}}
    \put(72,94){${}^r$}
    \put(0,90){
      \bezier{50}(47,94)(90,94)(144,94)
      \put(48,94){\vector(-1,0){1}}
      \put(144,94){\vector(1,0){1}}
      \put(90,94){${}^n$}
      }
    \bezier{\dimten}(40,1)(146,292.5)(252,1)
    \put(12,94){${}^{\ket{E(n)}_{\rm ch}}$}
    \put(31,59){${}_{{}^{\ket{S(r,s)}^-}}$ }
    \put(64.3,60){\circle{4}}
    \put(194,59){${}_{{}^{\ket{S(r,s)}^+}}$ }
    \put(228,60){\circle{4}}
    \put(43,100){$\bullet$}
    \put(-80,-10){
      \put(32.3,0){\bezier{50}(148.3,150)(148.3,125)(148.3,100)}
      \put(178.2,127.5){$\bullet$}
      }
    \bezier{\dimfifteen}(46,103)(156,280)(270,1)
    \bezier{\dimfifteen}(99,117)(150,170)(200,1)
  \end{picture}
  \message{done}
  \label{thethird}
\end{equation}\pagebreak[3]$\!$
Here, the vector $\ket{S(r,s)}^+\equiv\ket{S(r,s, \hcm^-(r, s, n, t),
  t)}^+$ is a representative of the massive singular vector in the
sense of Definition~\ref{defsingmass}, since its \ddsc s generate the
entire extremal diagram of the massive Verma submodule~\,$\mU'$.
According to~\req{innerhw}, the twisted topological \hw{} state
$\ket{T}^-$ is the embedding of the topological singular vector
$\ket{E(r,s,t)}^-$ into the submodule built on the charged singular
vector $\ket{E(n)}_{\rm ch}\equiv\ket{E(n,\hcm^-(r, s, n, t),t)}_{\rm
  ch}$.  The diagram shows the case where~$n\leq r(s+1)$ and, thus,
the vector $\ket{S(r,s)}^-\equiv\ket{S(r,s, \hcm^-(r, s, n, t), t)}^-$
belongs to the topological Verma submodule $\mC'$ generated from the
\hw{} state~\req{innerhw}.  In particular, its \ddsc s do not generate
the same diagram as \ddsc s of $\ket{S(r,s,\hcm^-(r, s, n, t),t)}^+$.
Together, the vectors $\ket{E(n,\hcm^-(r, s, n, t),t)}_{\rm ch}$ and
$\ket{S(r,s,\hcm^-(r, s, n, t),t)}^+$ generate a maximal submodule.
The submodules generated by each of these vectors intersect over the
submodule generated from~$\ket{T}^-$.

In this case, when one wishes to work with only the conventional,
top-level, representatives of singular vectors, one has to pay the
price of considering subsingular vectors. Their positions and explicit
constructions are a direct consequence of the above analysis.  In the
following proposition, we thus assume the conventional definition of
singular vectors, demanding that these always satisfy the `untwisted'
\hw{} conditions. Then, the subsingular vectors are as follows:
\begin{prop}\label{prop:cm}
  Under the conditions of Theorem~\ref{thm:codim2}, the massive Verma
  module \,$\mU_{h,\ell,t}$ contains a subsingular vector if and only
  if
  \begin{enumerate}\addtolength{\parskip}{-6pt}

  \item either $r\geq n>0$ and $h=\hcm^-(r, s, n, t)$, $s\geq1$,

  \item or $n\leq0$, $r\geq|n|+1$, and $h=\hcm^+(r, s, n, t)$,
    $s\geq1$.
  \end{enumerate}
  In the first case, the subsingular vector is given by
  \begin{eqnarray}
    \ket{\rm Sub}\kern-6pt&=&\kern-6pt\cG_{0}\ldots
    \cG_{-n+r}q(1-r+n,n + t-1)\,
    \cE^{+,-n+r-t}(r,s,t)\,q(n - r + t,0)\,
    \ket{\hcm^-(n,r,s,t),\ellcm^-(n,r,s,t),t}\nonumber\\
    {}\kern-6pt&=&\kern-6pt
    \cG_0\,\ldots\cG_{r-n-1}\,\cG_{r-n+1}\,\ldots\cG_{rs-1}\,
    \ket{S(r,s,\hcm^-(n,r,s,t),t)}^+\,.
    \label{subsing:cm}
  \end{eqnarray}
  This vector (which has the relative charge $1$) becomes singular in
  the module obtained by taking the quotient over the submodule
  generated by the (top-level) singular vector
  \begin{equation}
    \ket{s}=
    \cG_{0}\ldots \cG_{r-n-1}\,
    \cE^{-,-n}(r,s,t)\,\cG_{-n}\ldots\cG_{-1}\,
    \ket{\hcm^-(n,r,s,t),\ellcm^-(n,r,s,t),t}\,.
  \end{equation}
  In the case where $n\leq0$, $r\geq|n|+1$, similarly,
  \begin{equation}
    \ket{\rm Sub}= \cQ_1\ldots\cQ_{r+n-1}\,\cQ_{r+n+1}\ldots\cQ_{rs}\,
    \ket{S(r,s,\hcm^+(n,r,s,t),t)}^-\,.
  \end{equation}
\end{prop}
Let us remind the reader that, as in the general
construction~\req{Sminus}, \req{Splus} of $\N2$ singular vectors, the
state \ $q(1-r+n,n + t-1)\,\cE^{+,-n+r-t}(r,s,t)\,q(n - r + t,0)\,
\ket{h,\ell,t}$ \ in \req{subsing:cm} evaluates as an element of
\,$\mU_{h,\ell,t}$ using the formulae of Sec.~\ref{AlgI}, see
also~\cite{[ST3]}. Recall also that, as before,
$\cE^{\pm,\theta}(r,s,t)$ are topological singular vector operators
transformed by the spectral flow with the parameter $\theta$.
\begin{prf}
  Consider, for definiteness, case~1 of the Proposition. Then, the
  module~$\mU_{h,\ell,t}$ contains only three submodules
  $\mC\subset\mU_{h,\ell,t}$, $\mU'\subset\mU_{h,\ell,t}$, and
  $\mC'\subset\mU'$, $\mC'\subset\mC$, where $\mC$ and $\mU'$ are as
  in Theorem~\ref{thm:codim2} and
  $\mC'\approx\smV_{\hcm^-(r,s,n,t)+\frac{2}{t}(n-r),t;r-n}$. The
  embeddings are given by the singular vectors described in
  Theorem~\ref{thm:codim2}.  The submodule~$\mC'$ is embedded by the
  singular vector
  \begin{equation}
    \ket{T}^-=\cE^{-,-n}(r,s,t)\,\cG_{-n}\ldots\cG_{-1}\,
    \ket{\hcm^-(n,r,s,t),\ellcm^-(n,r,s,t),t}\,.
    \label{T}
  \end{equation}
  Obviously, $\ket{T}^-$ is inside the submodule generated from the
  charged singular vector
  \begin{equation}
    \ket{E(n,\hcm^-(n,r,s,t),t)}_{\rm ch}=\cG_{-n}\,\ldots\,\cG_{-1}\,
    \ket{\hcm^-(n,r,s,t),\ellcm^-(n,r,s,t),t}\,.
    \label{OK}
  \end{equation}
  Further, the top-level representative
  $$
  \ket{c}=\cQ_{1}\,\ldots\,\cQ_{n-1}\,\cG_{-n}\,\ldots\,\cG_{-1}\,
  \ket{\hcm^-(n,r,s,t),\ellcm^-(n,r,s,t),t}
  $$
  of this charged singular vector generates the module~$\mC$.  In
  the conventional description, the existence of a subsingular vector
  in the module~$\mU_{h,\ell,t}$ depends on whether the top-level
  representative~$\ket{s}$ of the extremal diagram connecting
  $\ket{T}^-$ and $\ket{S(r,s,\hcm^-(n,r,s,t),t)}^+$ belongs to the
  submodule~$\mC'$.  It is clear that
  \begin{equation}\new
    \begin{array}{l}
      \ket{s}=\cG_{0}\ldots \cG_{r-n-1}\,\ket{T}^-\,,\qquad
      r\geq n>0\quad
      \Longrightarrow\quad\ket{s}\in\mC'\,,\\
      \ket{T}^-=\cG_{r-n}\ldots \cG_{-1}\,\ket{s}\,,\qquad
      n\geq r>0\quad
      \Longrightarrow\quad\ket{s}\not\in\mC',
    \end{array}
  \end{equation}
  whence we see that the quotient of~$\mU_{h,\ell,t}$ over
  conventional singular vectors is reducible in the case where~$r\geq
  n>0$.
\end{prf}

In terms of extremal diagrams, the conditions relating $n$ and $r$
mean that the twisted topological \hw{} state $\ket{T}^-$
in~\req{thethird} has gone past the top of the `massive' parabola
(i.e., past the conventional singular vector). Therefore, the extremal
diagram actually takes the following form:
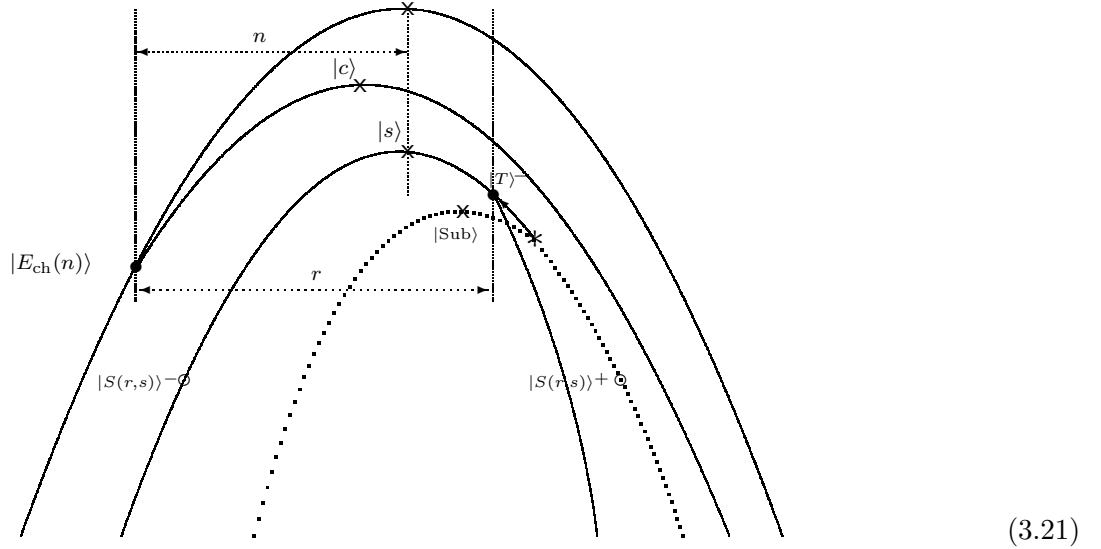
\begin{equation}
  \message{please wait...}
  \unitlength=1pt
  \begin{picture}(300,210)
    \put(120,172){${}^{\ket{c}}$}
    \put(137,148){${}^{\ket{s}}$}
    \bezier{\dimtwenty}(2,1)(146,400)(290,1)
    \put(146,198){${\ssf x}$}
    \bezier{50}(148.3,200)(148.3,165)(148.3,130)
    \put(-103,0){\bezier{100}(148.3,200)(148.3,144)(148.3,90)}
    \put(32.3,0){\bezier{100}(148.3,200)(148.3,144)(148.3,90)}
    \bezier{50}(47,94)(112,94)(178,94)
    \put(48,94){\vector(-1,0){1}}
    \put(178,94){\vector(1,0){1}}
    \put(112,94){${}^r$}
    \put(0,90){
      \bezier{50}(47,94)(90,94)(146,94)
      \put(47,94){\vector(-1,0){1}}
      \put(147,94){\vector(1,0){1}}
      \put(90,94){${}^n$}
      }
    \bezier{\dimten}(40,1)(116,215)(195.8,114)
    \put(-2,98){${}^{\ket{E_{\rm ch}(n)}}$}
    \put(179,136){${}_{{}^{\ket{T}^{\!-}}}$ }
    \put(185,126.5){\vector(-1,1){2}}
    \put(31,59){${}_{{}^{\ket{S(r,s)}^-}}$ }
    \put(158,114){${}_{{}^{\ket{\rm Sub}}}$ }
    \put(64,60){\circle{4}}
    \put(194,59){${}_{{}^{\ket{S(r,s)}^+}}$ }
    \put(229,60){\circle{4}}
    \put(146,144){${\ssf x}$}
    \put(128,169){${\ssf x}$}
    \put(193.1,110){\Large$\ast$}
    \put(167,121.3){${\ssf x}$}
    \put(43,100){$\bullet$}
    \put(178.2,127.5){$\bullet$}
    \bezier{\dimfifteen}(46,103)(156,280)(270,1)
    \bezier{400}(181,130)(212,65)(220,1)
    {\linethickness{1pt}
      \bezier{50}(90,1)(130,160)(195.8,113)
      \bezier{40}(198.8,109.3)(232,65)(252,1)
      }
  \end{picture}
  \label{onesided}
  \message{done}
\end{equation}
Here, the crosses denote the conventional, top-level, representatives,
the $\bullet$ states satisfy twisted topological \hw{} conditions,
$\ast$ is the state (a descendant of $\ket{S(r, s,
  \hcm^-(n,r,s,t),t)}^+$) such that $\ket{T}^-=\cG_{-n+r}\cdot(\ast)$,
and $\ket{E(n)}_{\rm ch}\equiv\ket{E(n,\hcm^-(n,r,s,t),t)}_{\rm ch}$
(and, as before, we consider the case where $r\geq n > 0$).
The arrow in the diagram, which represents the action of $\cG_{r-n}$,
cannot be inverted because of the twisted topological \hw{} conditions
at $\ket{T}^-$, therefore the dotted line cannot be reached by the
action of elements of the $\N2$ algebra on either $\ket{T}^-$ or the
top-level representative $\ket{s}$ (nor, in fact,~$\ket{c}$).
Instead, acting with the highest of modes of $\cQ$ that produces a
non-vanishing result, one spans out the lower branch originating at
$\ket{T}^-$, which is shown in the solid line.  After taking the
quotient with respect to the singular vector
$\ket{S(r,s)}^-\equiv\ket{S(r,s,\hcm^-(n,r,s,t),t)}^-$ (or, {\it
  equivalently\/}, $\ket{s}$), we are left with the submodule whose
extremal diagram is precisely the dotted line.  Then the state
$\ket{\rm Sub}$ (the top-level representative of this diagram) is a
{\it sub\/}singular vector.

However, rather than describing the structure of $\N2$ Verma modules
in terms of subsingular vectors, it is much more convenient to
construct those vectors that do generate maximal submodules.
In~\req{onesided}, this is the canonical representative
$\ket{S(r,s)}^+\equiv\ket{S(r,s,\hcm^-(n,r,s,t),t)}^+$
from~\req{Splus}.

\subsection{Codimension-$3$ cases.}\lvm
Now we are going to analyze codimension-3 degenerations.\nopagebreak

Let us begin with the case when a further degeneration occurs in
Theorem~\ref{twomassive}, as described in case~\ref{codim3rat} of the
list on page~\pageref{thelist}.  Namely, one more massive Verma
submodule appears in the diagram of the type of~\req{thethird}, with
its own topological point similar to $\ket{T}^-$. All such topological
points are at the same time topological singular vectors in the
submodule generated from a charged singular vector
(Lemma~\ref{chargemass:lemma}). In this way, the structure of
submodules in the massive Verma module is still essentially described
by that of its topological Verma submodule generated from a charged
singular vector (while the structure of the topological Verma module,
in turn, is the same as for the corresponding $\tSL2$ Verma module).
\begin{thm}\label{thm:3rat}The \hw{} of the massive Verma module
  \,$\mU_{h,\ell,t}$ belongs to the set ${\cal O}_{\rm cmm}$ if and
  only if $h=\hcm^\sigma(r,s, n, t)$, $\ell=\ellcm^\sigma(r,s, n, t)$,
  where $\sigma\in\{-,+\}$ and
  \begin{equation}
  (\sigma,r,s,n,t)\in
  \Bigl((\{\pm\}\times\oN\times\oN\times\oZ\times\oQ)\,
  \bigcup\,\oA'\Bigr)
  \setminus\,\oB\,,
  \label{Ocmmsolut}
\end{equation}
  where $\oB$ is as in~\req{oAoB},
  $$\oA'=\Bigl\{(\{+\},r,0,n,t)\!\Bigm|\!r\in\oN,~n\in\oN,~t\in\oQ
  \Bigr\}
  \bigcup\,\Bigl\{(\{-\},r,0,n,t)\!\Bigm|\!r\in\oN,~n\in-\oN_0,~t\in\oQ
  \Bigr\}\,,$$
  and, with $t=\frac{p}{q}$,
  \begin{equation}
    \left|r-\frac{ps}{q}\right|\not\in\{|n|,|n|+1,|n|+2,\ldots\}\,.
    \label{ineq}
  \end{equation}
  Then, the structure of \,$\mU_{h,\ell,t}$ is described as follows:
  \begin{enumerate}\addtolength{\parskip}{-6pt}
  \item there exists a twisted topological Verma
    submodule~$\mC=\smV_{\htop^\pm(r,s,\frac{p}{q}),\frac{p}{q};-n}
    \hookrightarrow \mU_{\hcm^\pm(r, s, n, \frac{p}{q}),\ellcm^\pm(r,
      s, n, \frac{p}{q}),\frac{p}{q}}$, where the embedding is given
    by singular vector~\req{ECh};
    
  \item any other primitive submodule in \,$\mU_{\hcm^\pm(r, s, n,
      \frac{p}{q}),\ellcm^\pm(r, s, n, \frac{p}{q}),\frac{p}{q}}$
    satisfies one of the following:
    \begin{enumerate}
    \item it is a submodule in $\mC$ (hence, a twisted topological
      Verma module);

    \item it is a massive Verma module $\mU'$ generated from the
      representative \ $\displaystyle\ket{S(r,s,\hcm^\pm(r, s, n,
        \frac{p}{q}),\frac{p}{q})}^+$ \ (if $n\geq1$) or \
      $\displaystyle\ket{S(r,s,\hcm^\pm(r, s, n,
        \frac{p}{q}),\frac{p}{q})}^-$ \ (if $n\leq0$) of the massive
      singular vector, where $r,s\geq1$.  Then there exists a vector
      $\widetilde{\ket{T}}^\pm\in\mU'$ that satisfies twisted massive
      \hw{} conditions with the twist parameter $\theta=\mp r-n$ (if
      $n\leq0$) or $\theta=\mp r-n+1$ (if $n>0$) and such that the
      vector
      \begin{equation}
        \ket{T}^\pm=\left\{\kern-6pt\new
          \begin{array}{rcll}
            \cQ_{n\pm r}\widetilde{\ket{T}}^\pm\kern-7pt&=&\kern-6pt
                \cE^{\pm,-n}(r,s + \half \mp \half,\frac{p}{q})\cQ_{n}
                \ldots\cQ_{0}
                \ket{\hcm^\pm(n,r,s,\frac{p}{q}),
                  \ellcm^\pm(n,r,s,\frac{p}{q}),\frac{p}{q}},\kern-7pt
            & n\leq0,\kern-4pt\\
            \cG_{-n\mp r}\widetilde{\ket{T}}^\pm\kern-7pt&=&\kern-6pt
                \cE^{\pm,-n}(r,s + \half \pm \half,\frac{p}{q})\cG_{-n}
                \ldots\cG_{-1}
                \ket{\hcm^\pm(n,r,s,\frac{p}{q}),
                  \ellcm^\pm(n,r,s,\frac{p}{q}),\frac{p}{q}},\kern-7pt
            & n\geq1,\kern-4pt
          \end{array}\right.
        \label{ketT}
      \end{equation}
      satisfies twisted topological \hw{} conditions and generates the
      twisted topological Verma module \,$\mC'
      =\mU'\cap\,\mC$.
    \end{enumerate}
    
  \item\label{item:last} For any twisted topological Verma submodule
    \,$\mC'\subset\,\mC$, there exists a massive Verma submodule
    \,$\mU'\subseteq\mU_{h,\ell,t}$ such that
    \begin{enumerate}
    \item\label{work01} $\mC'\subset\mU'\cap\mC$ is a submodule in
      \,$\mU'$ corresponding to a charged singular vector;

    \item there exists a vector $\widetilde{\ket{T}}\in\,\mU'$ that
      generates \,$\mU'$ such that the vector $\ket{T}$ defined as
      in~\req{ketT} generates a twisted topological Verma module that
      either coincides with \,$\mC'$ or contains \,$\mC'$ as a
      submodule generated from the topological singular vector
      $\ket{E(a, 1, \frac{p}{q})}^{+,-n}$, $a\in\oN$, or $\ket{E(a, 1,
        \frac{p}{q})}^{-,-n}$, $a\in\oN$, for $n\geq1$ and $n\leq0$
      respectively.
    \end{enumerate}
  \end{enumerate}
\end{thm}
In case~\ref{item:last} of the Theorem, $\mU'=\mU_{h,\ell,t}$ occurs
only when~$s=0$ (when there are two charged singular vectors on one
side of the \hw{} state), otherwise $\mU'\mathrel{
  \raisebox{-4pt}{\mbox{$\stackrel{\textstyle\subset}{\scriptstyle\neq}$}}}
\mU_{h,\ell,t}$. For negative rational $t$ and small $r$ and $s$, the
excluded cases where only one massive Verma submodule exists are those
covered by Theorem~\ref{thm:codim2}.  As in the above,
$\cE^{\pm,\theta}(r,s,t)$ denote topological singular vector operators
transformed by the spectral flow with the parameter~$\theta$.
\begin{prf}
  {}From the definition of ${\cal O}_{\rm cmm}$, follow the equations
  on $h$, $\ell$, and $t$ with the
  solutions~\req{Ocmmsolut}--\req{ineq}.  Item~1 of the Theorem is a
  part of the definition of ${\cal O}_{\rm cmm}$.  Further, by
  Lemma~\ref{chargemass:lemma}, each twisted topological Verma
  submodule is embedded into the module~$\mC$ by a topological
  singular vector.  Each massive submodule has a non-empty
  intersection with~$\mC$.  This intersection is generated from the
  topological singular vector written on the RHS of~\req{ketT}.

  The crucial point in the Theorem is the existence of the
  $\widetilde{\ket{T}^\pm}$ states.  For definiteness, we choose $n>0$
  and the `$-$' case in~\req{ketT}.  We then apply the $g$ operator of
  length $-1$ to the twisted topological \hw{} state $\ket{T(r, s, n,
    t)}\equiv\ket{T}^-$ from~\req{ketT},
  \begin{eqnarray}
     \widetilde{\ket{T}^-}\equiv\ket{\widetilde{T}(r,s, n,t)} &=&
      g(r-n+1,r-n-1)\,\ket{T(r, s, n, t)}\nonumber\\
      {}&=&g(r-n+1,r-n-1)\,{\cal E}^{-,-n}(r,s,t)\,
      \ketM{E(n,\hcm^-(r, s, n, t),t}_{\rm ch}\,,
    \label{tildeSigma}
  \end{eqnarray}
  in accordance with the rules of Sec~\ref{AlgI}.  The condition for
  the $\ket{\widetilde{T}(r,s, n,t)}$ state to exist in the module
  $\,\mU_{\hcm^-(n,r,s,t),\ellcm^-(n,r,s,t),t}$ is given by the next
  Lemma, from which we see that, under the conditions of the Theorem,
  $f(r,s,n,t)\neq0$ and therefore the $\widetilde{\ket{T}}$ state does
  exist.
\end{prf}

\vspace{-8pt}

\begin{lemma}\label{lemma:continue}
  The state $\ket{\widetilde{T}(r,s, n,t)}$, Eq.~\req{tildeSigma},
  exists in the massive Verma module~$\mU_{\hcm^-(r, s, n,
    t),\ellcm^-(r, s, n, t), t}$ with $1\leq n\leq r(s+1)$ if and only
  if $f(r,s,n,t)\neq0$, where
  \begin{equation}
    f(r,s,n,t)=\left\{
      \begin{array}{ll}
        \prod\limits_{i=0}^{2r-n}(s t + n - r + i)\,,&n\leq2r\,,\\
        1\,,&n\geq2r+1\,.
      \end{array}\right.
    \label{f}
  \end{equation}
  This state is then a representative of the massive singular vector;
  further, the \dQdsc\ of $\ket{S(r,s,\hcm^-(r, s, n, t),t)}^+$ that
  lies in the same grade as~\req{tildeSigma} is proportional to that
  vector:
  \begin{equation}
    \cG_{r-n+1}\,\ldots\,\cG_{rs-1}\,
    \ket{S(r,s,\hcm^-(r, s, n, t),t)}^+=
    a(r,s,n,t)\,
    \ket{\widetilde{T}(r,s, n,t)}\,,
  \end{equation}
  where $a(r,s,n,t)$ is $\left(\frac{2}{t}\right)^{rs}$ times a
  polynomial of the order $r(s+1)$ in~$t$.
\end{lemma}
\begin{prf}
  To evaluate~\req{tildeSigma}, one uses formulae~\req{LLeft} and
  then, as negative-length $g$ operators reach the \hw{} state, one
  applies the formula,
\begin{equation}
  g(\theta_1,\theta-1) \,\ket{h, \ell, t; \theta}=
  {1\over 2 \hell(h, \ell, t, \theta - \theta_1 + 1)}
  \cQ_{-\theta_1 + 1}\, g(\theta_1 - 1, \theta - 1) \ket{h, \ell, t; \theta}\,,
\end{equation}
(which also follows from Sec.~\ref{AlgI}) with
\begin{equation}
  \hell(h, \ell, t, N) = \ell - \ellch(N,h,t)\,.
  \label{hell}
\end{equation}
In the case at hand, we further use the fact that
\begin{equation}
  \widehat{\ell}(\hcm^-(r, s, n, t), \ellcm^-(r, s, n, t), t, -i)
  = -\frac{1}{t}(i + n) (s t + n - r + i)\,.
\end{equation}
A simple analysis of the relative charge of
$\ket{{\widehat{S}}^-(r,s,\hcm^-(r, s, n, t),t)}$ shows that
$f(r,s,n,t)$ is precisely the function responsible for the existence
of $\ket{\widetilde{T}(r,s, n,t)}$ because, for $t\neq0$, the relevant
factors from the denominators are precisely the above~$f(r,s,n,t)$,
whence the lemma follows.
\end{prf}

While this case is rather straightforward when described in terms of
singular vectors that generate maximal submodules, the analysis of the
same structure in terms of top-level singular vectors and subsingular
vectors that become necessary then is quite lengthy when it comes to
listing all possible occurrences of subsingular vectors.  Comparing
\req{thethird} and \req{onesided} we have seen that the appearance of
subsingular vectors in the conventional setting is due to the fact
that the twisted topological \hw{} state is shifted to a certain side
(depending on the sign, etc., of the parameters) of the top-level
vector in the extremal diagram of the submodule. In the present case,
however, there are two independent massive subdiagrams in the extremal
diagram, each with its own `topological point'. The description in
terms of conventional singular vectors and subsingular vectors would
then amount to classifying all possible relative positions of the
topological points and top-level vectors of the parabolas.  Although
this presents no conceptual difficulties and can be carried out
similarly to Proposition~\ref{prop:cm}, yet there are a large number
of different cases. We omit this analysis, since it does not add
anything to Theorem~\ref{thm:3rat} as regards the structure of
submodules, while at the same time is too long to serve as an example.

\medskip

It remains to consider the case where, in addition to the conditions
of Theorem~\ref{twocharge}, the submodules generated from the charged
singular vector, in their own turn, admit topological singular
vectors. Then, the corresponding twisted topological Verma submodules
may be such that a given massive Verma submodule may be embedded into
a direct sum of two such twisted topological Verma submodules.  The
corresponding massive singular vector then `splits' into a pair of
singular vectors, {\it each of which belongs to the respective twisted
  topological Verma submodule}. In the restricted setting with only
top-level representatives allowed to generate submodules, this can be
observed as the occurrence of two linearly independent singular
vectors in the same grade\footnote{The occurrence of linearly
  independent singular vectors in the same grade was noticed for the
  first time in~\cite{[Doerr2]}.  As we are going to see, they are
  necessarily elements of twisted topological, not massive, Verma
  submodules.} or as the appearance of a singular vector and a
subsingular vector in the same grade.

As in Theorem \ref{twocharge}, we assume for definiteness that the
integers labelling the charged singular vectors in the massive Verma
module~$\mU_{h,\ell,t}$ are such that $n>0$ and $m\leq0$.  By the {\it
  distance\/} between any two vectors on the same extremal diagram we
mean the difference of their $U(1)$ charges.  Then the distance
between the charged singular vectors in $\,\mU_{\hcc(n,m,t),
  \lcc(n,m,t),t}$ is equal to~$-m+n+1$.

Now we are ready to describe case \ref{codim3last} of the list on
page~\pageref{thelist}, i.e., the coexistence of a massive singular
vector with two charged singular vectors on different sides of the
\hw{} vector:
\begin{lemma}\label{lemma:3last}The \hw{} of the massive Verma module
  \,$\mU_{h,\ell,t}$ belongs to the set ${\cal O}_{\rm ccm}$ if and
  only if
    \begin{equation}
    h=\hcc(n,m,t)\,,\qquad\ell=\lcc(n,m,t)\,,\quad
    t= \frac{r \pm (n-m)}{s}\,,\quad
    r,s,n\in\oN\,,\quad m\in-\oN_0\,.
  \end{equation}
  Then, the massive Verma module \,$\mU_{h,\ell,t}$ contains twisted
  topological submodules \,$\mC_1$ and \,$\mC_2$ generated from the
  charged singular vectors \ $\displaystyle\ket{E(n, \frac{(1 - m -
      n)s}{r \mp (m - n)}, \frac{r \pm (n-m)}{s})}_{\rm ch}$ and \ 
  $\displaystyle\ket{E(m, \frac{(1 - m - n)s}{r \mp (m - n)}, \frac{r
      \pm (n-m)}{s})}_{\rm ch}$ respectively.  Each of the modules
  \,$\mC_1$ and \,$\mC_2$ admits a singular vector; moreover, a
  singular vector $\ket{E(a,b,t)}^{\mp,-n}$ exists in \,$\mC_1$ if and
  only if\, $\ket{E(a,b,t)}^{\pm,-m}$ exists in \,$\mC_2$ with the
  same $a,b\in\oN$ (and with the above $t$). Any other primitive
  submodule \,$\mU'\subset\,\mU_{h,\ell,t}$ satisfies one of the
  following:
  \begin{enumerate}\addtolength{\parskip}{-6pt}
  \item it is a twisted topological Verma module, in which case it is
    a submodule of either $\mC_1$ or $\mC_2$;\pagebreak[3]
    
  \item it is a massive Verma module, in which case the non-empty
    intersections \,$\mC'_1=\mU'\cap\,\mC_1$ and
    \,$\mC'_2=\mU'\cap\,\mC_2$ are generated each from a topological
    singular vector in $\mC_1$ and $\mC_2$ respectively. If, then,
    \,$\mC'_1$ is generated from the singular vector
    $\ket{E(a,b,t)}^{\pm,-n}$, then $\mC'_2$ is generated from the
    singular vector $\ket{E(a,b,t)}^{\mp,-m}$ with the same
    $a,b\in\oN$ (and with the above $t$). Each of the submodules
    \,$\mC'_1$ and \,$\mC'_2$ is at the same time generated from a
    charged singular vector in~\,$\mU'$.
  \end{enumerate}
\end{lemma}
\begin{prf}
  The definition of~${\cal O}_{\rm ccm}$ means that both solutions of
  Eq.~\req{ell} are integers of different signs and, in addition, the
  states~\req{toverma} admit a topological singular vector,
  whence~\req{lemma:3last} follows. Further, each of the modules
  $\mC_1$ and~$\mC_2$ generated from the charged singular vectors
  contains a topological singular vector.  Now assuming that the \hw{}
  vector of~$\mC_1$ is~$\kettop{\htop^{\pm}(a,b,t),t;-n}$, we see that
  the \hw{} vector of~$\mC_2$ is~$\kettop{\htop^{\mp}(a,b,t),t;-m}$.
  Therefore the assertion that a singular vector
  $\ket{E(a,b,t)}^{\mp,-n}$ exists in \,$\mC_1$ if and only if\,
  $\ket{E(a,b,t)}^{\pm,-m}$ exists in \,$\mC_2$ with the same
  $a,b\in\oN$ is obvious.

  Note further that the quotient of~\,$\mU_{h,\ell,t}$ over $\mC_1$ or
  $\mC_2$ is a twisted topological Verma module $\mQ_1$ or $\mQ_2$
  respectively.  Let us assume that there exists a twisted topological
  \hw{} state~$\ket{t}$ such that $\ket{t}\not\in\mC_1$,
  $\ket{t}\not\in\mC_2$.  Then this state is a topological singular
  vector in $\mQ_1$ and, likewise, in~$\mQ_2$.  However, the Verma
  module generated from~$\ket{t}$ contains states in the gradings
  where there are no states from either the $\mQ_1$ or the $\mQ_2$
  module. Therefore, the state~$\ket{t}$ belongs to a twisted
  topological Verma module and, at the same time, generates a
  submodule which is isomorphic to the quotient of a twisted
  topological Verma module.  This contradicts the structure of the
  topological Verma modules described in Theorem~\ref{sl2n2}.  Part~2
  follows immediately from Lemma~\ref{chargemass:lemma}.
\end{prf}

In the case described in the Lemma, therefore, a given massive Verma
submodule \,$\mU'$ in \,$\mU_{h,\ell,t}$ necessarily has two charged
singular vectors lying on the different sides of the \hw{} vector
of~\,$\mU'$. It may be useful to recall the diagram~\req{thethird},
where the topological singular vector $\ket{T}^-$ is, at the same
time, a charged singular vector in the massive Verma module whose
extremal diagram is the parabola connecting $\ket{S(r,s)}^-$ and
$\ket{S(r,s)}^+$. In the present case, we have two topological
points on the extremal diagram of any massive submodule, which are the
\hw{} states of the modules \,$\mC'_1$ and \,$\mC'_2$:
\begin{equation}
  \begin{picture}(200,70)
    \put(20,30){
      \put(0,0){$\displaystyle
        \begin{array}{ccccccccc}
          &&&&\kern-10pt\mU\kern-10pt\\
          &&&&\!\!\Bigm\uparrow\\
          \mC_1\kern-10pt&{}&{}&{}&\kern-10pt\mU'\kern-10pt&{}&{}&{}&
          \kern-10pt\mC_2\\
          {}&\nwarrow&{}&\nearrow&{}&\nwarrow&{}&\nearrow\\
          {}&{}&\kern-10pt\mC'_1\kern-10pt&{}&{}&{}&\kern-10pt\mC'_2
          \kern-10pt
        \end{array}$}
      \put(20,10){\vector(3,1){39}}
      \put(123,10){\vector(-3,1){39}}
      }
  \end{picture}
  \label{diamond}
\end{equation}

Conversely, let us be given any topological singular vector in
\,$\mC_1$,
\begin{equation}
  \ket{e_1} = {\cal E}^{\pm,-n}(a, b, t)\,
  \cG_{-n}\ldots\cG_{-1}\,
  \ketM{\hcc(n,m,t), \lcc(n,m,t), t}\,,
  \label{e1}
\end{equation}
(with $t$ as in the Lemma). Then we find the corresponding singular
vector in~\,$\mC_2$:
\begin{equation}
  \ket{e_2}= {\cal E}^{\mp,-m}(a, b,t)\,
  \cQ_{m}\ldots\cQ_0\,
  \ketM{\hcc(n,m,t), \lcc(n,m,t), t}\,.
  \label{e2}
\end{equation}
Now the question is whether a massive submodule $\,\mU'$ exists in the
Verma module under consideration such that \req{e1} and~\req{e2} {\it
  would be charged singular vectors in that massive Verma
  submodule\/}. This is not always the case.  In more detail, the
structure of the module $\mU$ is described in the following
\begin{thm}\label{thm:3last} Under the conditions of
  Lemma~\ref{lemma:3last},
  \begin{enumerate}\addtolength{\parskip}{-6pt}

  \item\label{I} Whenever $t=\frac{-m + n + r}{s}$, the module
    \,$\mC_1$ contains the state
    \begin{equation}
      \ket{e_1} = {\cal E}^{+,-n}(r, s+1, t)\,
      \cG_{-n}\ldots\cG_{-1}\,
      \ketM{\hcc(n,m,t), \lcc(n,m,t), t}\,,
      \label{e1plus}
    \end{equation}
    that satisfies twisted topological \hw{} conditions. Let then
    \begin{equation}
      \ket{e_2} = {\cal E}^{-,-m}(r, s+1, t)\,
      \cQ_{m}\ldots\cQ_{0}\,
      \ketM{\hcc(n,m,t), \lcc(n,m,t), t}
      \label{e2minus}
    \end{equation}
    be the singular vector in \,$\mC_2$ whose existence is claimed in
    the Lemma. There exist states $\widetilde{\ket{T}}^-$ and
    $\widetilde{\ket{T}}^+$ in $\mU_{\hcc(n,m,t),\lcc(n,m,t),t}$ such
    that
    \begin{equation}\new
      \begin{array}{rcl}
        \cG_{-r-n}\,\widetilde{\ket{T}}^+&=&\ket{e_1}\,,\\
        \cQ_{-r+m}\,\widetilde{\ket{T}}^-&=&\ket{e_2}\,.
      \end{array}\label{descendants}
    \end{equation}
    Each of these two states generates the same massive Verma
    submodule \,$\mU'\subset\,\mU_{h,\ell,t}$, in which $\ket{e_1}$
    and $\ket{e_2}$ are charged singular vectors.

  \item Whenever $t=\frac{m - n + r}{s}$, the module
    \,$\mC_1$ contains the state
    \begin{equation}
      \ket{e_1} = {\cal E}^{-,-n}(r, s, t)\,
      \cG_{-n}\ldots\cG_{-1}\,
      \ketM{\hcc(n,m,t), \lcc(n,m,t), t}
      \label{e1minus}
    \end{equation}
    that satisfies twisted topological \hw{} conditions. Let then
    $\ket{e_2}$ be a singular vector in \,$\mC_2$ whose existence is
    claimed in the Lemma. Then,

    \begin{enumerate}

    \item\label{II-i} if $2r+m-n\leq-1$, there exists a massive Verma
      submodule \,$\mU'\subset\mU_{h,\ell,t}$ generated from any of
      the states $\widetilde{\ket{v_1}}$ or $\widetilde{\ket{v_2}}$
      such that
      \begin{equation}\new
        \begin{array}{rcl}
          \cG_{r-n}\,\widetilde{\ket{v_1}}&=&\ket{e_1}\,,\\
          \cQ_{r+m}\,\widetilde{\ket{v_2}}&=&\ket{e_2}\,,
        \end{array}
      \end{equation}
      and, further, $\ket{e_1}$ and $\ket{e_2}$ are charged singular
      vectors in $\,\mU'$, the distance between them
      being~$-m+n+1-2r$;

    \item\label{II-ii} if $2r+m-n\geq0$, there does not exist a
      massive submodule \,$\mU'$ in \,$\mU_{h,\ell,t}$ in which either
      $\ket{e_1}$ or $\ket{e_2}$ would be a charged singular vector.
      If, further, $2r+m-n\geq1$, then,
      \begin{itemize}
      \item for each $i$ from the range $i=0,\ldots,2r+m-n-1$, the
        states $\cQ_{r+m-i}\,\ldots\,\cQ_{r+m-1}\,\ket{e_2}$ and
        $\cG_{-m-r+i+1}\,\ldots\,\cG_{r-n-1}\,\ket{e_1}$ satisfy
        twisted massive \hw{} conditions, are in the same grade and
        are linearly independent;

      \item the modules \,$\mC'_1$ and \,$\mC'_2$ generated by
        $\ket{e_1}$ and $\ket{e_2}$ respectively, contain topological
        singular vectors $\cG_{-r-m}\ldots\cG_{r-n-1}\,\ket{e_1}$ and
        $\cQ_{-r+n}\ldots\cQ_{r+m-1}\,\ket{e_2}$, respectively.
      \end{itemize}
    \end{enumerate}
  \end{enumerate}
\end{thm}
\begin{prf}
  In case 1 of the Theorem, the states $\widetilde{\ket{T}}^-$ and
  $\widetilde{\ket{T}}^+$ can be written as
  \begin{equation}\new
    \begin{array}{rcl}
      \widetilde{\ket{T}}^+&=&g(-r-n+1,-r-n-1)\,\ket{e_1}\,,\\
      \widetilde{\ket{T}}^-&=&q(-r+m+1,-r+m-1)\,\ket{e_2}\,.
    \end{array}
  \end{equation}
  They exist as elements of \,$\mU_{h,\ell,t}$ in view of the argument
  similar to the one used in Lemma~\ref{lemma:continue}.  The fact
  that $\widetilde{\ket{T}}^-$ and $\widetilde{\ket{T}}^+$ are \ddsc s
  of each other (up to a nonzero factor) is checked by quotienting
  $\mU_{\hcc(n,m,t),\lcc(n,m,t),t}$ over $\mC_1$ or $\mC_2$.  The
  assumption that \ddsc s of $\widetilde{\ket{T}}^-$ and of
  $\widetilde{\ket{T}}^+$ in the same grading are linearly independent
  leads to the contradiction with the structure of the quotient
  $\mU_{\hcc(n,m,t),\lcc(n,m,t),t}/\mC_1$, which is a twisted
  topological Verma module.
  
  In case~2, Lemma~\ref{lemma:continue} assures that the states
  \begin{equation}\new
    \begin{array}{rcl}
      \widetilde{\ket{v_1}} &=&g(r-n+1,r-n-1)\,\ket{e_1}\,,\\
      \widetilde{\ket{v_2}} &=&q(r+m+1,r+m-1)\,\ket{e_2}
    \end{array}
  \end{equation}
  exist in case (a) and do not exist in~\,$\mU_{h,\ell,t}$ in
  case~(b).
\end{prf}

Different cases in the Theorem are thus distinguished by whether or
not there exists a {\it massive\/} Verma submodule \,$\mU'$ such that
its intersections with \,$\mC_1$ and \,$\mC_2$ coincide with
submodules $\mC'_1\subset\mC_1$ and $\mC'_2\subset\mC_2$ generated
from the singular vectors~$\ket{e_1}$ and~$\ket{e_2}$ respectively.
In cases~\ref{I} and~\ref{II-i}, one has the embeddings as
in~\req{diamond}, whereas in case~\ref{II-ii} there is no massive
Verma submodule $\mU'$ in which the \hw{} vectors of $\mC'_1$ and
$\mC'_2$ would be charged singular vectors. Case \ref{II-ii} of the
Theorem can be illustrated~as
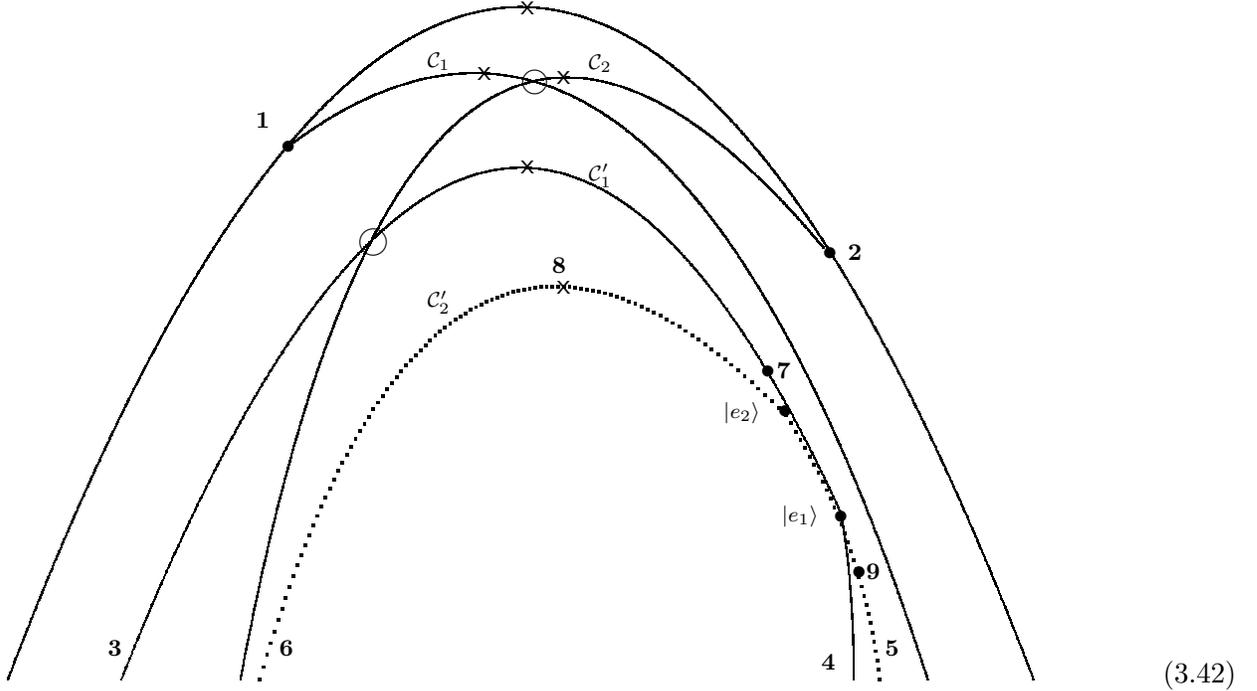
\begin{equation}
  \message{please wait...}
  \unitlength=1pt
  \begin{picture}(400,260)
    \bezier{\dimtwenty}(2,1)(196,510)(390,1)
    \put(196.2,253){${\ssf x}$}
    \put(161,229){$^{\mC_1}$}
    \put(201.5,227){\circle{9}}
    \put(180,228){${\ssf x}$}
    \put(222,228){$^{\mC_2}$}
    \put(222,186){$^{\mC'_1}$}
    \put(161,138){$^{\mC'_2}$}
    \put(210,226.5){${\ssf x}$}
    \put(196.2,192.5){${\ssf x}$}
    \put(210,147){${\ssf x}$}
    \put(105.5,200){$\bullet$}
    \bezier{\dimfifteen}(108.5,203)(250,310)(350,1)
    \put(310.5,160){$\bullet$}
    \bezier{\dimfifteen}(90,1)(155,350)(310.5,164.5)
    \put(140.5,166.5){\circle{10}}
    \put(25,0){
      \put(268.5,100){$\bullet$}
      \put(248,96){${}^{\ket{e_2}}$}
      \put(289.5,60){$\bullet$}
      \put(270,57){${}^{\ket{e_1}}$}
      \bezier{200}(292.5,62)(297,35)(297,1)
      \bezier{\dimten}(20,1)(160,355)(293,62)
      {\linethickness{1pt}
        \bezier{100}(72,1)(140,235)(271,101)
        \bezier{30}(271,101)(300,55)(306.5,1)
        }
      }
    \put(96,210){\footnotesize\bf 1}
    \put(320,160){\footnotesize\bf 2}
    \put(40,10){\footnotesize\bf 3}
    \put(310,5){\footnotesize\bf 4}
    \put(334,10){\footnotesize\bf 5}
    \put(105,10){\footnotesize\bf 6}
    \put(293,115){\footnotesize\bf 7}
    \put(208,155){\footnotesize\bf 8}
    \put(327,39){\footnotesize\bf 9}
    \put(321.5,39){$\bullet$}
    \put(287,115){$\bullet$}
  \end{picture}
  \message{done}
  \label{overlap}
\end{equation}
Here, {\bf 1} and {\bf 2} are the charged singular vectors in the
massive Verma module~$\mU$, which read
$\cG_{-n}\ldots\cG_{-1}\cdot\ket{\frac{(1 - m - n) s}{m - n + r},
  \frac{m n s}{n - m - r}, \frac{m - n + r}{s}}$ and
$\cQ_{m}\ldots\cQ_{0}\cdot\ket{\frac{(1 - m - n) s}{m - n + r},
  \frac{m n s}{n - m - r}, \frac{m - n + r}{s}}$, respectively. The
extremal diagrams of the corresponding twisted topological submodules
in $\mU$ are labelled by $\mC_1$ and $\mC_2$ respectively.  The
top-level representatives~\req{top-level-charged} are marked with
crosses.  The extremal diagrams of the twisted topological Verma
submodules generated from $\ket{e_1}$ and $\ket{e_2}$ respectively are
given by {\bf 3}--$\ket{e_1}$--{\bf 4}, with the cusp at $\ket{e_1}$,
and by {\bf 5}--$\ket{e_2}$--{\bf 6}, with the cusp at~$\ket{e_2}$.
Thus, neither $\ket{e_1}$ nor $\ket{e_2}$ {\it alone\/} generates all
of the states in {\bf 3}--$\ket{e_2}$--$\ket{e_1}$--{\bf 5}.  Those
states of the two submodules that lie between $\ket{e_1}$ and
$\ket{e_2}$ are in the same grade and are linearly independent. They
satisfy twisted massive \hw{} conditions and might thus be taken for
two linearly independent massive singular vectors in the same grade
(when, e.g., such $\ket{e_1}$ and $\ket{e_2}$ happen to lie on
different sides of the top of the parabola, one would observe the pair
of conventional singular vectors~\cite{[Doerr2]} at the top of the
parabola).  However, we have seen that each of the two linearly
independent states in the same grade belongs in fact to its own
twisted {\it topological\/} Verma submodule.

The state at {\bf 7}${}\in\mC'_1$ is yet another topological singular
vector from part \ref{II-ii} of the Theorem, in particular
$\cQ_{r+m}\,{\bf 7}=0$.  The state $\ket{s_1}\in \mC'_1$ (not
indicated in the diagram), which is in the same grade as
$\ket{e_2}\in\mC'_2$ but belongs to the other twisted topological
submodule, is such that $\cG_{-r-m}\,\ket{s_1}={\bf 7}$.  (Similarly,
${\bf 9}\in \mC'_2$ is a topological singular vector as well).  The
state {\bf 8} is the top-level representative of the extremal diagram
generated by the topological singular vector~$\ket{e_2}$, but is not
in the module generated from~$\ket{e_1}$.

\medskip

Similarly, in case \ref{II-i} of Theorem~\ref{thm:3last}, we have the
extremal diagram
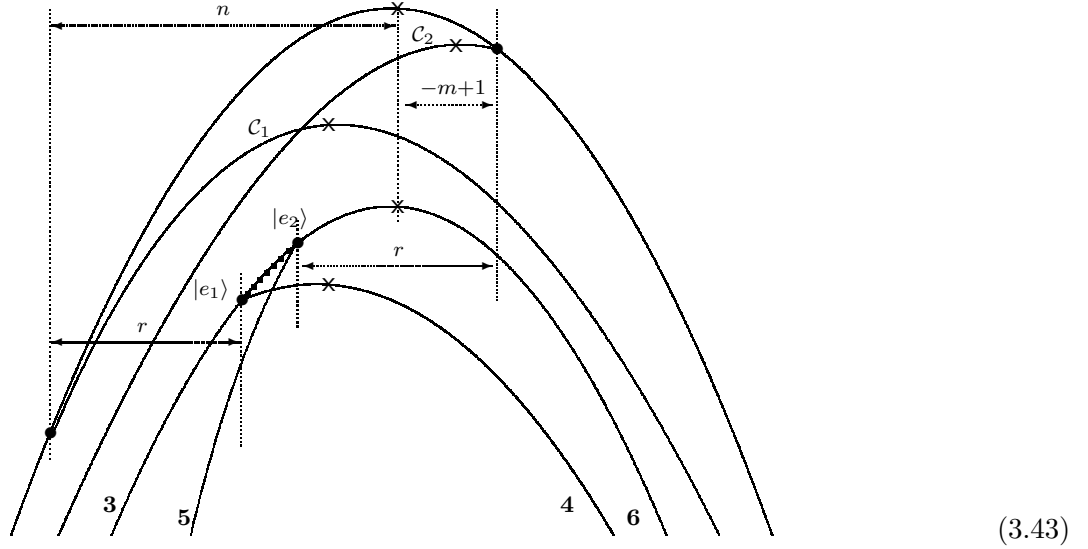
\begin{equation}
  \message{please wait...}
  \unitlength=1pt
  \begin{picture}(300,200)
    \bezier{\dimtwenty}(2,1)(146,400)(290,1)
    \put(146,198){${\ssf x}$}
    \put(146,123){${\ssf x}$}
    \put(120,154){${\ssf x}$}
    \put(120,93.4){${\ssf x}$}
    \put(168,184){${\ssf x}$}
    \put(14.5,37){$\bullet$}
    \put(183.5,182.5){$\bullet$}
    \put(108,109){$\bullet$}
    \put(87.0,87.5){$\bullet$}
    \bezier{\dimtwelve}(18,40)(125,291)(270,1) 
    \bezier{\dimtwelve}(20,1)(110,205)(185.5,185)
    \bezier{\dimten}(40,1)(146,250)(250,1)
    \bezier{200}(70,1)(86,70)(110,111.5)
    \bezier{\dimten}(89,90)(160,120)(230,1)  
    {\linethickness{1.8pt}
      \bezier{8}(89,90)(97,100)(109.5,111.5)}
    \bezier{50}(148.3,200)(148.3,165)(148.3,120)
    \put(-131.5,0){\bezier{100}(148.3,200)(148.3,124)(148.3,30)}
    \put(37.6,0){\bezier{80}(148.3,200)(148.3,144)(148.3,90)}
    \put(-38.0,0){\bezier{25}(148.3,120)(148.3,100)(148.3,80)}
    \put(-59.3,0){\bezier{40}(148.3,100)(148.3,64)(148.3,35)}
    \put(0,-20){
      \bezier{80}(18,94)(50,94)(87,94)
      \put(18.5,94){\vector(-1,0){1}}
      \put(87.5,94){\vector(1,0){1}}
      \put(50,94){${}^r$}
      }
    \put(0,100){
      \bezier{120}(18,94)(80,94)(146,94)
      \put(18.5,94){\vector(-1,0){1}}
      \put(146.5,94){\vector(1,0){1}}
      \put(80,94){${}^n$}
      }
    \put(67,9){
      \bezier{80}(47,94)(100,94)(116,94)
      \put(46.5,94){\vector(-1,0){1}}
      \put(116.5,94){\vector(1,0){1}}
      \put(80,94){${}^r$}
      }
    \put(67,70){
      \bezier{20}(86,94)(100,94)(116,94)
      \put(85.5,94){\vector(-1,0){1}}
      \put(116.5,94){\vector(1,0){1}}
      \put(90,94){${}^{-m+1}$}
      }
    \put(92,149){${}^{\mC_1}$}
    \put(154,184){${}^{\mC_2}$}
    \put(71,88){${}^{\ket{e_1}}$}
    \put(101,115){${}^{\ket{e_2}}$}
    \put(37,10){\footnotesize\bf 3}
    \put(210,10){\footnotesize\bf 4}
    \put(65,5){\footnotesize\bf 5}
    \put(235,5){\footnotesize\bf 6}
  \end{picture}
  \message{done}
  \label{gap}
\end{equation}
where {\bf 3}--$\ket{e_1}$--{\bf 4} is the extremal diagram of the
twisted topological Verma submodule $\mC'_1=\mC_1\cap\mU'$, and {\bf
  5}--$\ket{e_2}$--{\bf 6}, that of $\mC'_2=\mC_2\cap\mU'$.  There are
$-m+n-2r$ states between $\ket{e_1}$ and $\ket{e_2}$ that satisfy
twisted massive \hw{} conditions but {\it do not belong to either
  $\mC'_1$ or $\mC'_2$ submodules\/}, nor, in fact, {\it to either the
  $\mC_1$ or $\mC_2$ submodules\/} of~$\,\mU_{h,\ell,t}$.  Thus, these
states survive in the quotient module with respect to $\mC_1$ and
$\mC_2$.  It is these states that generate the entire massive
submodule {\bf 3}--$\ket{e_1}$--$\ket{e_2}$--{\bf 6}, in which
$\ket{e_1}$ and $\ket{e_2}$ are charged singular vectors.

\bigskip

As regards describing the above pictures in terms of only the
top-level representatives of singular vectors and the subsingular
vectors, one has to consider submodules of the submodules described
above generated by the conventional, top-level, representatives of the
extremal diagrams. The missing parts of the submodules will then be
generated by subsingular vectors.  For the `traditional' reasons, we
now briefly describe the subsingular vectors `hidden' in the above
pictures.

To begin with case~\ref{II-ii}, consider what happens in \req{overlap}
after factoring away the submodule generated from the top-level
($\times$) representative of the extremal diagram {\bf
  3}--$\ket{e_1}$--{\bf 4} of the~\,$\mC'_1$ submodule.  The vector
$\ket{s_1}$ (not shown in~\req{overlap}) that lies at the same grade
as $\ket{e_2}\in\mC'_2$ but belongs to \,$\mC'_1$ then satisfies
twisted topological \hw{} conditions, thus giving rise to (the
extremal diagram of) a twisted topological Verma submodule. The
top-level representative of this extremal diagram is then a
subsingular vector in the conventional sense.  This top-level
representative would be at the same point as {\bf 8}, the top-level
representative of the topological singular vector $\ket{e_2}$ in
$\mU$.

Thus, the top-level representative of $\ket{s_1}$ does not belong to
the submodule generated by the top-level representative of $\mC'_1$
because of the topological \hw{} conditions at~{\bf 7}.  We also see
that this subsingular vector lies in the same grade as~{\bf 8}. Thus
-- continuing with the conventional definition of singular vectors --
a singular vector and a subsingular vector are in the same grade in
the case under consideration. The crucial feature of this case is that
the entire sections {\bf 7}--$\ket{e_1}$ and $\ket{e_2}$--{\bf 9} of
the extremal diagrams of each of the topological submodules are on one
side of the top of the parabola.  Had these sections included the
top-level representative, one would conclude that two conventional
singular vectors exist in the same grade. Whether or not this is the
case is determined by parameters $r$, $s$, $m$, and $n$. As they
change, one of these conventional singular vectors `submerges' and
becomes {\it sub\/}singular.

{}From the above discussion we immediately obtain the following three
Propositions.
\begin{prop}\label{prop:II-ii}
  Let the conditions of case \ref{II-ii} of Theorem~\ref{thm:3last}
  hold. Then, subsingular vectors exist in \ $\mU\equiv\mU_{\frac{(1 -
      m - n) s}{m - n + r}, \frac{m n s}{n - m - r}, \frac{m - n +
      r}{s}}$ if and only if either $r<-m+1$ or $r<n$. In the first
  case, the subsingular vector reads
  \begin{eqnarray}
      \ket{\rm Sub}\kern-5pt&=&\kern-5pt
      \cG_0\,\ldots\cG_{-r-m-1}\cdot\cG_{-r-m+1}\ldots\cG_{r-n-1}\,
      \ket{e_1}\label{subsingIIii}\\
      {}\kern-5pt&=&\kern-5pt{}
      \cG_0\,\ldots\cG_{-r-m-1}\cdot\cG_{-r-m+1}\ldots\cG_{r-n-1}\,
      \cE^{-,-n}(r,s,\frac{m - n + r}{s})\,\cG_{-n}\ldots\cG_{-1}\,
      \ket{\frac{(1 - m - n) s}{m - n + r}, \frac{m n s}{n - m - r},
        \frac{m - n + r}{s}}\nonumber
  \end{eqnarray}
  (where $\cE^{-,\theta}(r,s,t)$ is the spectral flow transform of the
  topological singular vector operator read off from \req{Tminus}).
  This becomes singular in the quotient module $\mU/\mC''_1$, where
  $\mC''_1$ is the submodule generated by the top-level
  representative of the extremal diagram {\bf 3}--{\bf 7}, which reads
  $$
  \cG_{0}\ldots\cG_{r-n-1}\,
  \cE^{-,-n}(r,s,\frac{m - n + r}{s})\,\cG_{-n}\ldots\cG_{-1}\,
  \ket{\frac{(1 - m - n) s}{m - n + r}, \frac{m n s}{n - m - r},
    \frac{m - n + r}{s}}\,.
  $$
\end{prop}
In the other case, the subsingular vector (in the conventional
nomenclature) is given by an equally simple construction.

\medskip

Similarly, describing case \ref{II-i} of Theorem~\ref{thm:3last} in
terms of only top-level representatives of singular vectors (and then,
in terms of subsingular vectors) one would conclude that whenever
$r\geq|m|+1$ or $r\geq n$ the states between $\ket{e_1}$ and
$\ket{e_2}$ in the diagram~\req{gap} are not generated from the
top-level representatives. The top-level representative of the state
$\widetilde{\ket{v_2}}$ (such that
$\cQ_{r+m}\,\widetilde{\ket{v_2}}=\ket{e_2}$) gives a subsingular
vector:
\begin{prop}
  Under the conditions of case \ref{II-i} of Theorem~\ref{thm:3last},
  the state
  \begin{eqnarray}
      \ket{\rm Sub}\kern-6pt&=&\kern-6pt
      \cQ_1\ldots\cQ_{r+m-1}\widetilde{\ket{v_2}}\nonumber\\
      {}\kern-6pt&=&\kern-6pt\cQ_1\ldots\cQ_{r+m-1}\,
      q(r+m+1,r+m-1)\,\ket{e_2}\label{eq:last}\\
      {}\kern-6pt&=&\kern-6pt{}
      \cQ_1\ldots\cQ_{r+m-1}\,
      q(r+m+1, r+m-1)\,
      \cE^{+,-m}(r,s,\frac{n - m + r}{s})\,\cQ_m\ldots\cQ_0\,
      \ket{\frac{(1 - m - n) s}{n - m + r}, \frac{m n s}{m - n - r},
        \frac{n - m + r}{s}}\nonumber
  \end{eqnarray}
  is a subsingular vector in the module \,$\mU\equiv\mU_{\frac{(1 - m
      - n) s} {n - m + r}, \frac{m n s}{m - n - r}, \frac{n - m +
      r}{s}}$ with $r\geq|m|+1$. It becomes singular in the quotient
  module $\,\mU/(\mC_1\cup\,\mC_2)$, where \ $\mC_1$ and \ $\mC_2$ are
  submodules in $\,\mU$ generated by the charged singular vectors
  $\ket{E(n, \frac{(1 - m - n) s}{n - m + r}, \frac{n - m +
      r}{s})}_{\rm ch}$ and $\ket{E(m, \frac{(1 - m - n) s}{n - m +
      r}, \frac{n - m + r}{s})}_{\rm ch}$ respectively.
\end{prop}
The length-$(-1)$ operator $q(r+m+1, r+m-1)$ evaluates according to
the rules given in Sec.~\ref{AlgI}. When $r\geq n$, the subsingular
vector is built similarly starting from the vector
$\widetilde{\ket{v_1}}=g(r-n+1,r-n-1)\ket{e_1}$.

Finally, even if one insists on considering only top-level
representatives of singular vectors, there would be no subsingular
vectors in case \ref{I} of the Theorem, because topological states
in the extremal diagram of the submodule have relative charges of
different signs, hence the entire maximal submodule can be generated
from the top-level representative.
\begin{prop}
  Under the conditions of case~\ref{I} of Theorem~\ref{thm:3last}, no
  subsingular vectors exist in $\mU_{h,\ell,t}$.
\end{prop}

\section{Concluding remarks}\lvm
We have analyzed the structure of $\N2$ Verma modules and classified
their degeneration patterns. We considered singular vectors that
generate maximal submodules (and satisfy twisted \hw{} conditions),
which has allowed us to describe the structure of submodules of $\N2$
Verma modules in a setting which is free of subsingular vectors.
However, in order to make contact with the approach existing in the
literature, we have also shown how the description in terms of the
conventional, `untwisted', singular vectors and the subsingular
vectors, as well as general expressions for the subsingular vectors,
follow from our approach and the expressions for the singular vectors
satisfying the twisted \hw{} conditions.

As we have seen, an important point about the structure of massive
$\N2$ Verma modules is that there are submodules of exactly two
different types, the massive and the twisted topological ones (and,
obviously, arbitrary sums thereof).
The existence of two types of submodules shows up also in the
classification of the patterns describing possible sequences of
submodules {\it of submodules\/} of a given $\N2$ Verma module, which
we have not considered yet.  This amounts to finding embedding
diagrams of $\N2$ Verma modules. Using the singular vectors
constructed in this paper, these would be {\it embedding\/} diagrams,
i.e., those consisting only of mappings with trivial kernels. The
sought sequence in which submodules may follow one another is
determined by the degeneration patterns found in this paper. As
regards the topological Verma modules, the answer is already known,
since the corresponding embedding diagrams are isomorphic to those of
$\tSL2$ Verma modules.  Further, we have seen that some of the
degeneration patterns of massive $\N2$ Verma modules are such that,
again, the structure of submodules is determined by that of a certain
topological (hence, $\tSL2$-) Verma module. It thus remains to analyze
several cases where the known embedding diagrams are `glued together'
to produce somewhat more complicated structures~\cite{[SSi]}. The
classification of $\N2$ embedding diagrams is, thus, a refinement of
the classification of degeneration patterns presented in this paper.

\medskip

In view of the relation existing between $\tSSL21$ and $\N2$ singular
vectors~\cite{[S-sl21sing]}, it would also be interesting to see how
$\tSSL21$ {\it sub\/}singular vectors behave under the
reduction~\cite{[S-sl21sing]} to $\N2$ Verma modules.

\paragraph{Acknowledgements.} We are grateful to B.~Feigin and
V.~Sirota for many helpful discussions. We thank I.~Todorov for
pointing the paper~\cite{[Ga]} out to us and F.~Malikov for useful
remarks. AMS thanks D.~Leites, I.~Shchepochkina, and V.~Tolstoy for
useful remarks.  We are grateful to the Editor for the criticism that
helped us to improve the presentation. This work was supported in part
by the RFFI grant 96-01-00725. IYT was also supported in part by a
Landau Foundation grant, and AMS, by grant 93-0633-ext from the
European Community.

\addcontentsline{toc}{section}{Appendix}
\def\theequation{A.\arabic{equation}}
\subsection*{Appendix: The proof of part~iii) of
  Lemma~\protect\ref{chargemass:lemma}}\lvm This proof exploits
heavily the properties of extremal diagrams of both the twisted
topological and the massive Verma modules (in fact, the reader would
find the proof easier to read if he draws the parabolas, dense
descendants, etc., which we deal with in what follows). The idea of
the proof is to demonstrate that the converse leads to a contradiction
either with the `size' of twisted topological Verma modules (i.e., the
appearance of states with bigradings outside the extremal diagram) or
with the structure of submodules in twisted topological Verma modules
(Theorem~\ref{sl2n2}).  This argument is applied several times to the
topological Verma modules that are the quotients of the massive Verma
module with respect to the topological Verma modules whose
existence is established at previous steps of the proof.

To begin with, note that the module $\mC'$ cannot be embedded by
charged singular vectors in more than one massive submodule in $\mU$.
Indeed, the assumption that $\mU'\supset\mC'\subset\mU''$, where the
embeddings are given by charged singular vectors, leads to the
contradiction, because either $\mU'$ or $\mU''$ then necessarily has
states in the gradings outside the module $\mU$.  To see this, let
$\ket{v'}$ be the \hw{} vector of $\mC'$.  Let $\ket{v'}$ have the
twist parameter~$\theta'$ and lie in the bigrading $(\ell',h')$.
Then, the extremal states of the massive Verma submodule that contains
$\ket{v'}$ as one of its extremal states have to lie in bigradings
$(\ell,h)$, all of which satisfy one and only one of the following
equations:
\begin{eqnarray}
  \label{Qside}
  \half h^2 - (\half + h' + \theta') h+
  \half h' + \half{h'}^2 + \ell' + h' \theta'=\ell\,,\\
  \label{Gside}
  \half h^2 + (\half - h' - \theta') h
  -\half h' + \half {h'}^2+ \ell' + h' \theta'=\ell\,.
\end{eqnarray}
Now, the following alternative is satisfied:
\begin{itemize} \addtolength{\parskip}{-6pt}

\item[--] either infinitely many bigradings satisfying~\req{Qside} lie
  outside the module $\mU$, in which case none of the bigradings
  satisfying~\req{Gside} lie outside the module $\mU$,

\item[--] or infinitely many bigradings satisfying~\req{Gside} lie
  outside the module $\mU$, in which case none of the bigradings
  satisfying~\req{Qside} lie outside the module $\mU$
\end{itemize}
(the converse would contradict the fact that $\mU$ is freely
generated).
Thus, we can associate with each state $\ket{v'}$ two sets of
bigradings $(\ell^1_i,h^1_i)$ and $(\ell^2_i,h^2_i)$, each of which
satisfies one and only one of Eqs.~\req{Qside} and~\req{Gside}.
Infinitely many bigradings from one set lie outside~$\mU$, while all
bigradings from the other set lie inside~$\mU$.  We will call
bigradings from the latter set admissible with respect to the
state~$\ket{v'}$.

Thus, there may exist at most one massive submodule $\mU'$ into which
$\,\mC'$ is embedded by a charged singular vector and, moreover, all
of the extremal states of $\mU'$ have admissible bigradings with
respect to the \hw{} vector of $\,\mC'$ if such a $\mU'$ exists.  In
the case where there {\it is\/} such a massive Verma module, we are in
the situation described in Part~ii) of the Lemma, using which~iii) is
proved.  Consider, therefore, the case where
\begin{equation}
  \label{condition}
  \mbox{\parbox{.9\textwidth}{
      there does not exist a massive submodule $\mU'\supset\mC'$ such
      that the embedding is given by a charged singular vector.}}
\end{equation}
It is easy to see then that there exists a state $\ket{y}$ that
satisfies the following properties (the proof of this statement is
left to the reader as a useful exercise):
\begin{itemize} \addtolength{\parskip}{-6pt}

\item[] $\ket{y}$ has an admissible bigrading with respect
  to~$\ket{v'}$;

\item[] $\ket{y}$ satisfies twisted topological \hw{} conditions;

\item[] unless $\ket{y}=\ket{v'}$, the vector $\ket{v'}$ is a \ddsc{}
  of $\ket{y}$, while $\ket{y}$ is not a \ddsc{} of $\ket{v'}$;

\item[] there are no states $\ket{z}$ with admissible bigradings with
  respect to~$\ket{y}$ such that $\ket{y}$ is a \ddsc{} of $\ket{z}$,
  while $\ket{z}$ is not a \ddsc{} of $\ket{y}$.
\end{itemize}
It is clear that $\ket{y}$ generates a twisted topological Verma
module~$\mC''\supseteq\mC'$
\ ($\mC''=\mC'$ whenever $\ket{y}=\ket{v'}$) with the
condition~\req{condition} satisfied for~$\mC''$.  Now, we will have
proved~iii) for the module~$\mC'$ as soon as we prove~iii)
for~$\mC''$.  Let $\ket{y}$ have the twist parameter $\theta_y$ and
the bigrading~$(\ell_y,h_y)$.  Assume, for definiteness, that any
admissible bigrading with respect to~$\ket{y}$ satisfies~\req{Qside}
with $\theta'=\theta_y$ and $(\ell',h')=(\ell_y,h_y)$.  Then, the fact
that there are no states $\ket{z}$ with the properties as described
above is equivalent to the fact that the
expression~$g(\theta_y+1,\theta_y-1)\ket{y}$ cannot be evaluated as a
polynomial in the modes of $\cQ$, $\cG$, $\cL$, and $\cH$ acting on
the \hw{} vector of~$\mU$.  By lengthy but direct calculations with
formulae from Sec.~\ref{AlgI} one can show that
$g(\theta_y+1,\theta_y-1)\ket{y}$ cannot be evaluated in this way if
and only if there exists a twisted topological Verma module
$\mC_1\subset\mU$, where the embedding is given by a charged singular
vector and such that~$\mC_1$ is maximal
($\mU\supset\mC''\supset\mC\Longrightarrow\mC''=\mC$)
and~$\mC_1\cap\mC''=\{0\}$. \ Then, consider the quotient
$\mQ_1=\mU/\mC_1$. This is a twisted topological Verma module, which
contains the topological singular vector~$\ket{y}$.  It follows by
comparing the \hw{} parameters of $\mQ_1$ and $\mC_1$ that $\mC_1$
contains a topological singular vector~$\ket{x}$.  The bigrading of
$\ket{x}$ is~$(\ell_x,h_x)=(\ell_y+\theta_y,h_y-1)$ and any bigrading
admissible with respect to~$\ket{y}$ is admissible with respect
to~$\ket{x}$, and vice versa.  Let $\mC_1'$ be the twisted topological
Verma module generated from~$\ket{x}$.  We have two possibilities:
\begin{itemize} \addtolength{\parskip}{-6pt}
\item[a)] there does not exist a massive submodule
  $\mU'\supset\mC_1'$, where the embedding is given by a charged
  singular vector;

\item[b)] there exists a massive submodule $\mU'\supset\mC_1'$, where
  the embedding is given by a charged singular vector.
\end{itemize}
In case~a), we can apply to $\mC_1'$ the same reasoning as in the case
of the~$\mC''$ module.  In this way, we see that a module
$\mC_2\subset\mU$ exists, where the embedding is given by a charged
singular vector, $\mC_2$ is maximal
($\mU\supset\mC'''\supset\mC_2\Longrightarrow\mC'''=\mC_2$),
and~$\mC_2\cap\mC_1'=\{0\}$.  It is also easy to see
that~$\mC_1\cap\mC_2=\{0\}$.  Further, the quotient $\mU/\mC_2$ cannot
contain all of the extremal states of $\mC''$, since $\mU/\mC_2$ is a
twisted topological Verma module, which already contains all of the
extremal states of $\mC'_1$.  Therefore,
$\mC_2\cap\mC''=\mC_2'\neq\{0\}$ and the quotient $\mU/\mC_2$ contains
the submodule $\mC''/\mC_2'$.  However, this can happen only in the
case where $\mC''/\mC_2'=\{0\}$ or, equivalently,~$\mC''\subset\mC_2$,
from which~iii) follows.

To complete the proof, it remains to consider case b).  Let
$\mU'\subset\mU$ be a massive submodule such that $\ket{x}$ is the
charged singular vector in~$\mU'$.  This means that there exists a
state $\ket{z}$ with an admissible bigrading with respect to~$\ket{x}$
such that~$\ket{x}$ is a \ddsc{} of $\ket{z}$, whereas $\ket{z}$ is
not a \ddsc{} of $\ket{x}$.  Since~$\ket{z}$ has an admissible
bigrading with respect to~$\ket{x}$ as well as with respect
to~$\ket{y}$, and, also, $(\ell_x,h_x)=(\ell_y+\theta_y,h_y-1)$, the
state~$\ket{z}$
can be represented in the form $\ket{z}=\ket{w}+a\ket{u}$, where
$\ket{u}$ is a \ddsc{} of $\ket{y}$ and $a\in\oC$.  Further,
$\ket{w}=0$ in the quotient $\mU/\mC_1$, since otherwise we are in
contradiction with the structure of the twisted topological Verma
module~$\mU/\mC_1$. Thus, we see that either $\ket{w}\in\mC_1$ or
$\ket{w}=0$.  However, $\mC_1$ cannot contain all of the bigradings
that are admissible with respect to~$\ket{x}$, therefore there exists
$\ket{z'}\in\mC''$ such that it is a \ddsc{} of $\ket{z}$. We now see
that $\mU'\cap\mC''\neq\{0\}$ or, equivalently, there exists
$\bar{\mC}\subseteq\mC''$ such that $\bar{\mC}\subset\mU'$, where the
embedding is given by a charged singular vector.  {}From part~ii) of
the Lemma, it follows that there exists \,$\mC_2\subset\mU$, where
\,$\mC_2$ is a submodule generated from a charged singular vector in
\,$\mU$, \,$\mC_2$ is maximal
($\mU\supset\mC'''\supset\mC_2\Longrightarrow\mC'''=\mC_2$) and
\,$\bar{\mC}\subset\mU'\cap\mC_2$.  We now have
$\mC_2\cap\mC''\neq\{0\}$, which allows us to repeat the arguments
regarding taking the quotient and, thus, to obtain~iii).

\small


\paragraph{Note added.} The fact that the charged singular vectors do
not generate the massive Verma modules was used in the recent
paper~[D2]. As we saw in Theorem~\ref{thm:charged}, one can be
considerably more precise by saying that this is a twisted topological
Verma module with the twist parameter~$-n$, where $n$ labels the
charged singular vector.  Similarly with the statement of~[D2]
regarding the degenerate case with two linearly independent singular
vectors in the same grade: as we have seen, such vectors generate a
direct sum of two twisted topological Verma modules, which makes the
``fermionic uncharged singular vectors'' introduced in~[D2] excessive.
The conditions for the absence of subsingular vectors applied in that
paper to the derivation of characters of unitary representations, are
a particular case of conditions of Proposition~\ref{prop:II-ii}.

\begin{itemize}\itemindent0pt
\item[{[}D2{]}] M.~D\"{o}rrzapf, {\it The embedding structure of
    unitary $N=2$ minimal models}, hep-th/9712165.
\end{itemize}

\end{document}